\documentclass[%
    superscriptaddress,
    unsortedaddress,
    preprintnumbers,
    amsmath,amssymb,
    prb,
    floatfix,
]{revtex4-2}

\usepackage[dvipsnames]{xcolor}
\usepackage{graphicx}
\usepackage{rotating}
\usepackage{physics}

\usepackage{tikz}
\usetikzlibrary{calc}
\usetikzlibrary{arrows.meta}
\usetikzlibrary{positioning}
\usetikzlibrary{backgrounds}
\usetikzlibrary{shapes.arrows}
\usetikzlibrary{shapes.geometric}
\usetikzlibrary{decorations.markings}
\usetikzlibrary{3d}
\tikzset{fontscale/.style = {font=\relsize{#1}}}

\definecolor{MainColor}{HTML}{2574A9}
\definecolor{CompColor}{HTML}{FFA12B}
\definecolor{MyOrange}{HTML}{f5a55a}
\definecolor{MyRed}{HTML}{ff6969}
\definecolor{MyGreen}{HTML}{28798c}
\definecolor{MyBlue}{HTML}{70c7ff}

\newcommand{\rvline}{\hspace*{-\arraycolsep}\vline\hspace*{-\arraycolsep}}
\newcommand{\ttimes}{\ensuremath{\setlength{\medmuskip}{0mu}\times}}

\graphicspath{{figures/}}

\begin{document}

% Author information 
% ==================
\author{Nils Wittemeier}
  \affiliation{Catalan Institute of Nanoscience and Nanotechnology - ICN2 (CSIC, BIST), Campus UAB, 08193, Bellaterra, Spain}%
  \email{nils.wittemeier@icn2.cat}
\author{Nick Papior}%
  \affiliation{DTU Computing Center, Technical University of Denmark, DK-2800 Kongens Lyngby, Denmark}%
\author{Mads Brandbyge}%
  \affiliation{Department of Physics, Technical University of Denmark, DK-2800 Kongens Lyngby, Denmark}%
  \affiliation{Center for Nanostructured Graphene (CNG), DK-2800 Kongens Lyngby, Denmark}%
\author{Zeila Zanolli}%
  \affiliation{Chemistry Department and, Debye Institute for Nanomaterials Science, Condensed Matter and Interfaces, Utrecht University and ETSF, PO Box 80.000, 3508 TA, Utrecht, the Netherlands}%
\author{Pablo Ordejón}%
  \affiliation{Catalan Institute of Nanoscience and Nanotechnology - ICN2 (CSIC, BIST), Campus UAB, 08193, Bellaterra, Spain}%

% Title 
% ==================
\title[Quantum Transport with Spin Orbit Coupling]{Quantum Transport with Spin Orbit Coupling: New Developments in TranSIESTA} 

\begin{abstract}
We present the implementation of spinor quantum transport within the non-equilibrium Green’s function (NEGF) code TranSIESTA based on Density Functional Theory (DFT).
First-principles methods play an essential role in molecular and material modelling, and the DFT+NEGF approach has become a widely-used tool for quantum transport simulation. 
Exisiting (open source) DFT-based quantum transport codes either model non-equilibrium/finite-bias cases in an approximate way or rely on the collinear spin approximation. 
Our new implementation closes this gap and enables the TranSIESTA code to use full spinor-wave functions. Thereby it provides a method for transport simulation of topological materials and devices based on spin-orbit coupling (SOC) or non-collinear spins. These materials hold enormous potential for the development of ultra-low energy electronics urgently needed for the design of sustainable technology. 
The new feature is tested for relevant systems determining magnetoresistance in iron nanostructures and transport properties of a lateral transition metal dichalcogenide heterojunction. 

\end{abstract}
\maketitle

\section{Introduction}
The energy consumption of information and communication technology has increased steadily over the last decades and is projected to exceed 20\% of the global energy consumption by 2030\cite{JonesHowStop2018}. This makes the development of smaller and more efficient devices for memory storage and information processing a crucial challenge in the pursuit of sustainability.

A possibility to address this issue is to design devices that exploit the spin degrees of freedom (spintronics), as magnetic phenomena occur on an energy scale which is an order of magnitude smaller than electronic ones.
\,\cite{ZuticSpintronicsFundamentals2004,TrassinLowEnergy2016,HirohataReviewSpintronics2020,PueblaSpintronicDevices2020}
Spintronics has lead to the discovery of novel devices such as nonvolatile magnetic random-access memories or spin-polarized field effect transistors.\,\cite{SlonczewskiCurrentdrivenExcitation1996,BergerEmissionSpin1996,KatineCurrentDrivenMagnetization2000,WolfPromiseNanomagnetics2010,ZuticSpintronicsFundamentals2004,FertNobelLecture2008,WolfPromiseNanomagnetics2010,KawaharaSpintransferTorque2012,KhvalkovskiyBasicPrinciples2013,GajekSpinTorque2012,KentNewSpin2015} However, neither first-generation (based on manipulation of spins via magnetic fields) nor second-generation (based on spin-transfer torque) spintronics devices have reached efficiencies required to stop or slow the rapid increase in energy consumption of global data processing. 
The ongoing ``second quantum revolution'' promises to bring about a new breakthrough by exploiting quantum physics directly, rather than using it as a tool to observe and understand the world. 
Quantum materials, i.e. materials which manifests quantum effects over large length scales, are at the forefront of this development. This class of materials includes Dirac materials, topological insulators (TI), and superconductors.
Developing spintronics devices based on quantum materials may lead to energy-efficient applications beyond dissipation-less charge transport. \cite{HeTopologicalSpintronics2022}.
Accurate simulations of transport characteristics of these materials and derived devices must include spin-orbit effects or non-collinear spin configuration. 
Today, electronic structure calculations are widely used to design new devices and investigate novel materials.
Density functional theory (DFT) can reliably model a wide range of systems without any system dependent parameters.\,\cite{JonesDensityFunctional2015,KapilReviewArticle2020,GeerlingsConceptualDensity2003}
These approaches are limited to periodic or finite systems in equilibrium. 
However, modeling electronic transport generally requires non-periodic geometries: a device connected to one or more electrodes. Moreover, the transport problem often includes non-equilibrium conditions. In ground-state DFT, the electron density is simply obtained by filling the Kohn–Sham states up to the Fermi level. Such a simple relation between occupations and states does not hold in the non-equilibrium situation. Hence, the ground-state DFT approach needs to be extended to describe quantum transport or open quantum systems in general.
The non-equilibrium Green's function (NEGF) formalism offers a way to make this extension possible. The DFT+NEGF approach\cite{BrandbygeDensityfunctionalMethod2002,RochaTheoreticalComputational2007}
has been implemented in large number of 
codes\,\cite{TaylorInitioModeling2001,BrandbygeDensityfunctionalMethod2002,PalaciosFirstprinciplesApproach2002,WortmannInitioGreenfunction2002,RochaMolecularSpintronics2005,Garcia-LekueElasticQuantum2006,WohlthatInitioStudy2007,PecchiaNonequilibriumGreen2008,SahaFirstprinciplesMethodology2009,OzakiEfficientImplementation2010,ChenInitioNonequilibrium2012,FerrerGOLLUMNextgeneration2014,GomezInitioElectronic2021,GarciaNonequilibriumMagnetoConductance2023}
and has been employed successfully to wide range of systems\,\cite{TaylorInitioModeling2001a,CorbelInitioCalculations1999,JoonChoiInitioPseudopotential1999}.
The majority of open source, quantum transport codes either rely on the collinear spin-approximation or can only model equilibrium transport, covering the non-equilibrium/finite-bias cases in an approximate way. 
However, full spinor wave function are required to model topological materials and advanced spintronics devices exploiting non-collinear spin configurations and SOC. 

In this work we 
%review the state of the art in first-principles electronic quantum transport and 
present the implementation of the DFT+NEGF approach for general spinors in the open-source code TranSIESTA\,\cite{BrandbygeDensityfunctionalMethod2002,PapiorImprovementsNonequilibrium2017}. TranSIESTA combines the NEGF formalism with the SIESTA DFT-method\,\cite{SolerSIESTAMethod2002,GarciaSiestaRecent2020}, including the solution of electrostatic equations, self-consistency of the NEGF density matrix and support for multi-terminal devices. 
Until now, TranSIESTA was implemented for nonequilibrium spin-unpolarized or collinear (independent) spin-channels in the calculations. 
In the collinear spin approximation, the Hamiltonian splits into two independent spin-blocks and transport calculations can be performed for each block separately.
In presence of SOC\,\cite{CuadradoFullyRelativistic2012} off-diagonal terms arise in the Hamiltonian, which couple the two spin channels, and an independent treatment is no longer possible. 
We address this issue by extending the NEGF matrices into $2\times 2$ blocks in spin-space and implementing support for all spin configurations supported.
The new features of the code are tested for three systems with strong spin orbit coupling and/or non-collinear spin configuration. Our new implementation enables TranSIESTA to simulate multi-electrode devices spin(orbit)-tronics devices and topological materials.

%Spin-orbit coupling is a relativistic correction to the Schrödinger equation. It causes the electron spin to no longer be a good quantum number and give rise to off diagonal terms in the Hamiltonian.
% ==============================================================================================
\section{Spin-orbit coupling}
% ==============================================================================================

% Spin-orbit coupling (SOC) is an important ingredient in spintronics applications\cite{ManchonNewPerspectives2015}. 
% %SOC is a relativistic correction to the Schrödinger equation, which originates from the interaction between an electrons spin magnetic moment, its orbital moment, and the electrostatic field of the positively charged nucleus.
% SOC couples the majority and minority electrons, and gives rise to various fundamental phenomena including the anomalous Hall effect and topological insulators. 
% Importantly, SOC can induce a conversion between spin and charge (spin conversion).\,\cite{OtaniSpinConversion2017} This provides a mechanism to manipulate electron spins without need for external magnetic fields.
% Therefore SOC is an important ingredient for future spintronics devices.
Spin-orbit coupling (SOC) is a relativistic effect originating from the interaction between the spin and orbital moment of an electron, and electric fields: 
In a simplified picture, an electron moving with momentum ($\mathbf{p}$) in an electric field ($\mathbf{E}$) experiences an effective magnetic field ($\mathbf{B}_{eff} \sim \mathbf{E}\times\mathbf{p}$).
This effective field gives rise to a momentum dependent Zeeman energy, the SOC, $\mathbf{H}_{SO} \sim (\mathbf{E}\times\mathbf{p}\cdot\mathbf{\sigma})$. 

Spin-orbit coupling is fundamental in understanding the fine structure of atoms\cite{WoodgateElementaryAtomic1992} and a variety of phenomena in solids. In solids the SOC is responsible for the coupling between magnetic moments (spin space) and the underlying crystal structure (real space).
It gives rise to magnetic anisotropy \cite{JohnsonMagneticAnisotropy1996}, 
spin relaxation \cite{WuSpinDynamics2010}, 
magnetic damping \cite{MillsSpinDamping2003}, 
anisotropic magnetoresistance \cite{McGuireAnisotropicMagnetoresistance1975}, 
and the anomalous Hall effect \cite{NagaosaAnomalousHall2010}.

\section{Spin degrees of freedom}
In the collinear spin case the spins are eigenstates of the spin operator $\mathbf{S}_z$. The Hamiltonian, overlap matrix and all other operators take on a block diagonal form
\begin{equation}
    A=\begin{pmatrix}
        A^\uparrow & \rvline & 0 \\
        \hline
        0 & \rvline & A^\downarrow
    \end{pmatrix}\label{eq:operator-collinear}
\end{equation}
where $A^\sigma$ is the submatrix for spin $\sigma\in\{\uparrow,\downarrow\}$. Majority ($\uparrow$) and minority ($\downarrow$) electrons are completely decoupled. In this case, the formalism can be applied to the two spin channels independently.
The total transmission and total current are given by sum of the two spin components.
\begin{align}
    T = T^\uparrow + T^\downarrow \\
    I = I^\uparrow + I^\downarrow.
\end{align}

In non-collinear spin-systems, electron spins are not aligned along a common axis.
Their wavefunctions %of these systems 
are no longer eigenstates of the spin operator $\mathbf{S}_z$ (note, that we will use the symbol $\mathbf S$ for both the overlap and spin matrices as their intent is clear from the context). In this case, the off-diagonal spin-blocks of any operator can be non-zero.
\begin{equation}
    A=\begin{pmatrix}
        A^{\uparrow\uparrow} & \rvline & A^{\uparrow\downarrow} \\
        \hline
        A^{\downarrow\uparrow} & \rvline & A^{\downarrow\downarrow}
    \end{pmatrix}.\label{eq:operator-noncollinear}
\end{equation}
The overlap matrix retains the form in Eq. \ref{eq:operator-collinear}.
In this case, the Green's function formalism has to be extended from independent spin channels to spinor wave functions. The case of non-collinear spins is essential for fully-relativistic DFT calculations, where SOC introduces non-zero terms in the off-diagonal spin-block of the Hamiltonian.
It should be noted that in non-collinear spin calculations without spin-orbit coupling the Hamiltonian and density matrix are ``spin-box Hermitian'':
% \begin{align}
    $H^{\sigma\sigma'}_{ij} = (H^{\sigma'\sigma}_{ij})^*$
    and
    $\boldsymbol{\rho}^{\sigma\sigma'}_{ij} = (\boldsymbol{\rho}^{\sigma'\sigma}_{ij})^* $
.
% \end{align}

% ==============================================================================================
\section{Non-equilibrium Green's function formalism}
% ==============================================================================================
\begin{figure}
    \centering
    \includegraphics[width=0.65\textwidth]{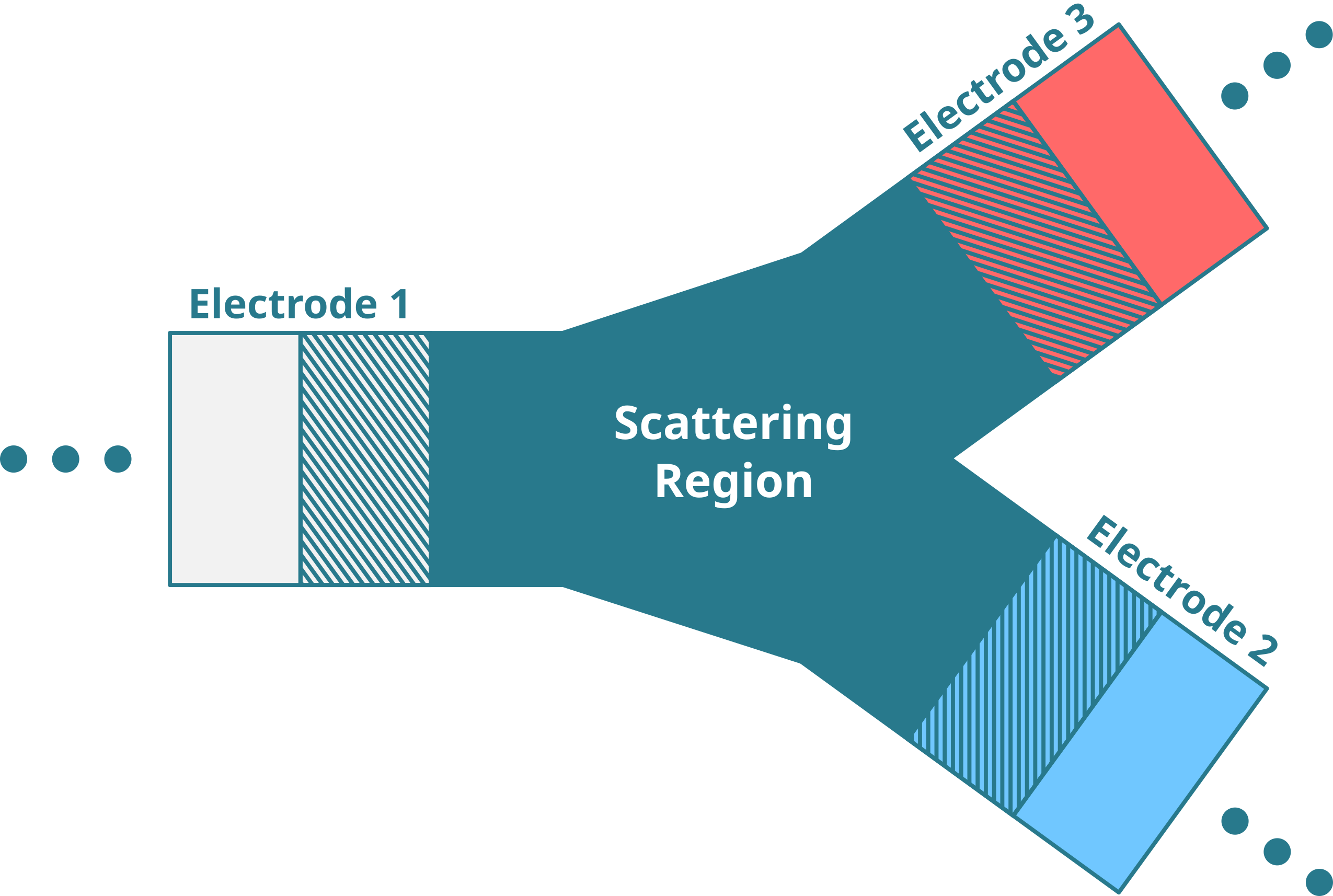}
    \caption{Sketch of a setup for 3-electrode transport simulation. The three electrodes are depicted in three different colors (grey, blue, red) these electrodes extend infinitely in the direction indicated by the dots. The striped regions denote the electrode screening regions, which are used to screen the perturbation of the scattering region. Screening is required to ensure an effective potential similar to the bulk potential of the electrodes at the edge of the scattering region.}
    \label{fig:transport_setup}
\end{figure}

The non-equilibrium Green's function (NEGF) formalism describes an open quantum system consisting of a scattering region in contact with one or more electrodes.\,\cite{DattaElectronicTransport1995} These electrodes act as reservoirs in which electrons are at (local) equilibrium. Combining the NEGF formalism with the framework of DFT, allows us to obtain self-consistent solutions for the electron density of such an open system from first-principles. Furthermore, the NEGF formalism can be used to calculate the non-equilibrium effects in the scattering region. These include electric currents between the electrodes, which can emerge when the electrodes are not in equilibrium with each other.

Figure~\ref{fig:transport_setup} depicts a sketch of a generic multi-electrode transport setup consisting of the scattering and electrode regions. We assume that the electrode regions extend regularly and infinitely in the direction of the dots. We further assume that electrode regions are bulk-like, i.e. unperturbed by the presence of the scattering region.
To achieve bulk-like behavior it is necessary to introduce screening regions between each electrode and the scattering region. While the atomic structure in the screening regions may be equal to the electrodes they should be thought of as part of the scattering region, for which the electronic structure is unknown. The Hamiltonian describing such a system is an infinite, hermitian matrix. 

A central idea of the NEGF formalism is to effectively re-normalise this Hamiltonian by eliminating the degrees of freedom of the electrodes and describe the transport system in terms of the retarded Green's function. The re-normalised Hamiltonian of the scattering region includes the effective interaction with electrodes in form of the surface self-energies $\Sigma_{\mathfrak{e}}$. While the self-energies take into account the the periodicity of the electrodes along the transport directions, periodic boundary conditions may also apply in the non-transport directions, e.g. in Fig.~\ref{fig:transport_setup} perpendicular to the depicted plane. One must utilize k-points in cases where the electrodes has periodic boundary conditions transverse to the transport direction. 

The retarded Green's function at energy $\epsilon$, $\mathbf{G}_\mathbf{k}(z)$, is
\begin{align}
    \mathbf{G}_\mathbf{k}(z) ={} &
        \Big[
            z \mathbf{S}_\mathbf{k}
            - \mathbf{H}_\mathbf{k}
            - \sum_{\mathfrak{e}}
                \mathbf{\Sigma}_{\mathfrak{e}, \mathbf{k}}(z)
        \Big]^{-1}
        , \quad \mathrm{with}\, z = \epsilon + i 0^+
    \label{eq:NEGF-G}
\end{align}
where $\mathbf{H}_\mathbf{k}$ ($\mathbf{S}_\mathbf{k}$) is the Hamiltonian (overlap) matrix of the central region at point $\mathbf{k}$ in reciprocal space corresponding to periodic boundary conditions in the directions transverse to the transport direction\cite{BrandbygeDensityfunctionalMethod2002}. Various approaches to calculating surface self-energies have been devised, and we refer to these works for further details.\,\cite{SanchoHighlyConvergent1985,ChangComplexBand1982,AllenGreenFunction1979,AllenGreenFunctions1979,WuGeneralRecursive1994,GalperinNumericalComputation2002,UmerskiClosedformSolutions1997,DyExactSolution1979,LeeSimpleScheme1981,TomfohrComplexBand2002}
It is important to note that the surface self-energies only depend on the bulk Hamiltonian and overlap matrices. For this reason the surface self-energies can be calculated from a converged, standard DFT Hamiltonian. 

Finally, the NEGF formalism expresses the electron density from the Green’s function of transport system and electrode self-energies
\begin{align}
    \boldsymbol{\Gamma}_{\mathfrak{e},\mathbf{k}}(z) ={} &
        i \Big[
            \boldsymbol{\Sigma}_{\mathfrak{e}, \mathbf{k}}(z)
            - \boldsymbol{\Sigma}^{\dagger}_{\mathfrak{e}, \mathbf{k}}(z)
        \Big]
    \label{eq:NEGF-Gamma}\\
    \mathcal{A}_{\mathfrak{e},\mathbf{k}}(z) ={} &
        \mathbf{G}_\mathbf{k}(z)
        \boldsymbol{\Gamma}_{\mathfrak{e},\mathbf{k}}(z)
        \mathbf{G}^{\dagger}_\mathbf{k}(z) 
    \label{eq:NEGF-A}\\
    \boldsymbol{\rho} ={} &
        \frac{1}{2\pi}
        \iint_{\mathrm{BZ}}\!\!\!\!\mathrm{d}\mathbf{k}\mathrm{d}\epsilon
            \sum_{\mathfrak{e}}
                \mathcal{A}_{\mathfrak{e}, \mathbf{k}}(e)
                n_{F,\mathfrak{e}}(\epsilon),
    \label{eq:NEGF-rho}
\end{align}
where $\mathbf{\Gamma}_{\mathfrak{e},\mathbf{k}}(z)$, and $\mathcal{A}_{\mathfrak{e},\mathbf{k}}(z)$ denote the broadening matrix, and spectral function of electrode $\mathfrak{e}$ respectively.
% Every electrode is characterized by a chemical potential $\mu_{\mathfrak{e}}$ and a temperature $T_{\mathfrak{e}}$.
The electrode itself is in equilibrium and occupation of states in the electrode is given by the Fermi distribution ($n_{F,\mathfrak{e}}$) with corresponding chemical potential ($\mu_{\mathfrak{e}}$) and temperature ($T_{\mathfrak{e}}$) of electrode $\mathfrak{e}$. Lastly, $\mathbf{\rho}$ is the non-equilibrium density matrix of the scattering region.  

It now becomes clear how the NEGF formalism can be combined with the DFT method to find self-consistent solutions for the density matrix. Starting from an initial guess for the density in the scattering region we determine the Kohn-Sham potential and construct the Hamiltonian in the scattering region. We then use Eq.~\ref{eq:NEGF-rho} to calculate the density matrix corresponding to the Hamiltonian. If the initial and final density matrices (and/or the Hamiltonian) coincide within a given threshold the calculation is converged. Otherwise, we mix the density matrices or the Hamiltonian to create an new guess, and repeat the procedure until convergence is achieved.

After finding a self-consistent solution for the density matrix,
the transmission functions ($T_{\mathfrak{e},\mathfrak{e}'}$) 
between any pair of electrodes ($\mathfrak{e},\mathfrak{e}'$) can be calculated using the scattering matrix formalism. 
\begin{align}
    T_{\mathfrak{e},\mathfrak{e}'}(\epsilon)
     ={}&\mathrm{Tr}\Big\{
     \mathbf{s}^{\dagger}_{\mathfrak{e},\mathfrak{e}',\mathbf{k}}\mathbf{s}_{\mathfrak{e},\mathfrak{e}',\mathbf{k}}
     \Big\}\\
     ={}&\int_{BZ}\!\!\!\!\mathrm{d}\mathbf{k}\,
     \mathrm{Tr}\Big\{
        \mathbf{\Gamma}_{\mathfrak{e},\mathbf{k}}(z)
        \mathbf{G}_\mathbf{k}(z)
        \mathbf{\Gamma}_{\mathfrak{e'},\mathbf{k}}(z)
        \mathbf{G}^{\dagger}_\mathbf{k}(z) 
    \Big\}
\end{align}
Here, the scattering matrix $\mathbf{s}_{\mathfrak{e},\mathfrak{e}',\mathbf{k}}$ at a given $\mathbf{k}$ is obtained from the Green’s function using the generalized Fisher-Lee relation~\cite{FisherRelationConductivity1981,DattaElectronicTransport1995} or the Lippmann-Schwinger equation~\cite{LippmannVariationalPrinciples1950,wangguo09}:
\begin{align}
    \mathbf{s}_{\mathfrak{e},\mathfrak{e}',\mathbf{k}} = 
        i\mathbf{\Gamma}^{1/2}_{\mathfrak{e},\mathbf{k}}(z)
        \mathbf{G}_\mathbf{k}(z)
        \mathbf{\Gamma}^{1/2}_{\mathfrak{e}',\mathbf{k}}(z)
        -\delta_{\mathfrak{e},\mathfrak{e}'} \mathbf{1}
        \label{eq.LippmannSchwinger}
\end{align}

Under non-equilibrium conditions the currents ($I_{\mathfrak{e},\mathfrak{e}'}$) can be calculated from the transmission function
\begin{align}
    I_{\mathfrak{e},\mathfrak{e}'} ={}
    \frac{G_0}{2|e|} &\int\mathrm{d}\epsilon
        \Big[n_{F,\mathfrak{e}}(\epsilon)
             - n_{F,\mathfrak{e}'}(\epsilon)\Big]
        T_{\mathfrak{e},\mathfrak{e}'}(\epsilon),
\end{align}
where $G_0 = 2e^2/h$ is the conductance quantum. We divide here by a factor of 2 to account for spin channels.

It is important to note that the NEGF formalism assumes that currents passing through the central region are not perturbing the electrodes which is assumed to be in equilibrium.

\section{Implementation details}
TranSIESTA\cite{BrandbygeDensityfunctionalMethod2002,PapiorImprovementsNonequilibrium2017} implements the DFT+NEGF approach based on the SIESTA\cite{SolerSIESTAMethod2002,GarciaSiestaRecent2020} method for density function theory simulations. 

\subsection{The SIESTA method}

The SIESTA method\,\cite{SolerSIESTAMethod2002} is an ideal starting point for the implementation of the DFT+NEGF approach. SIESTA is based upon strictly localized basis sets, which make it easy to defined local Hamiltonians and local Green's functions required in the NEGF formalism. 

In SIESTA, the basis orbitals are the product of real, spherical harmonics and numerically tabulated radial functions. The shape of the radial part is in principle arbitrary, but the solutions of the Schrödinger equation of isolated (pseudized) atoms are most common and often a good choice in terms of accuracy and efficiency.\,\cite{ArtachoLinearScalingAbinitio1999,AngladaSystematicGeneration2002,JunqueraNumericalAtomic2001}. 
The choice of these basis orbitals sets the SIESTA method apart from other DFT approaches. The atom-like character of the basis orbitals often means that the number of basis functions required is smaller compared to other methods. In addition, the strict localization of the orbitals leads to sparse matrices and sparse methods can be exploited to calculate the density matrix efficienctly.

\subsection{Spin degrees of freedom}
In the SIESTA code there are four different possible spin methods: unpolarized, polarized, non-collinear and spin-orbit. Spin unpolarized  calculations do not consider the spin degrees of freedom. Spin polarized calculations consider two independent spin channels, and non-collinear and spin-orbit calculations treat the spin channels as coupled. Spin-orbit calculations differ from non-collinear calculations in two aspects: (1) They are fully relativistic, i.e. they include SO interactions in addition to the Darwin and velocity correction terms. (2) They impose different symmetries on the Hamiltonian or the density matrix: no symmetries in the spin-orbit case, and ``spin-box Hermiticity'' ($A^{\sigma,\sigma'} = (A^{\sigma',\sigma})^{*}
$) in the non-collinear case\cite{GarciaSiestaRecent2020}.

We have now extended the TranSIESTA implementation to support simulations with non-collinear spins and spin orbit coupling. 
For this purpose we re-implement the core TranSIESTA routines. We expand the matrix element $A_{i,j}$ between orbitals $i$ and $j$ into a $2\ttimes2$ blocks rather than expanding the matrices in the form of Eq.~\ref{eq:operator-noncollinear}
\begin{equation}
    A =
    % Matrix 1
    \begin{pmatrix}
        A_{1,1} & A_{1,2} & \dots    \\
        \vdots  & \ddots  &          \\
        A_{N,1} & \dots   &  A_{N,N}
    \end{pmatrix}
    \longrightarrow \begin{pmatrix}
    % Matrix 2 -- row 1
    \begin{bmatrix}
        A^{  \uparrow\uparrow}_{1,1} & A^{  \uparrow\downarrow}_{1,1} \\
        A^{\downarrow\uparrow}_{1,1} & A^{\downarrow\downarrow}_{1,1} 
    \end{bmatrix} & 
    \begin{bmatrix}
        A^{  \uparrow\uparrow}_{1,2} & A^{  \uparrow\downarrow}_{1,2} \\
        A^{\downarrow\uparrow}_{1,2} & A^{\downarrow\downarrow}_{1,2}
    \end{bmatrix}
    & \dots \\
    % Matrix 2 -- row 2
    \vdots & \ddots & \\
    % Matrix 2 -- row 3
    \begin{bmatrix} 
        A^{  \uparrow\uparrow}_{N,1} & A^{  \uparrow\downarrow}_{N,1} \\
        A^{\downarrow\uparrow}_{N,1} & A^{\downarrow\downarrow}_{N,1}
    \end{bmatrix} & 
    \dots &
    \begin{bmatrix}
        A^{  \uparrow\uparrow}_{N,N} & A^{  \uparrow\downarrow}_{N,N} \\
        A^{\downarrow\uparrow}_{N,N} & A^{\downarrow\downarrow}_{N,N}
    \end{bmatrix}
    \end{pmatrix}
\end{equation}
This approach retains the smallest possible bandwidth of the Hamiltonian matrix, which is beneficial for the performance in sparse inversion algorithms.

% For our new implementation, we have added a set of new modules to the \textsc{Siesta} code base. These modules are labeled with the \texttt{\_spinor} postfix to indicate that they specifically address the spinor implementation.

% Every \textsc{TranSIESTA} calculation follows the same steps:
% \begin{enumerate}
%     \item For every energy point $z$ in a given contour and every $\vb{k}$ point:
%     \begin{enumerate}
%         \item Calculate the electrode surface self-energy: \texttt{m\_ts\_electrode\_spinor}
%         \item Calculate the Green's function and determine the contribution to the density matrix from current $\vb{k}$ and energy point: \texttt{m\_ts\_fullg\_spinor, }  \texttt{m\_ts\_fullk\_spinor}, \texttt{m\_ts\_trig\_spinor}, \texttt{m\_ts\_trik\_spinor}
%         \item If non-equilibrium conditions: weight contributions to density matrix:  \texttt{m\_ts\_dm\_update}
%     \end{enumerate}
%     \item Update the \textsc{Siesta} density matrix
%     \item Check convergence and mix: \textit{remains unchanged}
%     \item Repeat
% \end{enumerate}

After convergence of the self-consistency cycle in DFT-NEGF, the \textsc{TBTrans} code can be used to calculate the transmission, currents, density-of-states, and various other quantities. To have the same analysis tools available for spinor calculations, we have also extended the capabilities of \textsc{TBTrans}. 

\subsection{Calculating the Electrode Surface Self-Energy}
To calculate the electrode surface self-energy, we use the Sancho-Sancho-Rubio (SSR) algorithm, an iterative scheme for calculation surface self-energies by \textcite{SanchoHighlyConvergent1985}. This algorithm requires us to construct the Hamiltonian and overlap matrices for the principal layer and the corresponding interlayer matrices.
\begin{align}
    \mathbf{H}^{\mathfrak{e}}_{0,0}({\vb{k}}) &{}= \sum_{\vb{R}_{\perp}} \mathbf{H}^{\mathfrak{e}}_{0,0}(\vb{R}_{\perp}) e^{i\vb{k}\vb{R}_{\perp}} &
    \mathbf{S}^{\mathfrak{e}}_{0,0}({\vb{k}}) &{}= \sum_{\vb{R}_{\perp}} \mathbf{S}^{\mathfrak{e}}_{0,0}(\vb{R}_{\perp}) e^{i\vb{k}\vb{R}_{\perp}} \\
    \mathbf{H}^{\mathfrak{e}}_{0,1}({\vb{k}}) &{}= \sum_{\vb{R}_{\perp}} \mathbf{H}^{\mathfrak{e}}_{0,1}(\vb{R}_{\perp}) e^{i\vb{k}\vb{R}_{\perp}} &
    \mathbf{S}^{\mathfrak{e}}_{0,1}({\vb{k}}) &{}= \sum_{\vb{R}_{\perp}} \mathbf{S}^{\mathfrak{e}}_{0,1}(\vb{R}_{\perp}) e^{i\vb{k}\vb{R}_{\perp}},
\end{align}
where the sum runs over all lattice vectors perpendicular to the semi-infinite transport direction of the electrode $\mathfrak{e}$. At this stage, we convert the Hamiltonian matrix of the electrodes from their usual sparse matrix format to the dense matrix format in equation \ref{eq:operator-noncollinear}.
%A dummy code for the conversion between the sparse matrix and the dense matrix for a specific $\mathbf{k}$ point can be found in listing \ref{list:h-sparse-to-h-k}.
After determining the Hamiltonian and overlap matrices in k-space, we reuse the existing implementation of the SSR algorithm, which is agnostic to the difference of spin and orbitals indices. Therefore, it can be reused completely.
%by passing matrices of dimensions $2N\ttimes 2N$ to it, instead of $N\ttimes N$.
%This part of the workflow is implemented in \texttt{m\_ts\_electrode\_spinor}.  

By default, the surface self-energies for all energies and all $\mathbf{k}$ points are calculated during the initialization of \textsc{Siesta}, written to a file and later read when needed. Alternatively, the self-energies can be calculated \textit{on demand}. The latter option significantly reduces the amount of disk space used by \textsc{TranSIESTA}. However, calculating the self-energies only \textit{on demand}, means that the same calculation must be repeated at every step of the iterative procedure.

\subsection{Calculating the Green's Function}
To calculate the Green's function, we first set up its inverse $z\mathbf{S}_{\vb{k}} - \mathbf{H}_{\vb{k}} - \sum_{\mathfrak{e}} \boldsymbol{\Sigma}_{\mathfrak{e},\vb{k}}(z)$. 
For unpolarized and collinear spin calculations \textsc{TranSIESTA} supports three inversion methods: full matrix inversion using (Sca)LAPACK~\cite{AndersonLAPACKUsers1999,BlackfordScaLAPACKUsers1997}, block-tridiagonal (BTD) inversion method~\cite{PapiorImprovementsNonequilibrium2017}, and MUMPS~\cite{AmestoyFullyAsynchronous2001,AmestoyPerformanceScalability2019}. For spinor calculations, we have only implemented support for full matrix inversion using (Sca)LAPACK and the BTD method, because the BTD method is by far the most efficient~\cite{PapiorImprovementsNonequilibrium2017} and the simplicity of the full inversion method makes it ideal as a reference. 

We convert the Hamiltonian and overlap matrices of the device region for a given $\mathbf{k}$ point using a similar procedure used to set up the $\mathbf{k}$ point matrices for the SSR algorithm. For the full inversion method, we set up the inverse Green's function as a dense matrix, and for the block-tridiagonal inversion method as a sparse matrix. In either case, the matrix is structured as 
\begin{equation}
    % \extrarowheight6pt
    A=\left(\begin{array}{cc|cc|c}
        A^{\uparrow\uparrow}_{1,1} & A^{\uparrow\downarrow}_{1,1} &
        A^{\uparrow\uparrow}_{1,2} & A^{\uparrow\downarrow}_{1,2} & \cdots\\
        A^{\downarrow\uparrow}_{1,1} & A^{\downarrow\downarrow}_{1,1} & 
        A^{\downarrow\uparrow}_{1,2} & A^{\downarrow\downarrow}_{1,2} & \cdots
    \rule[-1.8ex]{0pt}{0pt} \\ \hline \rule{0pt}{3.5ex}
        A^{\uparrow\uparrow}_{2,1} & A^{\uparrow\downarrow}_{2,1} &
        A^{\uparrow\uparrow}_{2,2} & A^{\uparrow\downarrow}_{2,2} & \cdots\\
        A^{\downarrow\uparrow}_{2,1} & A^{\downarrow\downarrow}_{2,1} & 
        A^{\downarrow\uparrow}_{2,2} & A^{\downarrow\downarrow}_{2,2} & \cdots
    \rule[-1.8ex]{0pt}{0pt} \\ \hline \rule{0pt}{3.5ex}
        \vdots & \vdots & \vdots & \vdots & \ddots \\
        
    \end{array}\right)
\end{equation}
to keep the matrix bandwidth as small as possible and increase the efficiency of the matrix inversion method. Subsequently, we add the self-energy to set up the inverse Green's function.

% \subsection{Calculating the Transmission Function and Currents}

\subsection{Complex contour integration}

The Green's function in Eq. \ref{eq:NEGF-G} is analytical in the whole complex plane except for its poles along the real axis. These poles make the integration over energies in Eq. \ref{eq:NEGF-rho} numerically challenging. 
However, the integral can be simplified under equilibrium conditions ($n_{F,\mathfrak{e}}(\epsilon)=n_F(\epsilon)$ for all $\mathfrak{e}$):
\begin{equation}
\boldsymbol{\rho} =
        \frac{i}{2\pi}
        \iint_{\mathrm{BZ}}\!\!\!\!\mathrm{d}\mathbf{k}\mathrm{d}\epsilon
            \Big[
                \mathbf{G}_\mathbf{k}(z)
                - \mathbf{G}^{\dagger}_\mathbf{k}(z)
            \Big] n_{F}(\epsilon),\label{eq:NEGF-rho-eq}
\end{equation}
In this form it becomes clear that the integrand is analytical and the integral along the real axis can be replace using the residual theorem. 
\begin{equation}
    \oint\mathrm{d}\epsilon
        \Big[
            \mathbf{G}_\mathbf{k}(z)
            - \mathbf{G}^{\dagger}_\mathbf{k}(z)
        \Big] n_{F}(\epsilon)
    = -2\pi i k_B T \sum_{z_{\nu}}
            \mathbf{G}_\mathbf{k}(z_{\nu})
            - \mathbf{G}^{\dagger}_\mathbf{k}(z_{\nu})
\end{equation}
TranSIESTA supports square and circle contours and various quadratures (Newton-Cotes, Legendre polynomials and Tanh-Sinh quadrature) as well as the continued fraction method proposed by Ozaki et al.\,\cite{OzakiEfficientImplementation2010}.

In the non-equilibrium cases, i.e. when there are differences in the temperatures and/or chemical potentials of the electrodes, the integrand in Eq.~\ref{eq:NEGF-rho} is generally not analytical. In this case the residue theorem can only be applied to part of the integrand. We can rewrite the non-equilibrium integral as
\begin{align}
    \boldsymbol{\rho} ={}& \boldsymbol{\rho}_\mathfrak{e}^\mathrm{eq} + \sum_{\mathfrak{e}'\neq\mathfrak{e}} \boldsymbol{\Delta}_{\mathfrak{e},\mathfrak{e}'}
    \label{eq:NEGF-rho-neq}
    \\
    \boldsymbol{\rho}_\mathfrak{e}^\mathrm{eq} ={} &
    \frac{i}{2\pi}
        \iint_{\mathrm{BZ}}\!\!\!\!\mathrm{d}\mathbf{k}\mathrm{d}\epsilon
            \Big[
                \mathbf{G}_\mathbf{k}(z)
                - \mathbf{G}^{\dagger}_\mathbf{k}(z)
            \Big] n_{F,\mathfrak{e}}(\epsilon)
    \label{eq:NEGF-rho-neq_eq_part}
    \\
    \boldsymbol{\Delta}_{\mathfrak{e},\mathfrak{e}'} ={} &
    \frac{i}{2\pi}
        \iint_{\mathrm{BZ}}\!\!\!\!\mathrm{d}\mathbf{k}\mathrm{d}\epsilon
            \mathcal{A}_{\mathfrak{e},\mathbf{k}}(z) 
            \Big[n_{F,\mathfrak{e}'}(\epsilon)-n_{F,\mathfrak{e}}(\epsilon)\Big]
    \label{eq:NEGF-rho-neq_neq_part}.
\end{align}
In this form the equilibrium part of the density matrix (Eq.~\ref{eq:NEGF-rho-neq_eq_part}) can still be calculated using the residual theorem, and the non-equilibrium part (Eq.~\ref{eq:NEGF-rho-neq_neq_part}) only requires integration along the real axis in a finite ``bias window'' around $[\mu_{\mathfrak{e}};\mu_{\mathfrak{e}'}]$. The two occupations function are equal outside the bias window from min($\mu_\mathfrak{e}$) to max($\mu_\mathfrak{e}$) (with some correction for finite temperature). Therefore the integral along the real axis can be limited to this window, which makes this recast form of Eq.~\ref{eq:NEGF-rho} very suitable for numerical integration.

The choices of the electrode $\mathfrak{e}$ in Eq.~\ref{eq:NEGF-rho-neq} are all mathematically equivalent, making the choice of any electrode as a reference electrode arbitrary. To reduce numerical errors and avoid this arbitrariness, \textsc{TranSIESTA} weights the non-equilibrium density matrices calculated for each $\mathfrak{e}$:
\begin{align}
    \boldsymbol{\rho} ={}& \sum_{\mathfrak{e}} w_{\mathfrak{e}} \Big(\boldsymbol{\rho}_\mathfrak{e}^\mathrm{eq} + \sum_{\mathfrak{e}'\neq\mathfrak{e}} \boldsymbol{\Delta}_{\mathfrak{e},\mathfrak{e}'}\Big)\\
    \theta_{\mathfrak{e}} ={}& \sum_{\mathfrak{e}'\neq\mathfrak{e}} \mathrm{Var}[\boldsymbol{\Delta}_{\mathfrak{e},\mathfrak{e}'}]\\
    w_{\mathfrak{e}} ={}& \prod_{\mathfrak{e}'\neq\mathfrak{e}}\theta_{\mathfrak{e}'} / \sum_{\mathfrak{e}'}  \prod_{\mathfrak{e}''\neq\mathfrak{e}'}\theta_{\mathfrak{e}''}.
    \label{eq:NEGF-w-neq}
\end{align}
For the spinor implementation of \textsc{TranSIESTA}, we only consider the charge density, i.e. the real part of the sum of the diagonal elements of the spin box of $\boldsymbol{\Delta}_{\mathfrak{e},\mathfrak{e}'}$, when calculating the weights $w_{\mathfrak{e}}$. We omit the spin-box off-diagonal elements following the same rationale as in Ref.~\cite{BrandbygeDensityfunctionalMethod2002} because the off-diagonal terms do not affect the charge density.

\subsection{Convergence of the self-consistent cycle}
As is the case for normal SIESTA calculation, we find it more efficient to apply the self-consistency scheme directly to the Hamiltonian matrix instead of the density matrix\,\cite{CuadradoFullyRelativistic2012}. For this purpose at the end of each iteration the Hamiltonian corresponding to the charge density $\boldsymbol{\rho}$ is calculated and mixed with the Hamiltonian of previous steps. We find that mixing of the Hamiltonian is particularly advantages for materials with non-collinear spins, where mixing of density matrix often leads to oscillations of the magnetic moments. 

\subsection{Parallelization and Scaling}
\begin{figure}
    \centering
    \includegraphics[width=0.45\textwidth]{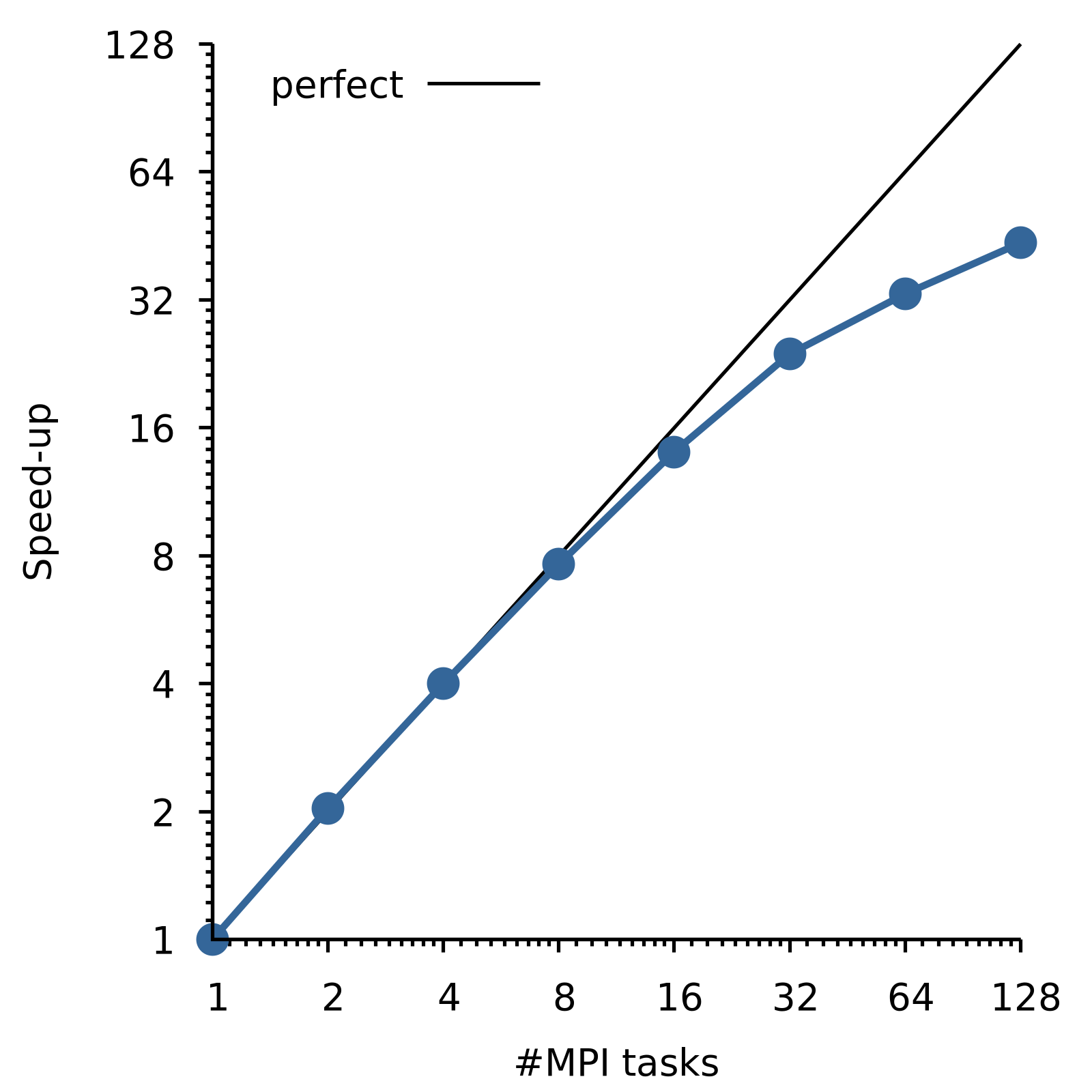}
    \includegraphics[width=0.45\textwidth]{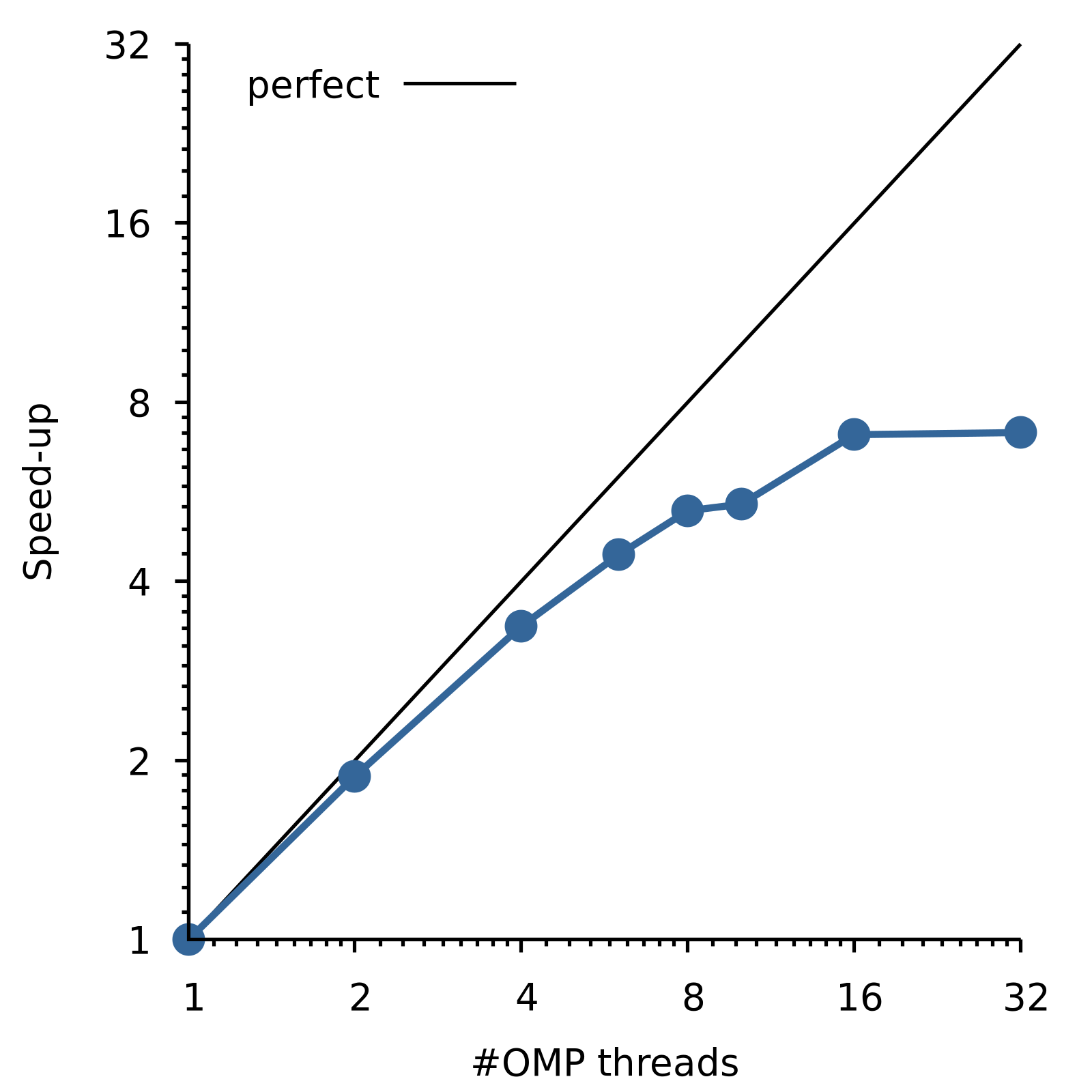}
    \caption{Performance characterization of \textsc{TranSIESTA} using a Fe/MgO/Fe tunneling junction with 816 atoms (11088 orbitals) }
    \label{fig:scaling}
\end{figure}

The newly implemented routines use the same to hybrid parallelization as the existing ones: each MPI task handles a distinct set of energy points, and OpenMP threading is used to speed up parallelized matrix operations. In Fig.~\ref{fig:scaling}, we show the scaling with the number of MPI tasks and OMP threads for a test system comprising 816 atoms (11088 orbitals) and a complex contour containing 256 energy points. The tests were performed using the BTD method (20 blocks; 200 to 720 orbitals each) with bias on a single computing node composed of 2 sockets holding 128 AMD Rome 7H12 CPUs each. We compiled SIESTA with GCC (10.3.0), OpenMPI (4.1.1), ScaLAPACK (2.1.0) and OpenBLAS (0.3.15) and enabled AVX2 instruction sets. We first vary the number of MPI tasks using a single OMP thread per task and then change the number of OMP thread using a single MPI task and the \textit{close} thread affinity policy. The calculations scale well up to 128 MPI tasks (with 1 OMP thread per task) and 8 OMP threads (using a single MPI task). For a larger number of OMP threads the performance scales sub-linear, but remains good up to 16 threads. The reported OMP performance characteristics show slightly better scaling with the number of OMP threads than previously reported\cite{PapiorImprovementsNonequilibrium2017}. Improved scaling results from the larger block sizes in the BTD matrix due to the spin blocks.

\section{Postprocessing}
\subsection{Spin Channel Projected Transmission}
There are several approaches to analyze transport properties by calculating the local density of states, transmission eigenchannels~\cite{PaulssonTransmissionEigenchannels2007}, bond-currents~\cite{NakanishiTsukada01,SolomonExploringLocal2010} or molecular state projection transmission~\cite{PapiorImprovementsNonequilibrium2017}. With these methods, it is possible to gain insight into which states or regions contribute to the transmission. 
Similarly, it might be desirable to decompose the transmission into contributions of different spin channels and transitions between them. 
In the collinear spin case, this decomposition arises naturally since all operators are diagonal in spin space. In the noncollinear case, the spin-up and spin-down transmission can be recovered discarding the off-diagonal spin terms of the scattering matrix $\mathbf{s}^{\uparrow\downarrow}$ and $\mathbf{s}^{\downarrow\uparrow}$. We can then calculate the transmission probably between states with the same spin $\sigma$. 
\begin{align}
T_{\mathfrak{e},\mathfrak{e}'}^{\sigma\sigma}(z) 
    ={}& \int_{\mathrm{BZ}}\!\!\!\!\dd{\vb{k}} \mathrm{Tr}\Big\{
         \mathbf{s}_{\mathfrak{e},\mathfrak{e}',\mathbf{k}}^{\sigma\sigma}
         \mathbf{s}^{\sigma\sigma\dagger}_{\mathfrak{e},\mathfrak{e}',\mathbf{k}}
     \Big\}
\end{align}

In addition to the spin-up and spin-down transmission, we might also find a non-zero probability for transmission between states with opposite spins 
\begin{align}
T_{\mathfrak{e},\mathfrak{e}'}^{\sigma\sigma'}(z) 
    ={}& \int_{\mathrm{BZ}}\!\!\!\!\dd{\vb{k}} \mathrm{Tr}\Big\{
         \mathbf{s}_{\mathfrak{e},\mathfrak{e}',\mathbf{k}}^{\sigma\sigma'}
         \mathbf{s}^{\sigma'\sigma\dagger}_{\mathfrak{e},\mathfrak{e}',\mathbf{k}}
     \Big\}
\end{align}
Such spin-flip transitions are possible because the Green's function matrix couples the different spin channels. 
The total transmission can be recovered as the sum of spin-up and spin-down and spin-flip transmissions.
\begin{align}
    T_{\mathfrak{e},\mathfrak{e}'}^{\mathrm{total}} =
        T_{\mathfrak{e},\mathfrak{e}'}^{\uparrow\uparrow}
        + T_{\mathfrak{e},\mathfrak{e}'}^{\uparrow\downarrow}
        + T_{\mathfrak{e},\mathfrak{e}'}^{\downarrow\uparrow}
        + T_{\mathfrak{e},\mathfrak{e}'}^{\downarrow\downarrow}    
\end{align}
More generally, we might be interested in the transmission probability between states with spin (up or down) along an axis $\vec{n}$ in electrode $\mathfrak{e}$ and along axis $\vec{m}$ in electrode $\mathfrak{e}'$. We can easily generalize the spin-projected transmission function to the case of arbitrary polarization axes
\begin{align}
 T_{\mathfrak{e},\mathfrak{e}'}^{\sigma\vec{n}, \sigma'\vec{m}}(z) 
    ={}& \int_{\mathrm{BZ}}\!\!\!\!\dd{\vb{k}} \mathrm{Tr}\Big\{
         \mel{\sigma\vec{n}}{\mathbf{s}^{\dagger}_{\mathfrak{e},\mathfrak{e}',\mathbf{k}}}{\sigma'\vec{m}}
         \mel{\sigma'\vec{m}}{\mathbf{s}_{\mathfrak{e},\mathfrak{e}',\mathbf{k}}}{\sigma\vec{n}}
     \Big\}\label{eq:ts-soc-TSpin-scattering}
\end{align}
where $|\sigma\vec{n}\rangle$ is the eigenstate of the spin operator oriented along $\vec{n}$ with eigenvalues $\sigma$, that is, $\sigma|\sigma\vec{n}\rangle = (\sum_{i=x,y,z}n_i\vu{\sigma}_i)|\sigma\vec{n} \rangle$. 

The spin-channel projection differs from other projection methods such as molecular state projection transmission in one key aspect. Rather than a projection of states in the device region, the spin-channel projection is projection on the states on electrode channels. In order to correctly calculate these projections it is necessary to apply the projector to scattering matrix rather than the Green's function matrix. To demonstrate the difference, we also implemented an alternative approach to calculating the spin-channel projected transmission, following  the same strategy employed to define the molecular state projection transmission~\cite{TodorovTightbindingSimulation2002}. Instead of projecting the broadening matrices $\Gamma_{\mathfrak{e}}$ onto molecular eigenstates, we project them onto different spin channels. In this light, we can define the transmission between a spin channel $\vec{n}$ in electrode $\mathfrak{e}$ and a spin channel $\vec{m}$ in electrode $\mathfrak{e}'$ as 
\begin{align}
T_{\mathfrak{e},\mathfrak{e}'}^{\sigma\vec{n}, \sigma'\vec{m}}(z) 
    ={} \int_{\mathrm{BZ}}\!\!\!\!\dd{\vb{k}}
    \mathrm{Tr}\Big\{ & 
        \mel{\sigma\vec{n}}{\mathbf{\Gamma}_{\mathfrak{e},\mathbf{k}}(z)}{\sigma\vec{n}}
        \mel{\sigma\vec{n}}{\mathbf{G}_\mathbf{k}(z)}{\sigma'\vec{m}}\nonumber\\
        &\mel{\sigma'\vec{m}}{\mathbf{\Gamma}_{\mathfrak{e'},\mathbf{k}}(z)}{\sigma'\vec{m}}
        \mel{\sigma'\vec{m}}{\mathbf{G}^{\dagger}_\mathbf{k}(z)}{\sigma\vec{n}}
    \Big\}\label{eq:ts-soc-TSpin-Gamma}.
\end{align}
If the spins in both electrodes are collinear and aligned with the projection axes $\vec{n}$ and $\vec{m}$, then the projector and the broadening matrix commute, and both approaches are equivalent. However, if the electrode states at a given energy are not collinear, the resulting transmission is incorrect. In section \ref{section:tests} we will see an example of this. Therefore, the latter approach is generally not applicable in fully relativistic calculations. It can find applications in calculations where both electrodes are perfectly collinear and noncollinear effects occur only in the scattering region.

Directly projecting the broadening matrix is less computationally demanding than the first approach because it avoids calculating the matrix square root of the broadening matrix (cf. Eq.~\ref{eq.LippmannSchwinger}). However, in most cases the number of electrode orbitals is small compared to the scattering device and in these cases calculating the matrix square root will not affect the computing time significantly. 
For these reasons, we have enabled the scattering matrix approach as the default way of calculating spin channel projected transmissions.

Instead of implementing equation \ref{eq:ts-soc-TSpin-scattering} directly and constructing the scattering matrices, we define a spin-channel selective broadening matrix and a spin-channel selective spectral density matrix:
\begin{align}
    \mathbf{\Gamma}_{\mathfrak{e},\mathbf{k}}^{\sigma\vec{n}}(z) &= 
        \mathbf{\Gamma}_{\mathfrak{e},\mathbf{k}}^{1/2}(z)
        \dyad{\sigma\vec{n}}
        \mathbf{\Gamma}_{\mathfrak{e},\mathbf{k}}^{1/2}(z) \\
    \mathcal{A}_{\mathfrak{e},\mathbf{k}}^{\sigma\vec{n}}(z) &= 
        \mathbf{G}_\mathbf{k}(z)
        \mathbf{\Gamma}_{\mathfrak{e},\mathbf{k}}^{\sigma\vec{n}}(z)
        \mathbf{G}^{\dagger}_\mathbf{k}(z).
\intertext{We can then rewrite the spin projected transmission as}
 T_{\mathfrak{e},\mathfrak{e}'}^{\sigma\vec{n}, \sigma'\vec{m}}(z)
    &= \int_{\mathrm{BZ}}\!\!\!\!\dd{\vb{k}} \quad\mathrm{Tr}\Big\{
        \mathbf{\Gamma}_{\mathfrak{e},\mathbf{k}}^{\sigma\vec{n}}(z)
        \mathcal{A}_{\mathfrak{e'},\mathbf{k}}^{\sigma'\vec{m}}(z)
     \Big\}
\end{align}
This allows us to reuse the matrices for multi-electrode calculation and avoids having to calculate projections for each pair of electrodes. 
To calculate the spin-projected broadening matrices $\mathbf{\Gamma}^{\sigma\vec{n}}$, we (1) diagonalize $\mathbf{\Gamma}$ to obtain the eigenvalues $d_i$ and the eigenvectors matrix $v$ ($v_j^{\dagger}\mathbf{\Gamma} v_i = d_i\delta_{ij}$), (2) construct the matrix square root of $\mathbf{\Gamma}$ ($\mathbf{\Gamma}^{1/2}_{ij}=\sum_k (v_{k})_i\sqrt{d_{k}}(v_{k})^*_j$), and (3) calculate the triple matrix product $\mathbf{\Gamma}^{1/2}\dyad{\sigma\vec{n}}\mathbf{\Gamma}^{1/2}$.

% \Nils{Add definition of bond currents. How to extend the definition to non-collinear spins.}

\section{Spin-orbit coupling in open systems}\label{section:tests}
In this chapter, we use our newly developed methods in the \textsc{TranSIESTA} code to study the effect of SOC and noncollinear spin configuration on transport properties. We determine the anisotropic magnetoresistance (AMR) in an infinite monatomic iron chain and study the domain wall resistance in the constraint domain wall between two semi-infinite ferromagnetic chain segments. We utilized Fe/MgO/Fe to test whether our new implementation can reproduce the previous prediction for tunneling magnetoresistance (TMR) in Fe/MgO/Fe junctions, and whether these junction exhibit tunnel anisotropic magnetoresistance (TAMR). As a third test case, we simulate a lateral MoS$_2$/WS$_2$ heterojunction, which exhibits strong intrinsic SOC.

The magnetic anisotropy energy (MAE) describes how the free energy of magnetic materials depends on the relative orientation of magnetic moments with respect to the crystal structure. MAE emerges as the result of Coulomb repulsion, SOC, and the broken rotational symmetry in the crystal. The anisotropic magnetoresistance (AMR) is the transport counterpart of MAE and describes the dependence of the resistance on the relative orientation between magnetization and current flow. While AMR is known to have a small effect in bulk materials ($<5$\% in 3d alloys~\cite{McGuireAnisotropicMagnetoresistance1975}), a variety of low-dimensional systems have a large AMR (20-50\% in 3d transition metal nanojunction~\cite{BagretsMagnetoresistanceAtomicsized2004}, $>$100,000\% in (Ga, Mn)As/GaAs/(Ga, Mn) stacks~\cite{RusterVeryLarge2005}). 
Previously simulations of AMR were only possible for collinear spin configuration. Now, with our new implementation it is possible to study AMR in materials with arbitrary spin configuration, e.g. in domain walls. 

In addition to AMR, magnetic transport devices may also exhibit tunneling magnetoresistance (TMR), which refers to changes in the resistance of a magnetic tunneling junction depending on the relative magnetic alignment of its magnetic components. Magnesium oxide (MgO) based tunneling junctions, such as Fe / MgO / Fe, have been predicted to yield TMR ratios between a few hundred and a few thousand percent~\cite{ButlerSpindependentTunneling2001,MathonTheoryTunneling2001}. 
We utilized Fe/MgO/Fe to test whether our new implementation could  obtained results for the  previously obta,ined results for the TMR in Fe/MgO/Fe, and whether they exhibit tunnel anisotropic magnetoresistance (TAMR).

As a third test case, we simulate a lateral MoS$_2$/WS$_2$ heterojunction, which has been shown to function as monolayer p-n junctions,~\cite{GongVerticalInplane2014}. Transition metal dichalcogenides (TMDs) are semiconductors of the type MX2, where M is a transition metal atom (such as Mo or W) and X is a chalcogen atom (such as S, Se, or Te). TMDs are promising materials for a wide range of possible applications, including energy conversion~\cite{WiEnhancementPhotovoltaic2014} and storage~\cite{DingFacileSynthesis2012} and hydrogen evolution reaction~\cite{LiMoSNanoparticles2011}, due to their unique combination of the direct band gap, strong spin-orbit coupling, and favorable electronic and mechanical properties. In addition, TMDs hold a lot of promise for spintronics applications because of the SOC-induced splitting of the valence and conduction bands into different spin states. This splitting makes manipulations of the electron spin possible through optical excitation~\cite{WangColloquiumExcitons2018}.

Our three test cases cover all different spatial dimensions of the electrodes: iron chain (1D), TMD monolayer (2D), and nanowire with the bulk electrode (3D). We include elements with strong SOC and enforce noncollinear spin configurations in order to demonstrate and test the new features of TranSIESTA. We will include the computational details after presenting the results in each case.

\subsection{Monatomic Iron Chain (1D)}
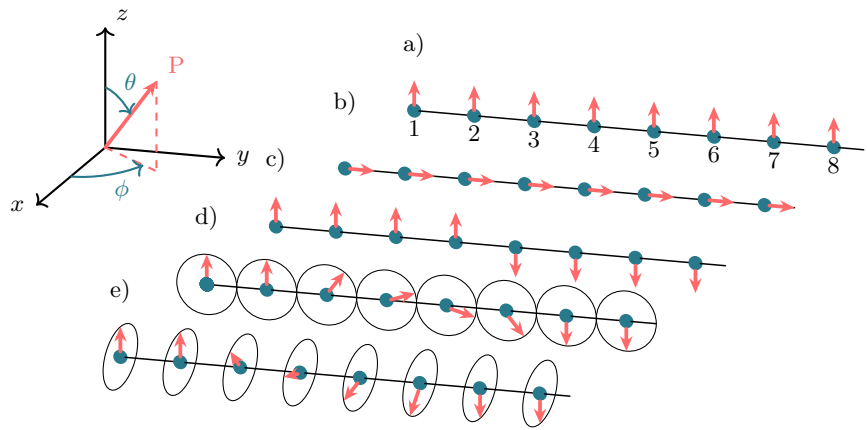
\begin{figure}
    \centering
    \begin{tikzpicture}[scale=1.0,x={(220:0.6cm)},y={(-5:0.8cm)},z={(90:0.8cm)},]
       \tikzstyle{spin}=[-{Stealth[scale=0.6,angle'=45]},line width=0.5mm,MyRed,line cap=round]
       \tikzstyle{chainax}=[line width=0.2mm,black,line cap=round]
       \tikzstyle{atm}=[line width=0.2mm,MyGreen]
      
      % AXES
      \coordinate (O) at (0,-2,0);
      \draw[thick,->] (O) -- (2,-2,0) node[left=0.1] {$x$};
      \draw[thick,->] (O) -- (0,0,0) node[right=0.1]{$y$};1
      \draw[thick,->] (O) -- (0,-2,2) node[above right=0.1]{$z$};
      
      % VECTORS 
      \coordinate (P) at (0.6,-0.8,1.5);
      \draw[spin] (O)  -- (P) node[above right=0.1] {P};
      \draw[dashed,thick,MyRed] (O)  -- (0.6,-0.8,0.0);
      \draw[dashed,thick,MyRed] (P)  -- (0.6,-0.8,0.0);
      
      % ARCS
       \begin{scope}[canvas is xy plane at z=0]
         \draw[->,thick,MyGreen] (1,-2) arc (0:63.43:1) node[below left=0.1] {$\phi$};
       \end{scope}
    
       \begin{scope}[plane origin={(0.,-2.,0.)},plane x={(0.,-2.,1.)},plane y={(0.447213595,-1.105572809,0.)}, canvas is plane]
         \draw[->,thick,MyGreen] (1,0.) arc (0:41.81:1) node[above=0.2] {$\theta$};
       \end{scope}
       \def\ylist{2,...,9}
       \def\labellist{"a)","b)","c)","d)","e)"}
       \def\mlength{0.5}
       
        \foreach \x [count=\c,evaluate=\c as \lab using {{\labellist}[\c-1]}] in {-2,0,2,4,6.5} {
            \begin{scope}[canvas is yz plane at x=\x, scale=1]
                \filldraw[atm] (2,0) circle (3pt) node[above=0.6,black] {\lab};
                \foreach \y in \ylist {
                    \ifnum \y>2
                        \draw[chainax] (\y-0.92,0.0) -- (\y,0.0);
                        \filldraw[atm] (\y,0) circle (3pt);
                    \fi
                };
                \draw[chainax] (9.08,0.0) -- (9.5,0.0);
            \end{scope}
        };
        
        % Chain 1
        \begin{scope}[canvas is yz plane at x=-2, scale=1]
            \foreach \y [count=\d] in \ylist {
                \draw[spin] (\y,0.09) -- (\y,\mlength);
                \node at (\y,-0.3) {\d};
        };
        \end{scope}
        
        % Chain 2
        \begin{scope}[canvas is yz plane at x=0, scale=1]
            \foreach \y [count=\d] in \ylist { 
                \draw[spin] (\y+0.08,0.) -- (\y+\mlength,0.); 
                % \node at (\y,-0.3) {\d};
                };
        \end{scope}
        
        % Chain 3
        \begin{scope}[canvas is yz plane at x=2, scale=1]
           \foreach \y [count=\d] in \ylist { 
                \ifnum \y<6
                    \draw[spin] (\y,0.09) -- (\y,\mlength);
                    % \node at (\y,-0.3) {\d};
                \else
                    \draw[spin] (\y,-0.13) -- (\y,-\mlength);
                    % \node at (\y, 0.3) {\d};
                \fi
            };
        \end{scope}
        
        % Chain 4
        \begin{scope}[canvas is yz plane at x=4, scale=1]
           \foreach \y [count=\d] in \ylist { 
                \draw (\y,0.0) circle (0.5); \filldraw[atm] (2.,0) circle (3pt);
                \ifnum \y<4
                    \draw[spin] (\y,0.09) -- (\y,\mlength); 
                \else \ifnum \y>7
                    \draw[spin] (\y,-0.13) -- (\y,-\mlength); 
                \else
                   \draw[spin] ({\y+0.08*sin(36*(\y-3))},{-0.02+0.11*cos(36*(\y-3))}) -- ({\y+\mlength*sin(36*(\y-3))},{\mlength*cos(36*(\y-3))});
                \fi\fi
            %     \ifnum \y<6
            %         \node at (\y,-0.3) {\d};
            %     \else
            %         \node at (\y, 0.3) {\d};
            %     \fi
            };
        \end{scope}
        
        % Chain 5
       \foreach \y [count=\d] in \ylist { 
            \begin{scope}[canvas is xz plane at y=\y, scale=1]
                \draw (6.5,0.0) circle (0.5);
                \ifnum \y<4
                  \draw[spin] (6.5,0.09) -- (6.5,\mlength); 
                \else \ifnum \y>7
                    \draw[spin] (6.5,-0.13) -- (6.5,-\mlength); 
                \else
                   \draw[spin] ({6.5+0.08*sin(36*(\y-3))},{-0.02+0.11*cos(36*(\y-3))}) -- ({6.5+\mlength*sin(36*(\y-3))},{\mlength*cos(36*(\y-3))});
                \fi\fi
                
            \end{scope}
        };
        % \begin{scope}[canvas is yz plane at x=6.5, scale=1]
        %     \foreach \y [count=\d] in \ylist { 
        %         \ifnum \y<6
        %             \node at (\y,-0.3) {\d};
        %         \else
        %             \node at (\y, 0.3) {\d};
        %         \fi
        %     };
        % \end{scope}
    \end{tikzpicture}
    \caption{Schematic of five different spin alignments in monoatomic chains: (a) parallel spins perpendicular to the chain axis ($\theta_i=0$), (b) parallel spin along the chain axis ($\theta_i=\pi/2$,$\phi_i=\pi/2$), (c) abrupt domain wall between two domains with $\theta_i=0$ and $\theta_i=\pi$, (d) N\'eel type domain wall ($\theta_i\in[0,\pi], \phi=\pi/2$) (e) Bloch-type domain wall ($\theta_i\in[0,\pi], \phi_i=0$).} 
    \label{fig:01-negf-soc-schematic-spin}
\end{figure}

In an infinite monatomic iron chain, the Hamiltonian of the scattering region is independent of a transverse k. With this system, we test the part of our implementation without $\mathbf{k}$ point sampling. We check whether the magnetic moments in the scattering region relax when the initial guess does not match the spin orientation of the electrodes. Furthermore, we calculate the transport properties of monatomic iron chains with different magnetic configurations: infinite periodic chains with collinear spin moments (Figure \ref{fig:01-negf-soc-schematic-spin} (a) and (b)) and constraint domain walls (Figure \ref{fig:01-negf-soc-schematic-spin} (c), (d), and (e)). We compare the contribution of different spin channels between the scalar relativistic (collinear spin approximation and no SOC) and fully relativistic (with SOC) simulations, using the previously induced concept of spin-channel projected transmission.
Before discussing these properties obtained with the NEGF formalism, it is important to study the electronic structure of the infinite periodic iron chain (Figure \ref{fig:01-negf-soc-schematic-spin} (a) and (b)).

\subsubsection{Magnetic Anisotropy in the Ideal Iron Chain}

\begin{figure}
    \centering
    \includegraphics[width=1.\textwidth]{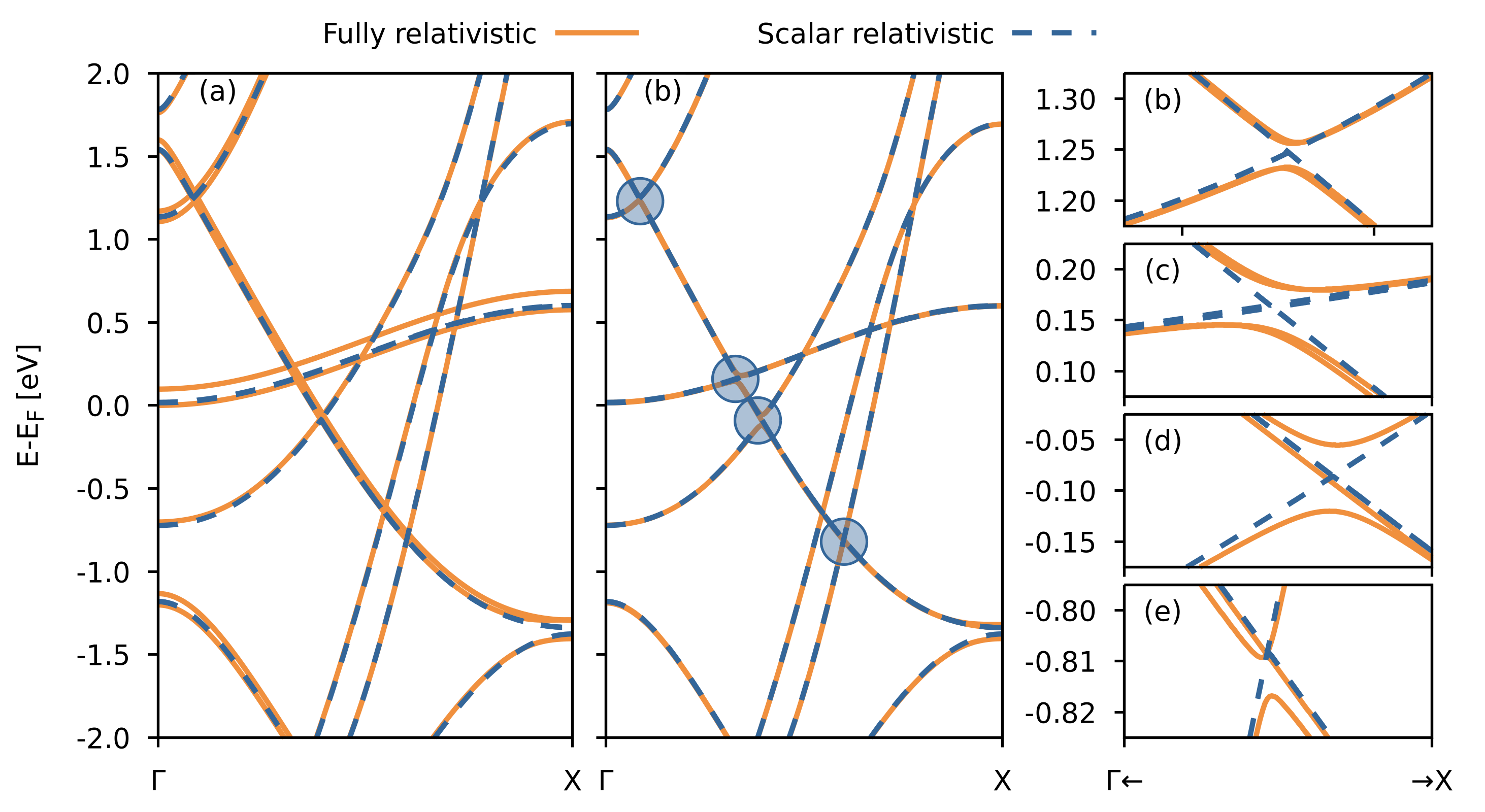}%
    \caption{Comparison of scalar relativistic (SR) and fully relativistic (FR) spin band structure of a monatomic iron chains with spin moment parallel to the chain axis (a) and perpendicular to the chain axis (b through f). For spins aligned with the chain axis SOC acts as an effective magnetic field, shifting bands proportionally to their orbital magnetic moment (a). For spins perpendicular to the chain axis (b) the two band structures are almost identical, except for SOC induced avoided band crossings, which are highlighted with blue circles. Zoomed-in views of each avoided band crossing in the energy range from -2~eV to 2~eV are displayed in panels (c) through (f).}
    \label{fig:01-fe-chain-bands}
\end{figure}

The ground state of an iron chain is characterized by an interatomic spacing of 2.26 \AA{} and ferromagnetic alignment of spin magnetic moments with 3.35$\mu_B$ per iron atom. We observe small differences in the total energy depending on the alignment of the spin moments relative to the chain axis. This MAE favors an alignment of the spin moments parallel to the chain axis ($E(\theta=\pi/2)-E(\theta=0)\approx 1$~meV per Fe).
Similarly, the band structure of the iron chain is anisotropic with respect to the direction of the spin magnetic moments. Like MAE, the anisotropy of bands arises because of the SOC in Fe. When the spin moments are parallel to the chain axis ($\theta=\pi/2$) SOC acts like an effective magnetic field and splits some of the bands, which are doubly degenerate in the scalar relativistic case (Figure \ref{fig:01-fe-chain-bands} (a)). Not all bands split, nor is the splitting uniform for all bands that are split.
\textcite{JacobAnisotropicMagnetoresistance2008} have shown that the same effect can be observed in nickel chains. They showed that the size of this splitting is proportional to the size of the orbital magnetic moment ($m_{\parallel}$) and spin ($\sigma_{\parallel}$) parallel to the chain axis. Effectively, the SOC introduces an orbital Zeeman effect along the spin-polarization axis. This same effect is responsible for band splitting in the iron chain with spin moments parallel to the chain axis~\cite{TungSystematicInitio2007}.
\begin{figure}
    \centering
    \includegraphics[width=1.\textwidth]{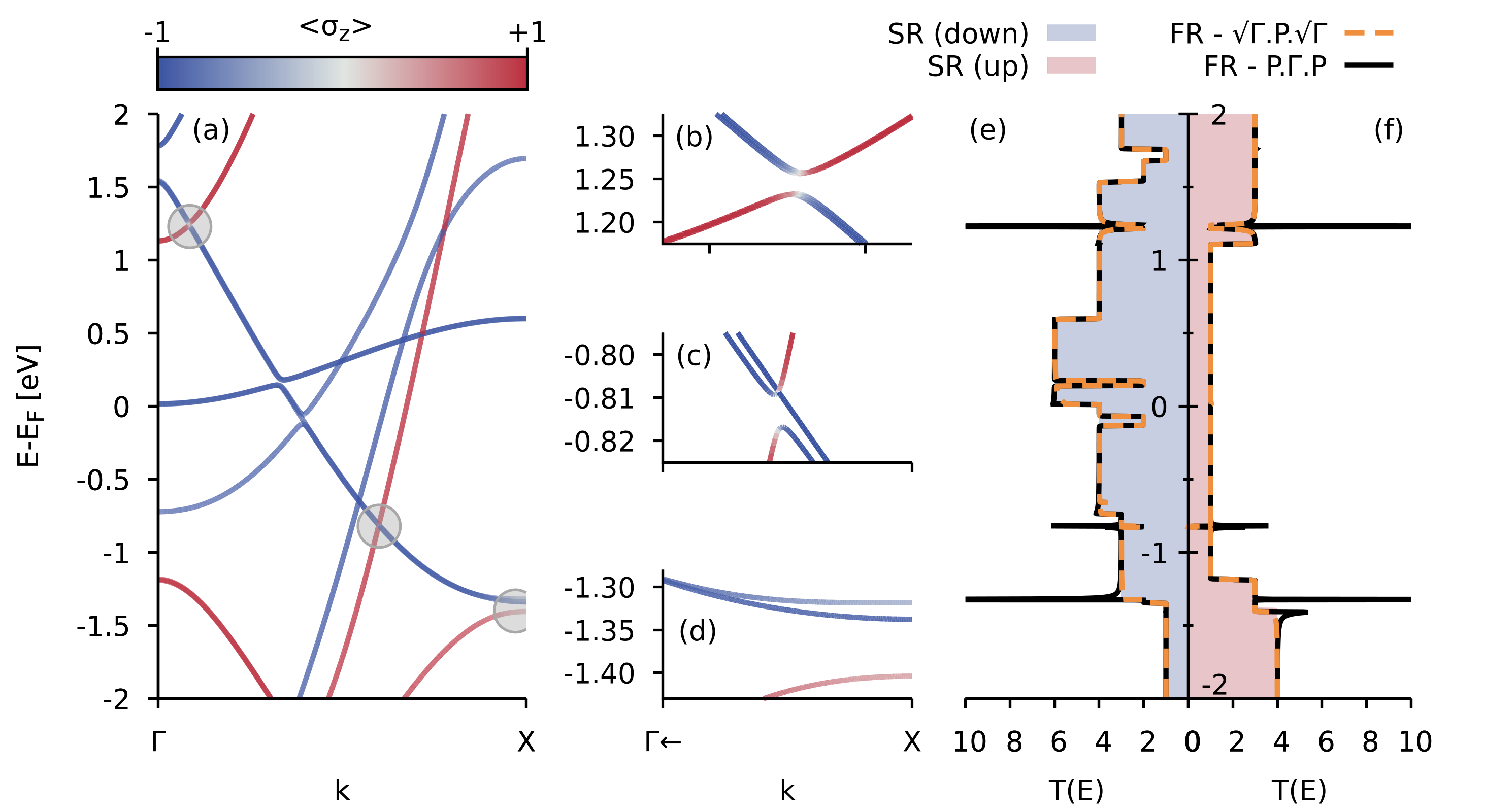}%
    \caption{Spin texture (a - d) and spin-channel projected transmission (e,f) of an iron chain with spin magnetic moments perpendicular to the chain axis ($\theta=0$). The spin texture is collinear to the magnetic axis in the -2~eV to 2~eV energy window except for three points (gray circles in (a) and zoomed-in graphs (b - d)). The spin channel projected transmission calculated with equation \ref{eq:ts-soc-TSpin-scattering} ($\sqrt{\Gamma}P\sqrt{\Gamma}$) reproduces the scalar relativistic (SR, FR is full relativistic) spin channel transmission except for the energy corresponding to avoided crossings (-0.81~eV, -0.10~eV, 0.15~eV, and 1.25~eV) where it is reduced according to the number of split bands. The spin channel projected transmission calculated with equation \ref{eq:ts-soc-TSpin-Gamma} ($P\Gamma P$)) also, reproduces the scalar relativistic case closely, but diverges at those energies where the spin texture is not collinear to the projection axis. The results obtained with $P\Gamma P$-method are unphysical and shown here only to demonstrate it's flaw of this approach.
    }
    \label{fig:01-fe-chain-t-perp}
\end{figure}

In contrast, the band structure of an iron chain with perpendicular moments features no shifted bands. It reproduces the scalar-relativistic case almost exactly. However, closer inspection of the intersection points of all bands reveals avoided band crossings for some bands (Figure \ref{fig:01-fe-chain-bands} (b to f)). These avoided band crossings are absent when the spin moments are parallel to the chain axis or without SOC. Whether two bands are repelled by SOC is determined by their respective spin, $\sigma$, and orbital magnetic quantum numbers, $m$, projected along the magnetic axis of the chain: crossings of bands with ($\sigma, m$) and ($\pm\sigma,\mp m$) are lifted~\cite{JacobAnisotropicMagnetoresistance2008}.
The results we obtain for the electronic and magnetic structure of infinite monatomic iron chains match previously published studies~\cite{Dorantes-DavilaMagneticAnisotropy1998,EdererMagnetismSystems2003,Autes2006, TungSystematicInitio2007}. The magnetic anisotropy energy we obtain is a factor of 2 lower compared to other studies~\cite{Wijn3d4d1986,OuAnisotropicMagnetism2009}, however this number is also strongly dependent on the exact bond-length in the chain\cite{Autes2006}.
Given that all other properties are reproduced well, and our focus is on the transport properties, we can proceed safely with our simulation parameters. 

\begin{figure}
    \centering
    \includegraphics[width=1.\textwidth]{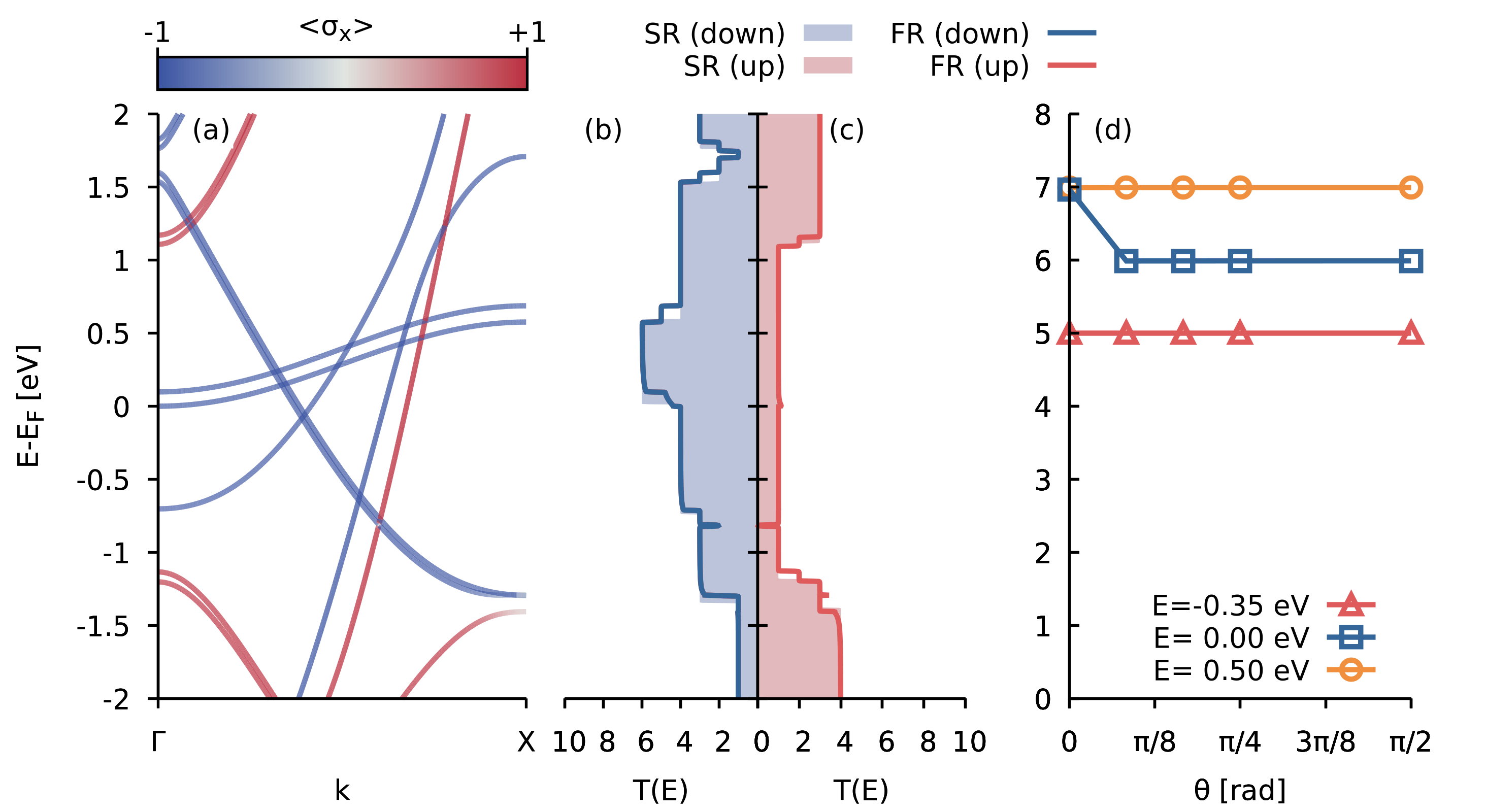}%
    \caption{Spin texture (a) and spin-channel projected transmission (b,c) of an iron chain with spin magnetic moments parallel to the chain axis ($\theta=\pi/2$) and magnetic anisotropy of the transmission. The spin texture is collinear to the magnetic axis ($x$). The fully and scalar relativistic transmission function match closely. Near the band edges of the bands shifted by SOC, the fully relativistic transmission function exhibits additional plateaus corresponding in width to the band splitting. In both cases, the transmission corresponds to the number of bands at any given energy. Depending on the angle between the spins and the chain axis the transmission at the Fermi level changes between 7 and 6 implying that iron chains exhibit anisotropic magnetoresistance at sufficiently low temperatures.}
    \label{fig:01-fe-chain-t-para}
\end{figure}

Regardless of the alignment of the spin relative to the chain axis, the bands of the iron chain are almost perfectly spin-polarized along the magnetic axis of the system (Fig.~\ref{fig:01-fe-chain-t-perp} (a) and \ref{fig:01-fe-chain-t-para} (b)). 
In the case of spins perpendicular to the chain axis ($\theta=0$) the spin texture becomes noncollinear at the two avoided band crossings where bands with opposite spin magnetic moment are repelled (Figure \ref{fig:01-fe-chain-t-perp} (b) and (c)), and near the band edges of the two parabolic bands with opposite spin moments at $X$ (Figure \ref{fig:01-fe-chain-t-perp} (d)). The fully relativistic spin channel projected transmission calculated using equation \ref{eq:ts-soc-TSpin-scattering} (orange dashed line in Fig.~\ref{fig:01-fe-chain-t-perp}) reproduces the scalar relativistic spin channel transmissions. It corresponds to the number of spin-up (spin-down) bands at any given energy and deviates from the scalar-relativistic calculation at the avoided band crossing. Depending on the number of bands that repel due to the SOC, the transmission is reduced by the same number, either by 2 or 4.  The spin-channel-projected transmission calculated using equation~\ref{eq:ts-soc-TSpin-Gamma} performs equally well for most energies but diverges at those points where the spin texture becomes noncollinear. This showcases that the two methods work as expected: where the spin texture is collinear, the projector commutes with the broadening matrix and the two methods become equivalent. At the points where the spin texture is noncollinear to the projection axis, the equivalence breaks down and the approximation in equation~\ref{eq:ts-soc-TSpin-Gamma} is not good enough to describe the transmission properly. From this point on, we only use Eq.~\ref{eq:ts-soc-TSpin-scattering} to calculate the spin-channel projected transmission.
In the case of spins parallel to the chain axis ($\theta=\pi/2$), the difference between the scalar relativistic and fully relativistic transmissions is slightly larger. Due to band splitting, new plateaus arise in the transmission function. This is particularly evident near the Fermi level, where the size of the band splitting is the largest. Overall, the transmission function still closely resembles the scalar relativistic case. The total transmission at the Fermi level is reduced by 1 even for small angles(Figure~\ref{fig:01-fe-chain-t-para} (d)). Further away from the Fermi level, the transmission is constant. The magnetic anisotropy of the transmission at the Fermi level implies that the anisotropic magnetoresistance predicted for nanowires~\cite{OuAnisotropicMagnetism2009,HuConditionsQuantized2015} extends down to the one-atom limit for sufficiently low temperatures.

\subsubsection{Domain Wall Conductivity}

\begin{figure}
    \centering
    \includegraphics[width=0.65\textwidth]{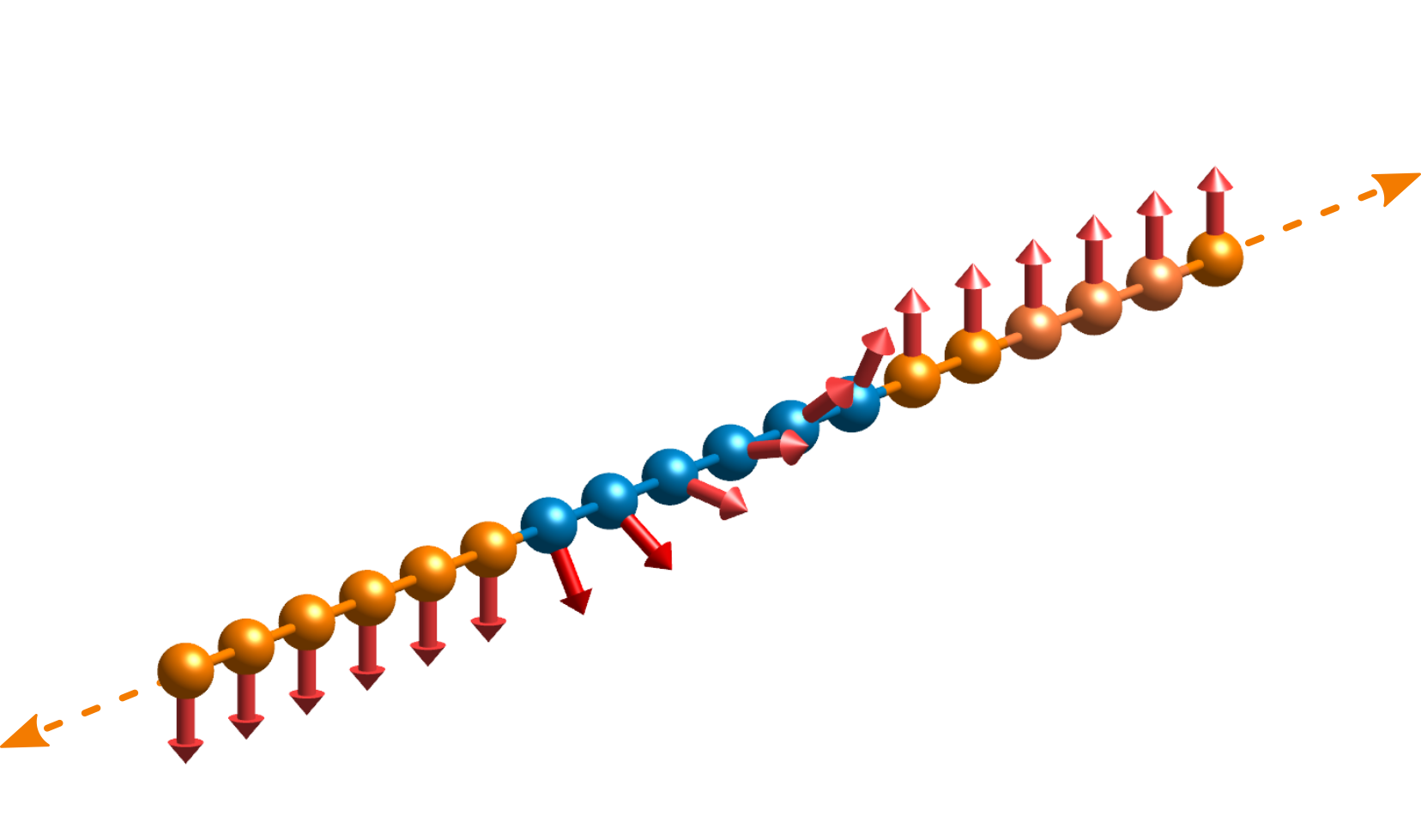}
    \caption{Infinite iron chain with a 6-atom wide domain wall (blue atoms) between two semi-infinite sections with opposite magnetic moments (orange atoms).}
    \label{fig:01-fe-chain}
\end{figure}

Next we set up a transport system with two electrodes (orange atoms in Fig.~\ref{fig:01-fe-chain}) with non-parallel spin moments to model a constrained domain wall.  
We then relax the direction and amplitude of the spin moments of the atoms between the electrodes (blue atoms in Fig.~\ref{fig:01-fe-chain}). The NEGF approach serves two purposes in these calculations: it provides a natural way of fixing the spin direction in parts of the system, and it allows us to simulate a domain wall without interactions between the periodic images that would occur in a standard DFT calculation. 

\begin{figure}
    \centering
    \includegraphics[width=\textwidth]{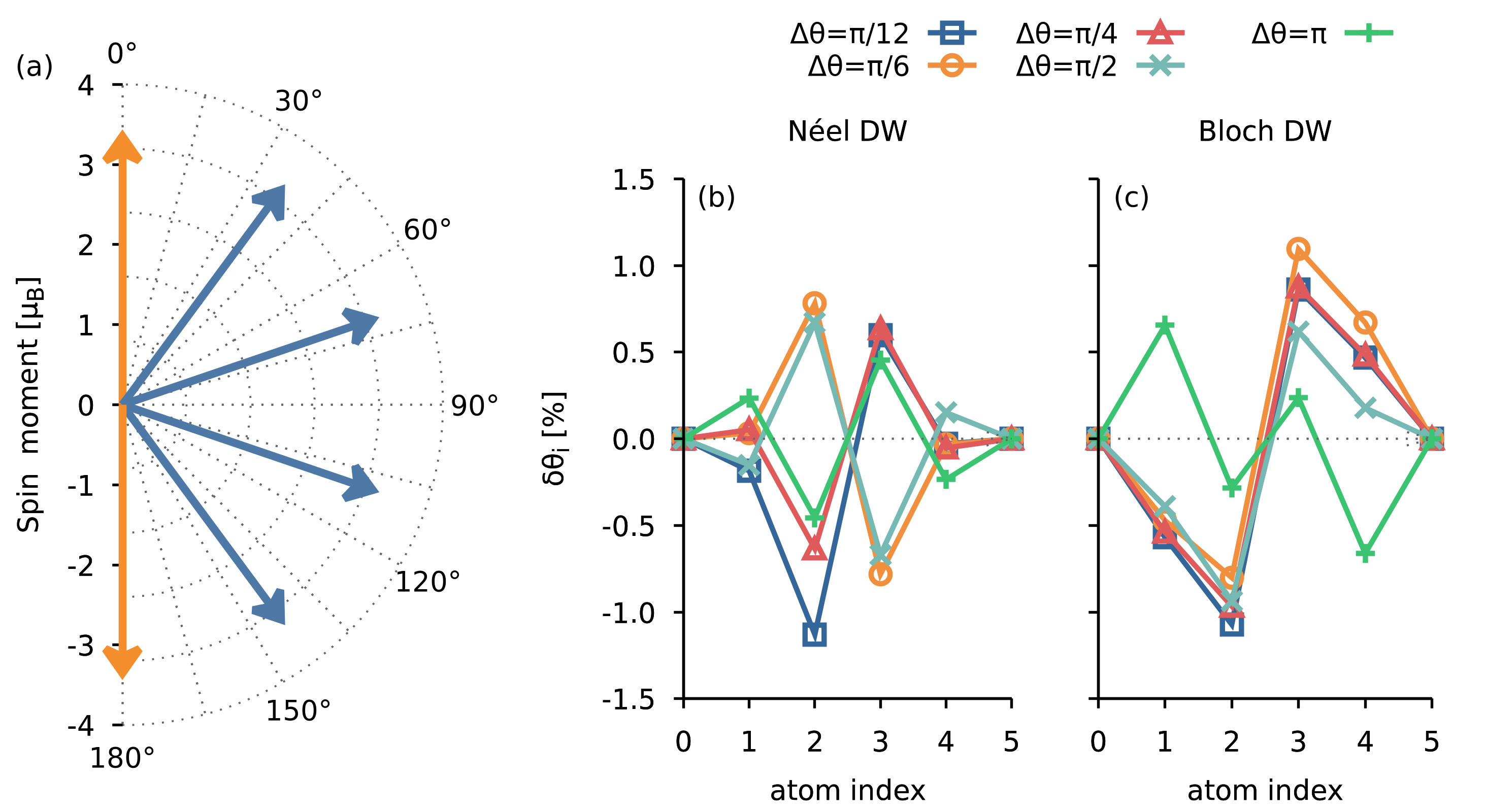}%
    \caption{Direction and size of spin moments in a 4-atom-wide, constrained domain wall with $\Delta\theta=\pi$ (a). Orange atoms indicate the fixed spin moments in the electrode and blue arrows the spin moments of the device atoms. Relative deviation from ideal positions $\delta\theta_i = \theta_i - i \Delta\theta/(N+1)$ for 4-atom-wide domain walls with $\Delta\theta\in\{\pi/12,\pi/6\pi/4,\pi/2,\pi\}$ for N\'eel (b) and Bloch domain walls (c).}
    \label{fig:01-fe-chain-spins-angle}
\end{figure}

The magnitude of the spin moments in the domain wall is constant at 3.35~$\mu_B$, irrespective of the angular offset between the spin moments in the two electrodes ($\theta$) (Figure \ref{fig:01-fe-chain-spins-angle}).
% and the size of the domain wall (Figure \ref{fig:01-fe-chain-spins-width}).
Depending on the initialization of the density matrix, the spin moments form different types of domain walls. If the density matrix is initialized with all spin moments in the $yz$-plane, i.e., in a plane containing the chain axis, then the final magnetic moments form a N\'eel-type domain wall (Fig.~\ref{fig:01-negf-soc-schematic-spin} (d)). Similarly, if all initial spin moments lie in the $xz$-plane, i.e., all perpendicular to the chain axis, then we obtain a Bloch-type domain wall (Figure \ref{fig:01-negf-soc-schematic-spin} (e)). We can produce these different types of domain walls because the initial guess determines the symmetry of the Hamiltonian, and there are no external fields that break the symmetry of the initial magnetic structure. There is one other notable high-symmetry case for electrodes with antiparallel spin moments: when the initial magnetic moments in the device are all collinear to the electrodes, we obtain an abrupt domain wall (Fig.~\ref{fig:01-negf-soc-schematic-spin} (c)). In principle, we could choose random initial spin moments to find the most stable domain wall. However, because of the small size of the MAE in the iron chain, this becomes a very challenging task. Even high-symmetry structures require up to a thousand SCF steps to converge. Instead, we focus on the three high-symmetry cases: abrupt, N\'eel, and Bloch domain walls.  

In N\'eel and Bloch domain walls we observe an equal change in the angle from one atom to the next (Figure \ref{fig:01-fe-chain-spins-angle}). The change in angle between two neighboring atoms in the domain wall is equal to $\Delta\theta/(N+1)$, where $N$ is the number of atoms in the domain wall and $\Delta\theta$ is the difference in the polar angle ($\theta$) that describes the spin direction in the left and right electrodes. 
The magnetic coupling between the atoms in the unit cell should be the same, considering that all species, coordination, and bond lengths in the device region are equal. Therefore, the constrained domain wall should be symmetric with equal change in angle. In N\'eel-type domain walls, the magnetic anisotropy could lead to a deviation from this uniformity. However, the Fe-Fe magnetic exchange constant ($\approx10$~meV~\cite{AntoniakCompositionDependence2010,Bezerra-NetoComplexMagnetic2013}) is approximately one order of magnitude larger than the MAE~\cite{Wijn3d4d1986,OuAnisotropicMagnetism2009}, and would thus push the system to a more uniform domain wall like the one we observe. We conclude that the relaxation of spin moments in our code is consistent with expectations based on the symmetry of the system and the comparison of MAE and magnetic exchange constant. 

\begin{figure}
    \centering
    \includegraphics[width=\textwidth]{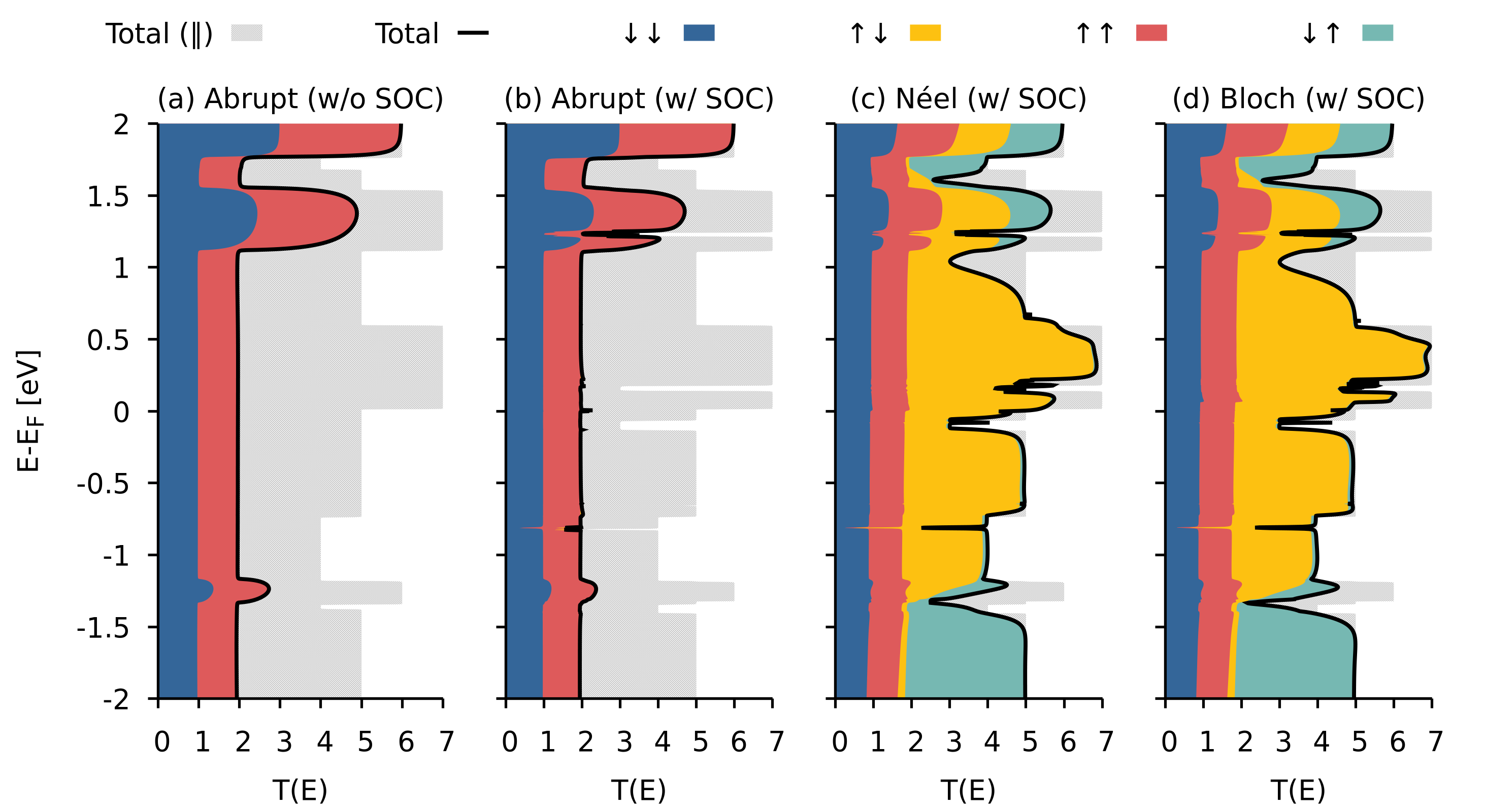}%
    \caption{Zero-bias transmission of 180$^\circ$ domain walls calculated with scalar relativistic formalism with collinear spin (a), fully relativistic formalism (b-d). Filled curves are stacked on top of each other to visualize the decomposition of the total transmission (black line) in terms of different spin channels and spin flips. The shaded gray area in the background displays the transmission of an iron chain without a domain wall (T$\parallel$) as a point of reference. For the abrupt domain wall (a) and (b), the total and spin channel projected transmission (blue (down) and red (up)) are significantly reduced compared to the iron chain without a domain wall. SOC has no significant effect on the transmission through the abrupt domain wall. For 4-atom-wide N\'eel (c) and Bloch (d) domain walls the transmission is much higher. The gradual change in the magnetic moment along the domain wall allows for significant spin flip transmission (yellow and green filled curves). The contributions of the pure spin-up and spin-down transmissions are comparable to the abrupt domain wall.} 
    \label{fig:01-fe-chain-t-dw}
\end{figure}

The non-periodic structure of the domain wall gives rise to electron scattering. The resulting transmission is not a step function as in the periodic parallel case (Figure \ref{fig:01-fe-chain-t-dw}). We find that the resistance of an abrupt domain wall is significantly higher than that of the N\'eel or Bloch domain walls. SOC in Fe is not strong enough to allow for spin-flips in an abrupt domain wall. As a result, the scalar and fully relativistic transmissions are almost identical and reduced by a factor of 3 compared to the parallel case near the Fermi level. In N\'eel or Bloch domain walls, a width of 4 atoms, spin-flip transmissions are significant. The pure spin-up and spin-down transmissions remain equally suppressed as in the abrupt domain-wall case. The reduction of these contributions in all four cases can be immediately understood from the band structure of the iron chain. In the left electrode with magnetic moments pointing along $z$ only one band with spin $\sigma_z=1$ exists. Therefore, the spin-up transmission can not exceed one. The same is true for the spin-down channel in the right electrode. Thus, both pure spin-channel transmissions are limited to 1. In N\'eel and Bloch domain wall the change in spin moments is gradual and allows for spin-flip, i.e. scattering between different spin channels. The contribution of the two spin-flip trans missions down into up (yellow) and up into down (green) are not equal. This allows spin-down states in the left electrodes to scatter into spin-up states in the right electrode (down-up spin-flip transmission; depicted as a yellow curve in Fig.~\ref{fig:01-fe-chain-t-dw}). Below 1~eV the availability of spin-down and spin-up channels in the left and right electrodes is reversed and up-down spin-flip transmission becomes dominant. We also performed calculations with 5 and 6-atom-wide domain walls which show qualitatively the same results. 
The low TAMR in domain walls with more than two atoms matches predictions by \textcite{AutesElectronicTransport2008}, and \textcite{VelevDomainwallResistance2004}.

\subsubsection{Computational Details}
Our transport setup consists of an infinite chain with Fe atoms. The simulation cell contains 12 atoms: 4 atoms on each side correspond to one principal electrode layer and the 4 atoms in the center represent the scattering device. In the electrode calculations, we sample the reciprocal space along the chain axis with 101 k points. We used a double-zeta-polarized basis set for the Fe atoms (energy shift 0.2~eV, split norm 0.15), a real space grid with a cut-off of 700~Ry, an electronic temperature of 8~meV, and the PBE exchange-correlation functional~\cite{PerdewGeneralizedGradient1996}.
The lattice constant is optimized up to a stress tolerance of 0.1~meV/\AA{}. 
Calculations were carried out within the fully relativistic pseudo-potential formalism~\cite{CuadradoFullyRelativistic2012}, i.e. including spin-orbit interactions, and the scalar relativistic collinear spin formalism. The pseudo potential for iron was generated using the approach of Troullier and Martins\cite{TroullierEfficientPseudopotentials1991} with a valence configuration of 3d$^{6}$4s$^{2}$, the PBE functional\cite{PerdewGeneralizedGradient1996} and  with non-local core-corrections.

The density matrix in the scattering region is initialized from a periodic SIESTA calculation. For the initial guess of the density matrix in the domain walls, we extend the scattering region by three extra iron atoms on each side and terminate the chain with hydrogen atoms. We then initialize the moments along the domain wall in one of the high symmetry configurations discusses above and perform a few SCF steps in SIESTA. It is important to terminate the initial SIESTA early enough for the spin moment to remain noncollinear, but late enough for the charge density to converge reasonably well. If the convergence criteria for this initial calculation are chosen too tight all magnetic moments will become parallel. On the other hand, if the convergence criteria are chosen too loosely the NEGF+DFT SCF loop diverges in the first few steps. We find that a convergence threshold of $10^{-3}$ for the density matrix and $10^{-2}$~eV for the Hamiltonian works well for this system. For the NEGF+DFT SCF loop, we choose much more stringent convergence thresholds: $10^{-6}$ for the density matrix and $10^{-4}$~eV for the Hamiltonian.
We perform the complex contour integral with a circle contour starting at ($-20+0.1i$)~eV which transitions into a line contour at $-10k_B$~eV including 10 poles of the Fermi-function in the contour.
The contour encircles 32 poles of the Fermi function.
The transmission function is calculated with \textsc{tbtrans} using a contour along the real axis with a spacing of 0.2~meV shifted by 0.1~meV in to the complex plane. 

The band structure and spin texture of the iron chains with different magnetic moments are calculated in the primitive unit cell. To resolve the SOC-induced band gaps we calculate the band structure and spin texture on 4800 points along the $\Gamma$-$X$ direction.

\subsection{Fe-MgO-Fe hetero structure (3D)}

\subsubsection{Magneto Resistance of Bulk Iron}
Fe/MgO/Fe is periodic in the two directions parallel to the interface and therefore requires sampling at $\mathbf{k}$ points in these directions. With this system, we check whether our code also works for a more realistic application with three-dimensional electrodes and a non-homogenous geometry in the scattering device. Fe/MgO/Fe tunneling junctions have previously been treated with scalar-relativistic DFT~\cite{BelashchenkoEffectInterface2005,WaldronFirstPrinciples2006,HeiligerThicknessDependence2008}. These calculations and studies of magnetic anisotropy in bulk iron~\cite{KhanAnisotropicMagnetoresistance2008,ZwierzyckiCalculatingScattering2008} will serve as a reference point.

\begin{figure}
    \centering
    \includegraphics[width=0.5\textwidth]{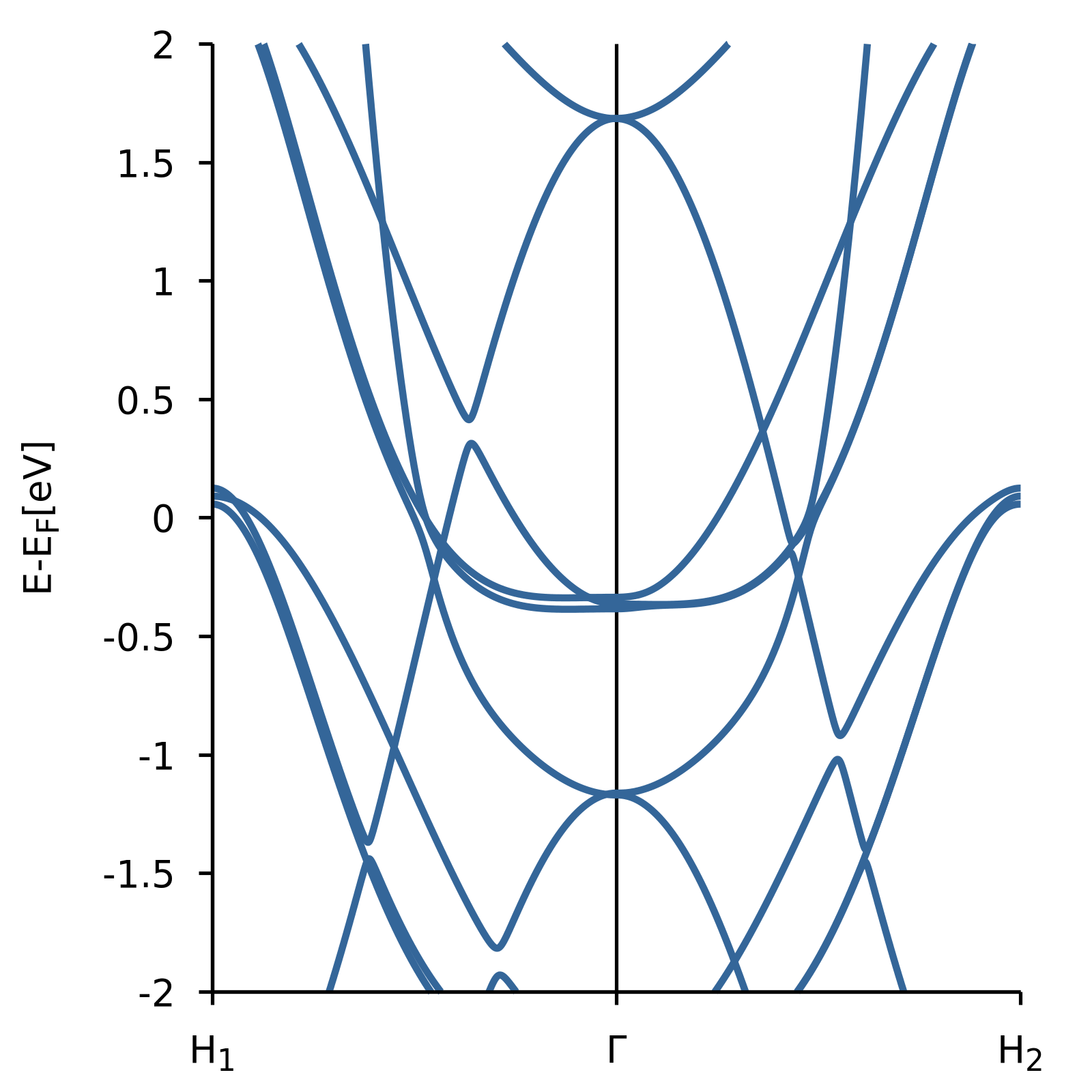}
    \caption{Band structure of bulk iron magnetized along the [001] direction. The band structure is calculated along a line parallel to the magnetization axis ($H_1$-$\Gamma$) and a line perpendicular to the magnetization axis $\Gamma$-$H_2$. Along the parallel direction, the degeneracy of the parabolic band crossing the Fermi level at $H_1$ is lifted due SOC. Along the $\Gamma$-$H_2$ the degeneracy is not lifted. Instead, multiple band crossings are opened up. The inequivalence the two band structures along these two lines demonstrates a magnetic anisotropy electronic structure of the bulk iron. } 
    \label{fig:05-band-structure-fe}
\end{figure}

\begin{figure}
    \centering
    \includegraphics[width=0.8\textwidth]{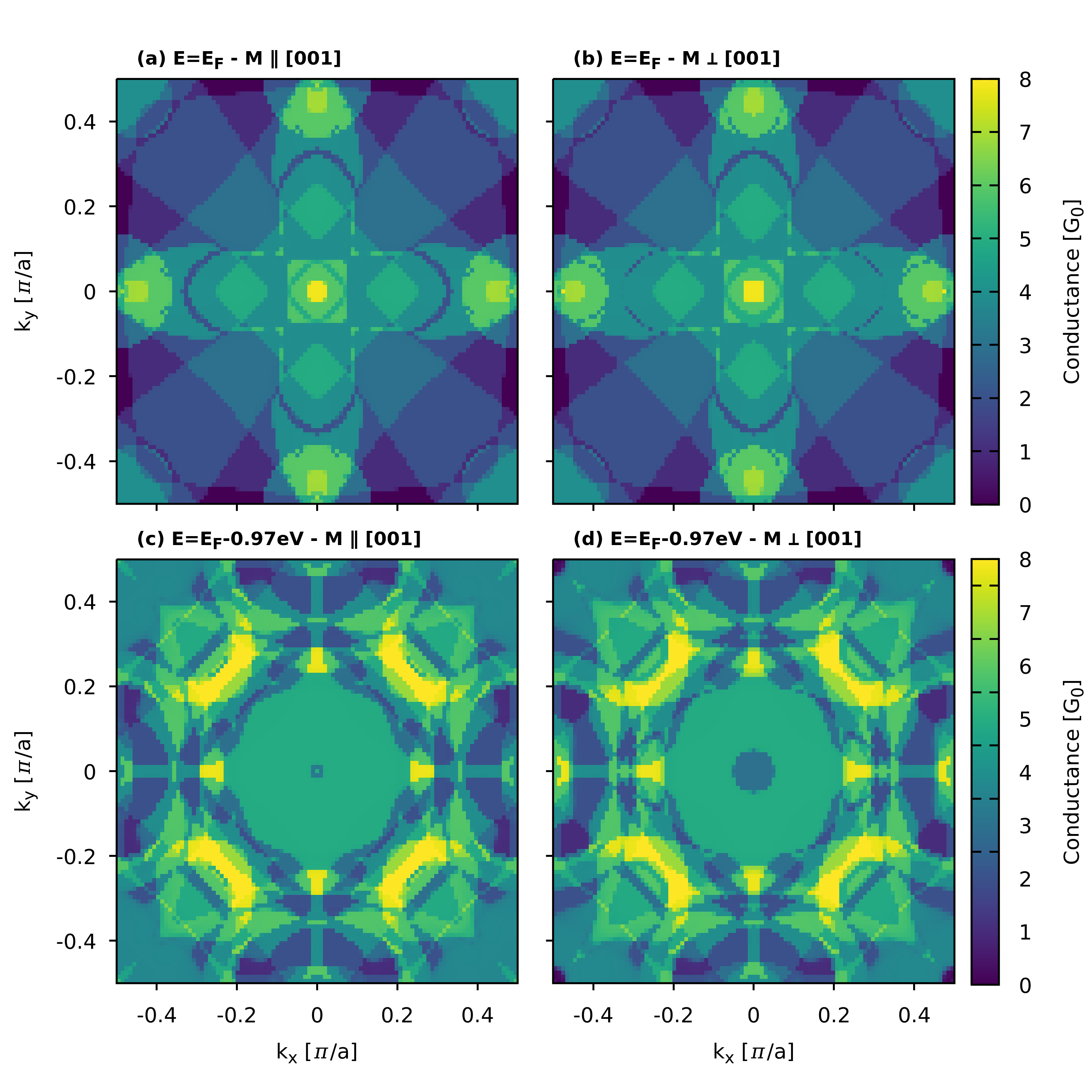}
    \caption{
    Zero-bias transmission of bulk iron for magnetization parallel (a and c) and perpendicular (b and d) to the transport direction ([001]) at the Fermi level ($E=E_F$) (a and b) and at $E=E_F$-0.97~eV (c and d). 
    % For magnetic moments parallel to the transport direction, the transmission function has an extended plateau at 5 around $\Gamma$. For magnetic moments perpendicular to the transport direction, an additional, circular plateau emerges at the center of the Brillouin zone. On this plateau, the transmission is reduced to 3. This reduction of the transmission function is caused by a band gap opening in the transport direction due to SOC in bulk iron. The average transmission is given by 0.4311 and 0.4257 for magnetization parallel and perpendicular to the transport direction, respectively. The spectral density exhibits the same symmetry as the transmission function.
    }
    \label{fig:05-Fe-T}
\end{figure}

Similarly to iron chains, bulk iron (bcc) exhibits magnetic anisotropy in its band structure. Without SOC, the Brillouin zone of iron contains 6 equivalent $H$ points. However, with SOC these 6 equivalent points split into 2 or 3 inequivalent groups depending on the alignment of the spin magnetic moments with respect to the crystal structure. For magnetic moments aligned along one of the high-symmetry axes [001] or [100] there are two inequivalent $H$ points ($H_1$ and $H_2$). Analogous to the iron chain, SOC causes band splitting or avoided crossing depending on the relative alignment of the spin magnetic and orbital magnetic moments of the bands (Fig.~\ref{fig:05-band-structure-fe}). The effects are different along the $\Gamma$-$H_1$ (parallel to the magnetic axis) and $\Gamma$-$H_2$ (perpendicular to the magnetic axis) directions. As a result, the transmission function for transport along the [001] direction also changes depending on the orientation of the magnetic moments. This effect is particularly strong at $E=-0.97$~eV, where a large gap opens in the direction perpendicular to the magnetic moment. This gap opening can also be observed in the zero-bias transmission function for bulk iron when the magnetic moment is perpendicular to the transport direction. In this case, the transmission function at $E=-0.97$~eV features a circular plateau at the center of the Brillouin zone (Fig.~\ref{fig:05-Fe-T} (d)). For magnetic moments parallel to the transport direction, this plateau is absent (Fig.~\ref{fig:05-Fe-T} (c)). Instead, a small ring appears at the center of the Brillouin zone, which  likely corresponds to a mini gap opening at the same energy at another point in the Brillouin zone perpendicular to the transport direction. To calculate the ballistic anisotropic magneto-resistance (BAMR) of iron we need to calculate the average of the transmission function of the whole BZ. For magnetic moments parallel and perpendicular to the transport direction, we obtain a total transmittance of 4.311 and 4.257 respectively. This corresponds to a BAMR of only 1\%. At the Fermi level (Fig.~\ref{fig:05-Fe-T} (a) and (b)), this effect is even further reduced to less than 0.1\%.  Although BAMR is small in bulk iron, we can understand, from the transmission function of bulk iron, why a nano-constriction in iron can lead to a large BAMR. The constriction acts as a phase space filter and restricts transmission to channels with vanish lattice momentum perpendicular to the transport direction ($k_x=k_y=0$). Thus, the reduction of BAMR due to k-space averaging is avoided and BAMR can be orders of magnitude larger than in bulk iron.
Our results for bulk iron are consistent with previous studies of magnetic anisotropy and BAMR in iron~\cite{KhanAnisotropicMagnetoresistance2008,ZwierzyckiCalculatingScattering2008}.

\subsubsection{Magneto Resistance in Fe/MgO/Fe Tunneling Junctions}

\begin{figure}
    \centering
    \includegraphics[width=0.8\textwidth]{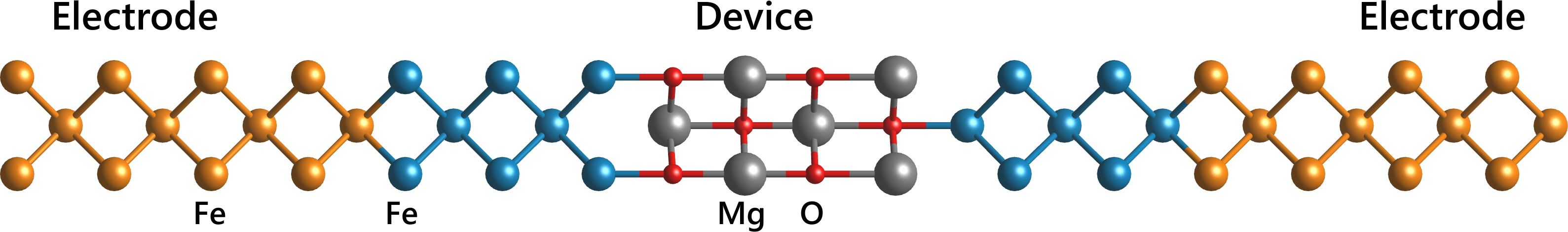}
    \hspace{0.09\textwidth}
    \includegraphics[width=0.08\textwidth]{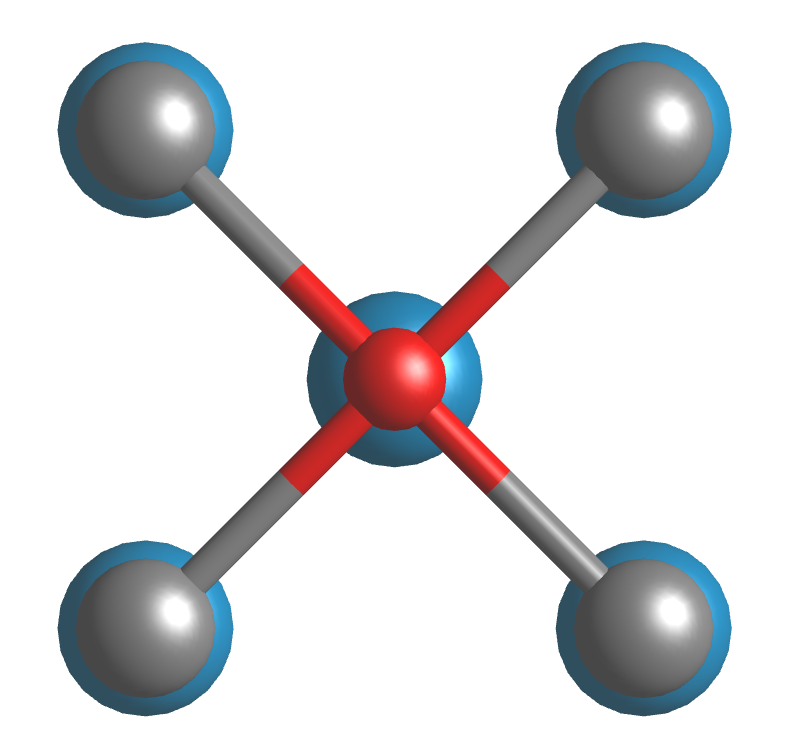}
    \caption{Ball and stick model of a Fe/MgO/Fe junction in side view (left) and a crosssection at the interface (right). At the interface the iron atoms sit directly on top of the oxygen atoms of the MgO and the Magnesium atoms sit in the hollow sites in between.} 
    \label{fig:05-crystal-structure-junction}
\end{figure}

Fe/MgO/Fe tunneling junction consists of a few layers of MgO sandwiched between multiple layers of iron (Figure \ref{fig:05-crystal-structure-junction}). The MgO[110] plane is parallel to the Fe[100], creating a clean interface between the two materials. The iron atoms of the contact layer sit on top of the oxygen sites in MgO.  The transmission of the junction decays exponentially with the number of MgO layers between the iron electrodes~\cite{HeiligerThicknessDependence2008}. Here we choose 4 MgO layers, for which we expect to observe a significant TMR effect,
and a transmittance which is large enough that it does not require excessive computational accuracy. 

\begin{figure}
    \centering
    \includegraphics[width=0.8\textwidth]{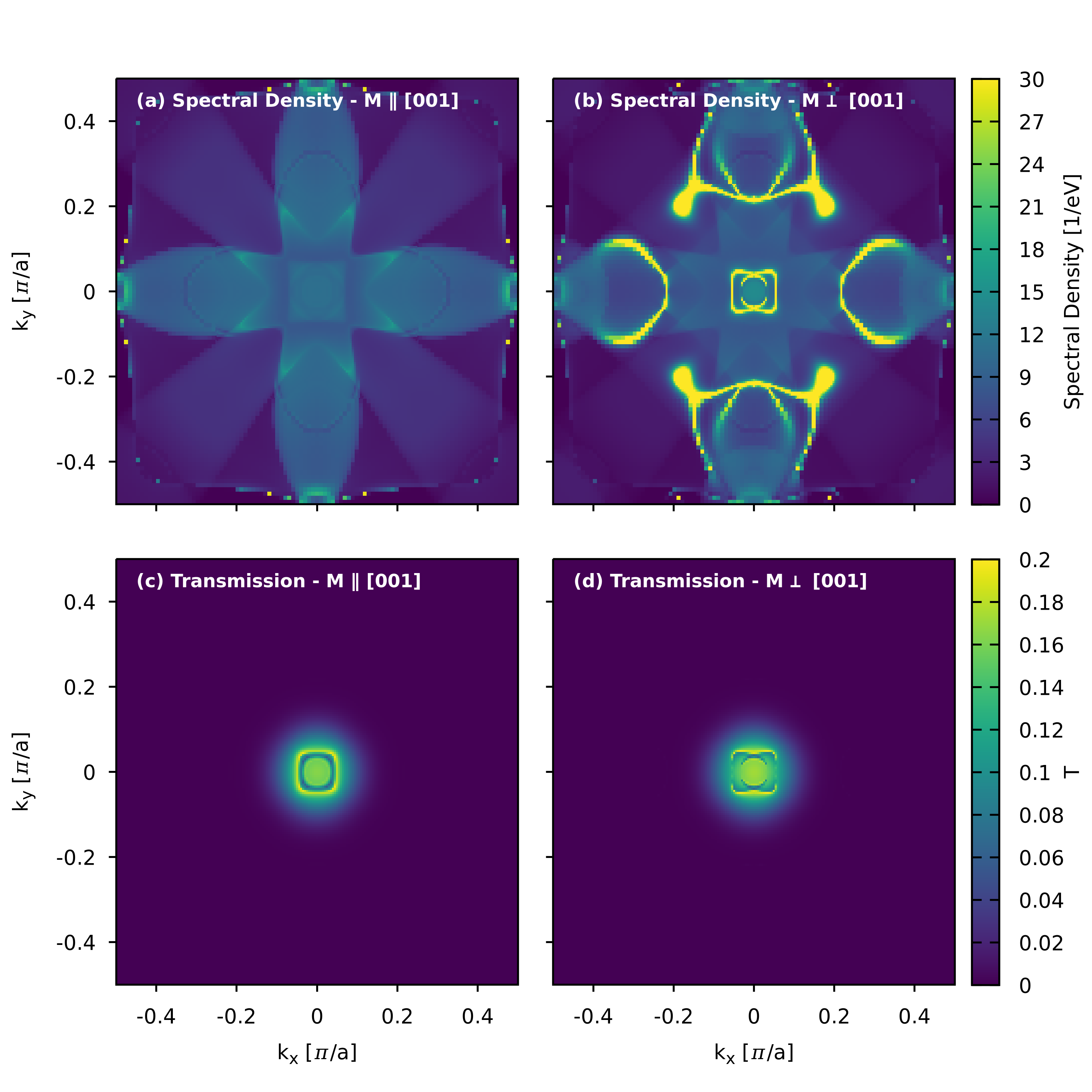}
    \caption{Spectral function and zero-bias transmission of a Fe/MgO/Fe tunneling junction with 4 MgO layers. 
    (a) Spectral function for magnetization in the iron layers parallel to the transport direction, and (b) perpendicular to the transport direction. (c) Transmission functions at $E=-0.97$~eV for magnetization parallel,
    and (d) perpendicular to the transport direction. 
    %For magnetization in the iron layers parallel to the transport direction, the spectral density is smooth throughout the Brillouin zone and resembles the spectral density of bulk iron closely. For magnetization perpendicular to the transport direction, sharp features emerge in the density, and new sharp features emerge in the density. For both orientations of the magnetization, the transmission function is zero in most of the Brillouin zone. Only around the center of the Brillouin zone do we observe significant transmission. Form magnetization perpendicular to the transport direction this feature in the transmission has cubic symmetry. For magnetization perpendicular, to the transport direction, the cubic symmetry is broken and an additional structure in the transmission function emerges. A feature of the same shape can be observed in the density of spectral density.
    }
    \label{fig:05-FeMgOFe-T-A}
\end{figure}

Figure \ref{fig:05-FeMgOFe-T-A} depicts the transmission function of this tunneling junction for electrodes with parallel magnetization at the Fermi level. The MgO tunneling barrier acts as a phase filter and the transmission function becomes strongly localized at the center of the Brillouin zone. The average transmission is reduced by a factor of 650 to 0.0044. Both aspects are consistent with the prediction of previous DFT studies for Fe/MgO/Fe junctions with 4 MgO layers. In contrast to previous studies we include spin-orbit coupling in our calculations and two additional features appears in the zero-bias transmission: a) a small ring at which the transmission is reduced, which corresponds to the ring of reduced transmission in the bulk electrode, and b) if the magnetization is perpendicular to the transport direction, the cubic symmetry of the transmission function is broken, and additional sharp features emerge in the transmission spectrum. 
The decomposition of the transmission function into spin-up and spin-down transmissions reveals that the primary contributions to the transmission stem from the spin channel parallel to the magnetization axis (up). The transmission of the spin-down channel is on average one order of magnitude smaller and features sharp peaks along the reciprocal lattice directions.  

\begin{figure}
    \centering
    \includegraphics[width=0.8\textwidth]{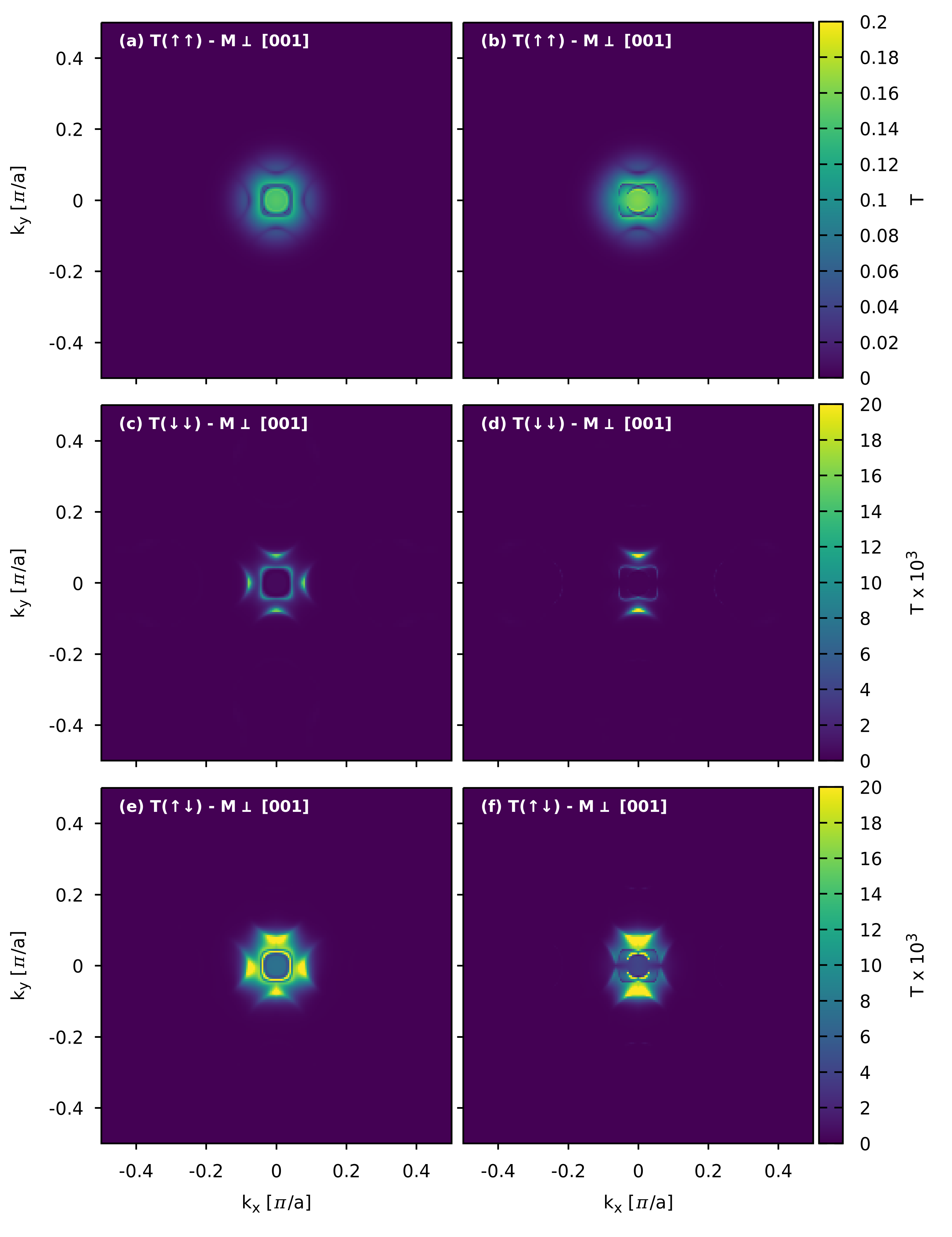}
    \caption{Spin channel projected zero-bias transmission of a Fe/MgO/Fe tunneling junction with 4 MgO layers.}
    \label{fig:05-FeMgOFe-T-spin}
\end{figure}

\subsubsection{Computation Details} 
\begin{figure}
    \centering
    \begin{tikzpicture}
        \node at (0,0) {%
            \includegraphics[height=0.49\textwidth]{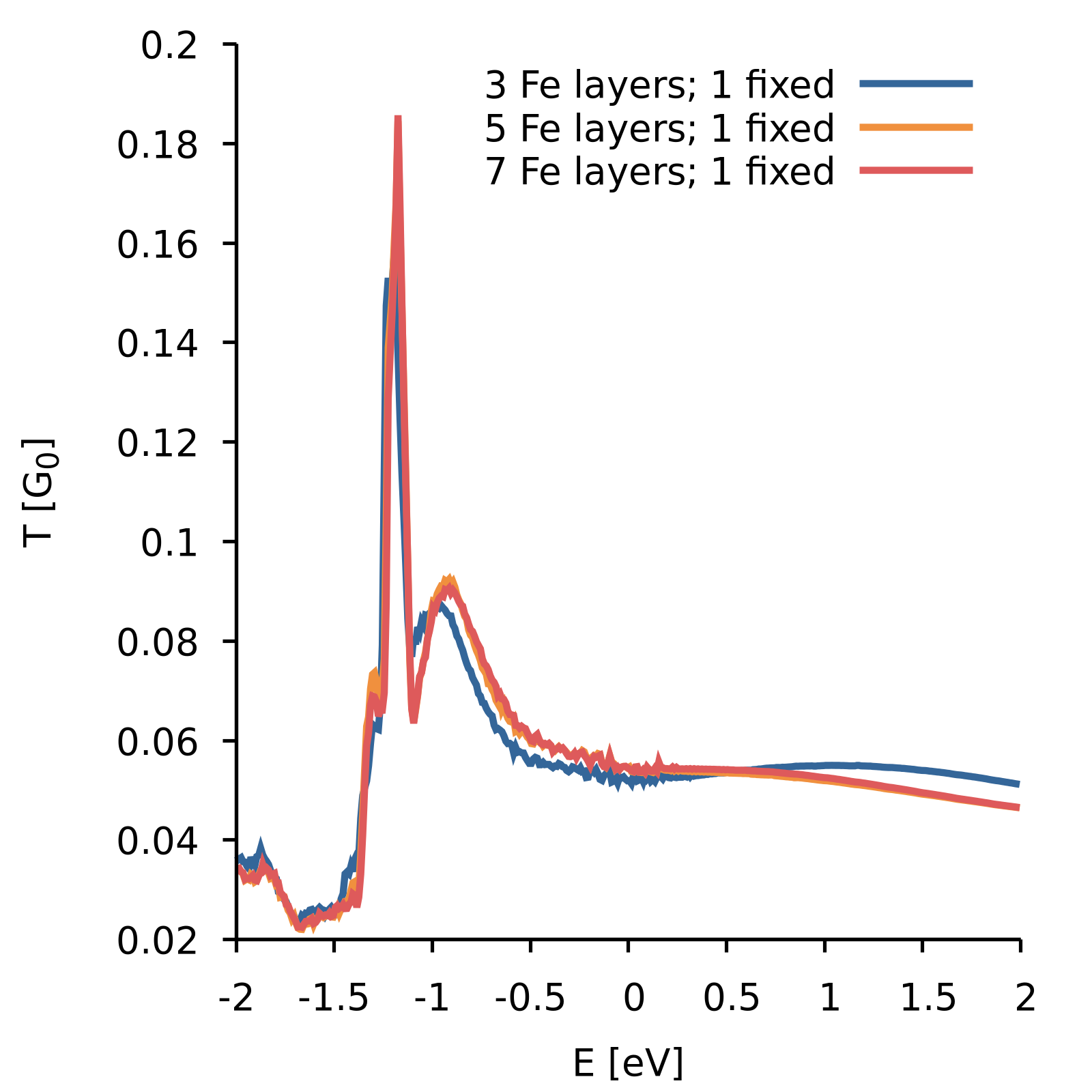}%
            \includegraphics[trim={11cm 0 0 0},clip,height=0.49\textwidth]{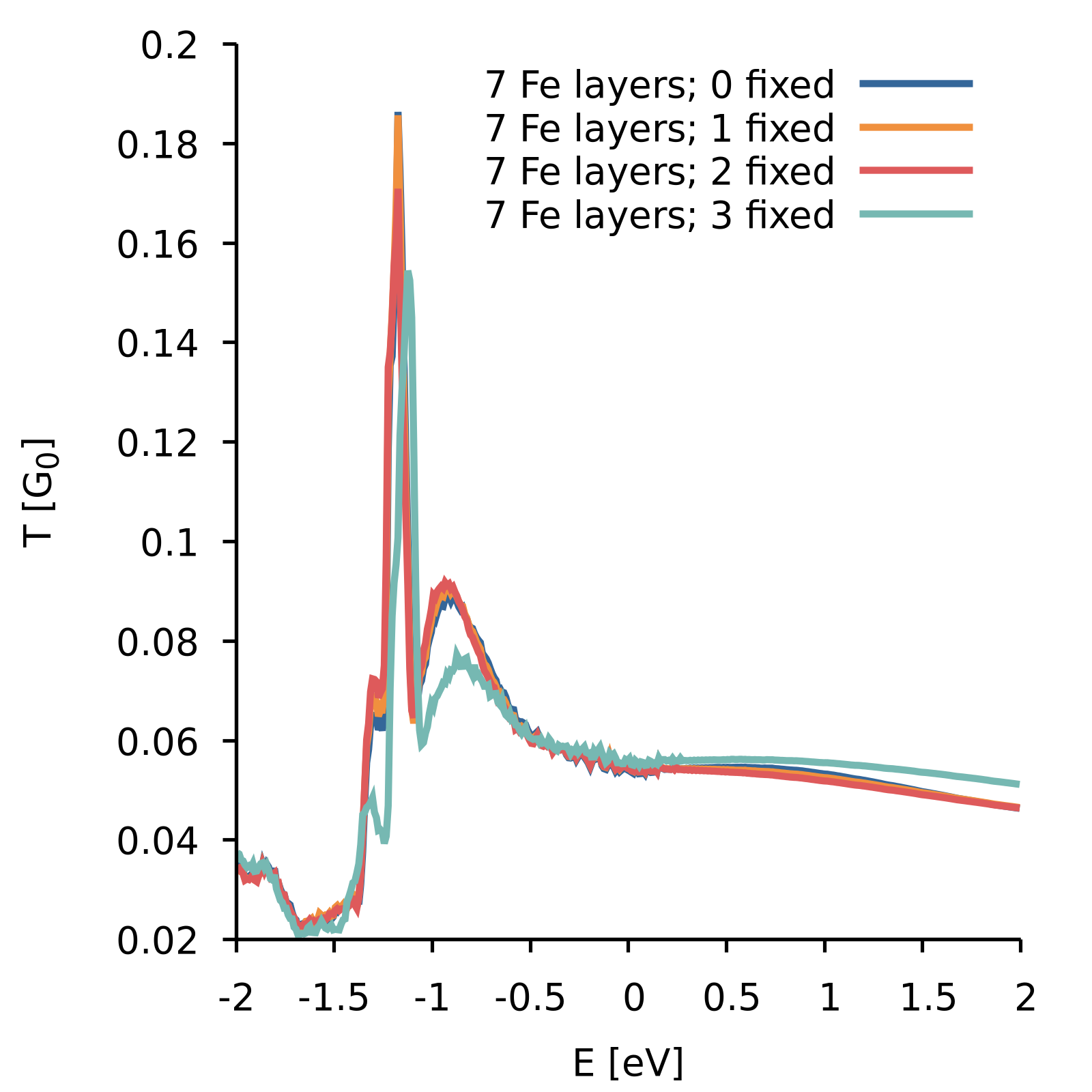}
        };
        \node at (-3.75,2.75) {(a)};
        \node at (1.15,2.75) {(b)};
    \end{tikzpicture}
    \caption{Convergence of the transmission function in a Fe/MgO/Fe junction with number of iron layers between the MgO layers and the electrodes (a) and with number of fixed layers (b).}
    \label{fig:05-FeMgOFe-converge}
\end{figure}

For bulk iron, our transport setup consists of 4 MgO layers sandwiched between 13 layers of iron on each side. 
In the electrode calculations, we sample the reciprocal space using a Monkhorst-Pack grid with  100 $\mathbf{k}$ points along the semi-infinite direction and $16\times 16$ transverse $\mathbf{k}$ points along the transversal direction. 
We use the same number of transversal $\mathbf{k}$ points in the NEGF+DFT calculation. 
The lattice constant of bulk iron is optimized up to a stress tolerance of 0.1~meV/\AA$^3$. 
We then relax the atoms in the central part of the transport device until the forces acting on them are smaller than 10~meV/\AA{} and stress along the direction perpendicular to the interface is smaller than 0.1~meV/\AA{}. During the relaxation, we keep the electrode regions (8 atoms per principal layer) and the lattice vectors in the transversal direction fixed. Comparison of the transmission function for Fe/MgO/Fe tunneling junctions with a different number of additionally fixed layers shows that relaxing all atoms between the two electrodes introduces additional scattering (Fig.~\ref{fig:05-FeMgOFe-converge}). Keeping one or two extra layers fixed, removes this contact scattering. Here we keep a single layer fixed. Similarly, we find that including less than 5 iron layers between the electrodes and the MgO layer introduces additional scattering. 

For the parallel magnetic configurations, we initialize the density matrix from a normal SIESTA calculation with the same geometry and magnetic structure. 
For the antiparallel magnetic configuration, we initialize the density matrix from a SIESTA calculation of 1\ttimes{} 1\ttimes{}  2 supercells with matching magnetic moments at the periodic boundary [ 13 layers Fe($\uparrow$), 4 layers MgO 26 layers Fe($\downarrow$), 4 layers MgO, 13 layer Fe($\uparrow$), repeat....]. In the \textsc{TranSIESTA} calculation the second half of this cell is discarded (``buffer'').

All calculations were carried out within the fully relativistic pseudo-potential formalism,~\cite{CuadradoFullyRelativistic2012} and the PBE functional~\cite{PerdewGeneralizedGradient1996}. 
We used a double-zeta-polarized basis set for the Fe atoms (energy shift 0.27~eV, split norm 0.15), a real space grid with a cut-off of 800~Ry, and an electronic temperature of 10~meV. The pseudo-potentials were generated using the approach of Troullier and Martins\cite{TroullierEfficientPseudopotentials1991} with a valence configuration of 3d$^{6}$4s$^{2}$ for iron, 2p$^6$3s$^2$ for magnesium, and 2s$^2$2p$^4$ for oxygen. The pseudo-potential for iron included non-local core-corrections. 

\subsection{Transition metal dichalcogenides (2D)}
In the example of iron chains, we have seen that our code works well for spins with noncollinear structure. However, the SOC in Fe is relatively weak. To ensure the correct treatment of systems with significant spin-orbit interactions, we simulate two infinite transition metal dichalcogenide (TMD) monolayers (MoS$_2$, W$_2$), construct a lateral hetero-junction of the two materials, and study the effect of a single sulfur vacancy on the transport properties of MoS$_2$. 

\subsubsection{Monolayer MoS$_2$ and WS$_2$}

% \begin{figure}
%     \centering
%     \includegraphics[width=0.9\textwidth]{02-TMD_junction-top-width-key.png}
    
%     % \includegraphics[width=0.096\textwidth]{TMD-Vacancy-side-a.png}\\
%     % \includegraphics[width=0.47\textwidth]{TMD-Vacancy-side-b.png}
%     \includegraphics[width=0.51\textwidth]{02-TMD_junction_bands+T.png}
%     \includegraphics[width=0.26\textwidth]{02-TMD_junction_VH_potentialDrop.png}
%     \caption{Lateral heterojunction between MoS$_2$ and WS$_2$: crystal structure (top), band structure of MoS$_2$ and WS$_2$ (bottom left),
%     {\textcolor{red}[correct the subscripts in the labels]} zero-bias transmission function of pristine MoS$_2$, WS$_2$, and heterojunction (bottom center) and electrostatic potential (bottom right) {\textcolor{red}[fonts not legible, line too thin. Figure too small, Macroscopic and Planar average are not labeled]}.
%     The potential drop in the averaged electrostatic potential is the signature of the built-in electrostatic field due to the p-n nature of the heterojunction.
%     }
%     \label{fig:02-tmd-heterojunction}
% \end{figure}

\begin{figure}
    \centering
    \includegraphics[width=\textwidth]{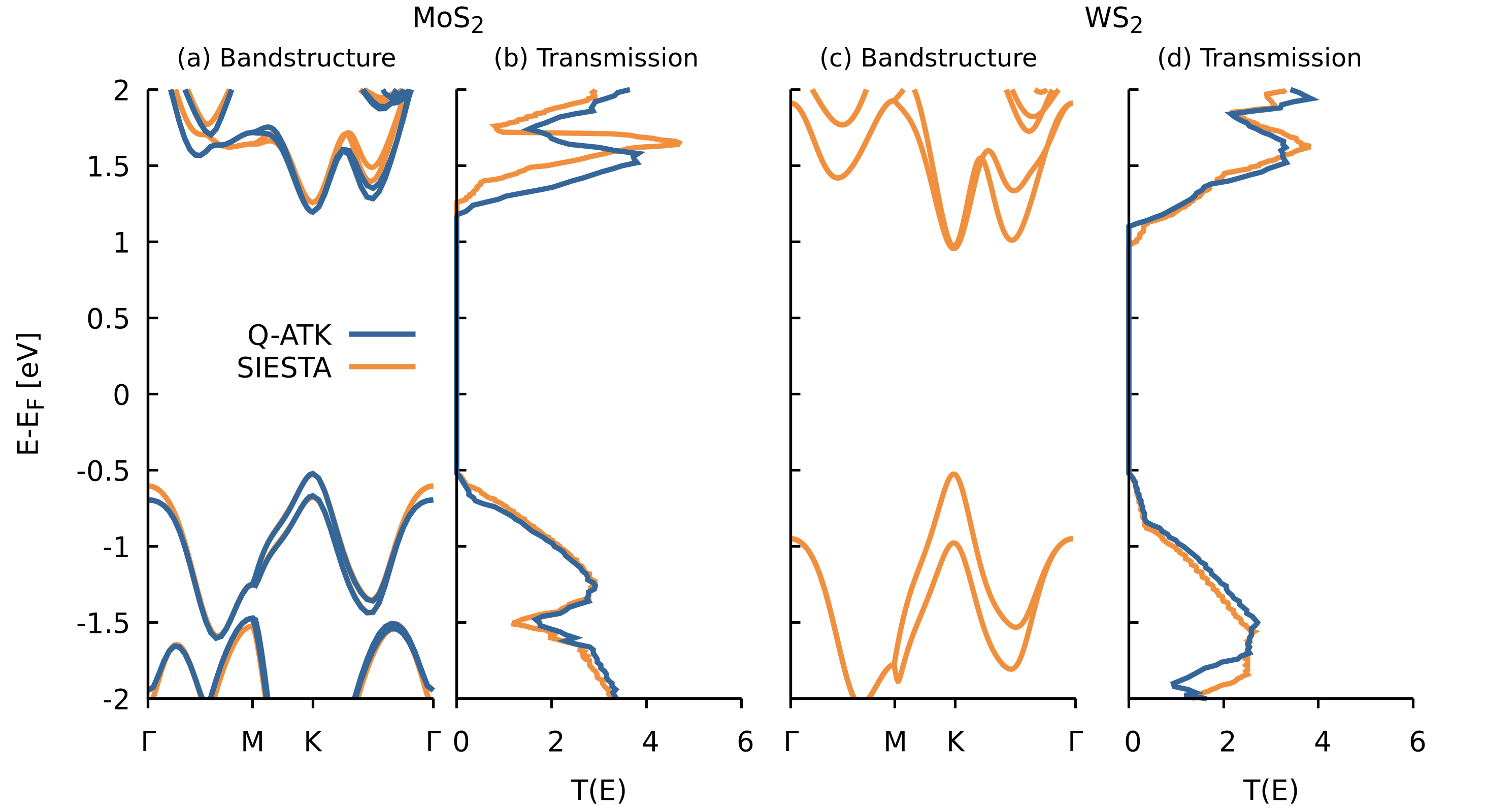}
    \caption{Band structure and zero-bias transmission function of monolayer MoS$_2$ (a and b) and WS$_2$ (c and d). Orange lines show band structure and transmission function calculated with (\textsc{Tran})SIESTA and blue, dashed lines those calculated with \textsc{QuantumATK}. The simulation results agree well within a few eV around the Fermi level. Along the $M-K$ and $K-\Gamma$ directions, the band dispersion of MoS$_2$ obtained with the two codes is slightly different resulting in a shift of the first peak of the transmission function above the transport gap.}
    \label{fig:02-TMD-bands-T}
\end{figure}

MoS$_2$ and WS$_2$ monolayers exhibit a direct band gap at $K$ and $K'$. In MoS$_2$ this band gap is 1.79~eV wide and WS$_2$ 1.48~eV. Around the valence band maximum (VBM) the bands of MoS$_2$ and WS$_2$ are split by 0.15~eV and 0.45~eV, respectively (Fig.~\ref{fig:02-TMD-bands-T}). SOC acts as a magnetic field perpendicular to the monolayer affecting the top valence bands near $K$ and $K'$ causing band splitting. The direction of this interaction is opposite for $K$ and $K'$, i.e. the spin moment of the upper split band is $\pm1$ for $K$ and $K'$ respectively.
These properties of the electronic ground state of MoS$_2$ and WS$_2$ and the overall band dispersion are comparable to our reference calculations with QuantumATK\cite{QuantumATK} and other DFT studies~\cite{ZahidGenericTightbinding2013,ZengOpticalSignature2013}.

In the pristine monolayers, the transport gap is equal to the band gap, and the electron conductance increases linearly at the edges of the band gap. 
% The two split bands at $K$ are spin-polarized perpendicular to the monolayer, and the order of the spin-up and spin-down bands is opposite for $K$ and $K'$. As a result, the k-averaged transmission function is the same for the spin-up and spin-down channels. 
% However, for individual k points, the spin-up and spin-down channels have different contributions.
The simulation results agree well within a few electronvolts around the Fermi level. Thus, we demonstrate that it is possible with \textsc{TranSIESTA} to correctly calculate the transmission of bulk materials self-consistently with strong spin-orbit coupling. The most noticeable deviation occurs for MoS$_2$ at approximately 1.6~eV, where the bands calculated with SIESTA are flatter and the two maxima along the $M-K$ and $K-\Gamma$ directions align. As a result, the maximum of the TranSIESTA transmission function is slightly higher compared to \textsc{QuantumATK}, and its position is shifted upward. These deviations are already present at the ground state DFT level and likely arises due to the use of different pseudopotential and basis set.

\subsubsection{Lateral MoS$_2$-WS$_2$ Heterojunction}

\begin{figure}
    \centering
    \includegraphics[width=\textwidth]{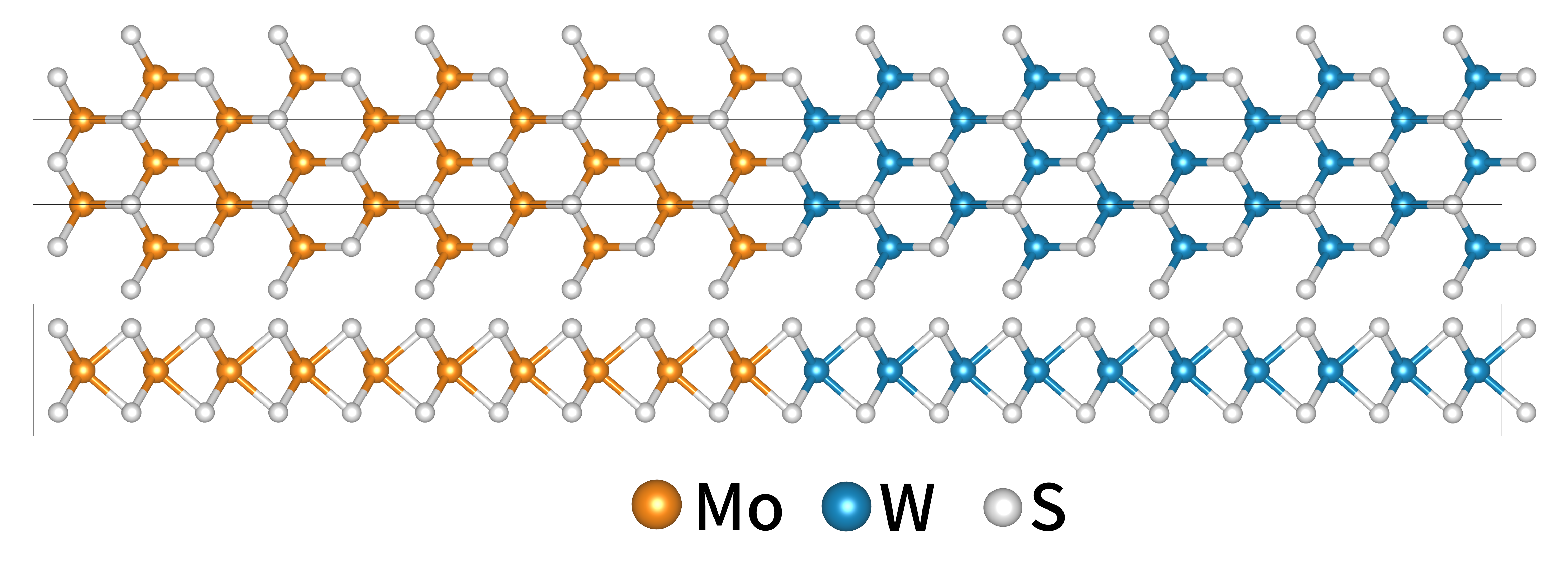}
    \caption{Crystal structure of a lateral MoS$_2$-WS$_2$ heterojunction top view (top) and side view (bottom)}
    \label{fig:02-TMD-junction-structure}
\end{figure}

Simulations of bulk semiconductors within the NEGF formalism are unproblematic. However, when two different semiconductors come into contact, problems arise because it is unclear how to match the potentials of the two materials to those of the different bulk electrodes. Furthermore, the low carrier density in semiconductors gives rise to very long electrostatic screening lengths, which would require extremely large device regions, to ensure proper screening of the electrode regions. The measurement of the built-in potential along WS$_2$-MoS$_2$ lateral junction (Figure \ref{fig:02-TMD-junction-structure}) shows a step between the two potentials of about 0.1~eV which is more than 2~µm wide~\cite{WangOneStepSynthesis2022,ChenLateralBuiltIn2015}. To avoid these issues, we uniformly gate the WS$_2$ and MoS$_2$ monolayers and the junction by adding a plane with constant charge density ($5.76\cdot10^{-3}$~e/\AA{}$^2$) 30~\AA{} below the monolayers\cite{C5CP04613K}. This plane of charge induces a countercharge of the same magnitude but of the opposite sign in the monolayer. Therefore, the valence bands of the two materials are partially depopulated, making them metallic. 
The transmission function of both materials is similar near the Fermi level ($T(E_F)\approx0.4$). In comparison to the ungated monolayer, the transmission functions are rigidly shifted, which indicates that this amount of doping has no significant effect on the electronic structure other than shifting the chemical potential. 

\begin{figure}
    \centering
    \includegraphics[width=\textwidth]{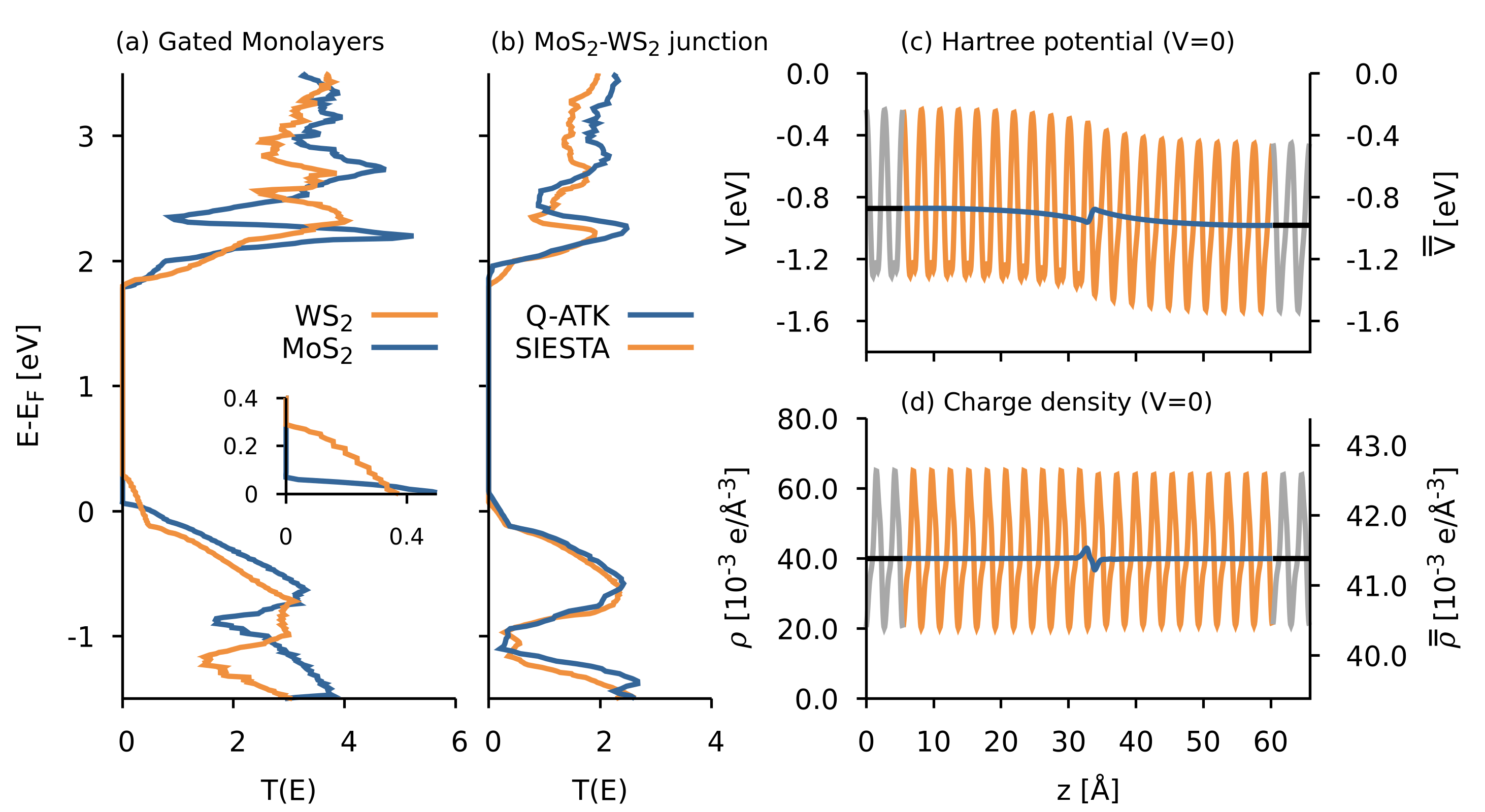}
    \caption{Zero-bias transmission of hole-doped ($5.76\cdot10^{-3}$~e/\AA{}$^2$) MoS$_2$ (a - blue line), WS$_2$ (a - orange line) monolayers and a lateral MoS$_2$-WS$_2$ heterojunction with same doping concentration (b); Planar- and macro averaged electrostatic potential (c) and charge density (d) of the same heterojunction. Comparison of the transmission function of the heterojunction in (b) calculated with \textsc{TranSIESTA} and {QuantumATK} shows very good agreement.} 
    \label{fig:02-TMD-bulk-T}
\end{figure}

\begin{figure}
    \centering
    \includegraphics[width=\textwidth]{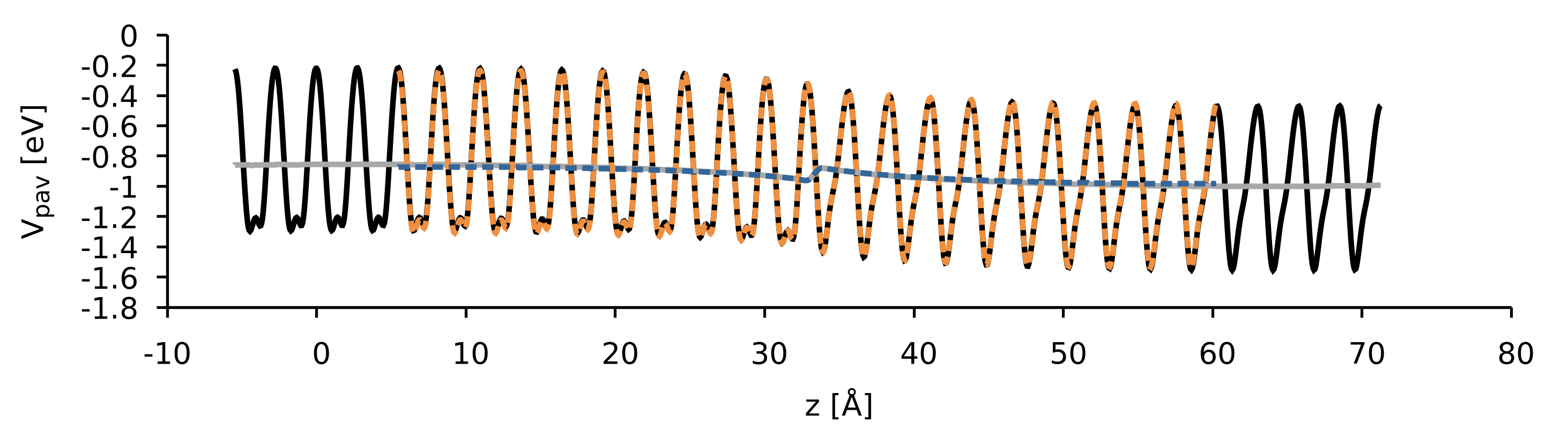}%
    \caption{Planar- and macro averaged electrostatic potential of an MoS$_2$-WS$_2$ heterojunction calculated with  two simulation cells containing 40 and 56 atoms respectively. The near identical behavior in both cases demonstrates that the electrodes are appropriately screened from the potential drop at interface.}
    \label{fig:05-TMD-converge}
\end{figure}

To evaluate whether this level of doping is sufficient to screen electrodes from the potential step at the interface, we calculate the planar and macroscopic average~\cite{BaldereschiBandOffsets1988} of the electrostatic potential ($V_H$): 
\begin{align}
    \textrm{Planar average:}&\quad\overline{V}_H(z) = \frac{1}{S}\iint_{S}\dd{x}\dd{y} V_H(x,y,z)\\
    \textrm{Macroscopic average:}&\quad\overline{\overline{V}}_H(z)= \iint\dd{z'} \frac{1}{a}\theta(\frac{a}{2}-\abs{z'})\overline{V}_H(z')
\end{align}
where $x$ is the position along the vacuum direction, $y$ is the position along the periodic direction, $S$ is the unit cell area in the transverse direction, $z$ is the position along the transport direction, $a$ is the strained lattice constant along $z$, and $\theta$ is the Heaviside step function. The convolution with a step function filters out the components of the planar average with a period of $a$. In Fig.~\ref{fig:02-TMD-bulk-T} the planar and macroscopic averages of the electrostatic potential and the charge density are displayed. The planar average is represented as orange and light gray lines in Fig.~\ref{fig:02-TMD-bulk-T} (a and b) for the scattering region and the electrode, respectively. The planar average oscillates throughout the system: close to the nuclei, the potential is the lowest, and the charge density is the highest, and vice versa in between the nuclei. These periodic oscillations are absent in the macroscopic average. Due to uniform doping, the macroscopic average of the charge density is equal in the two electrodes and close to constant in the scattering region.
Immediately at the interface layer, we observe a small depletion region with electron exchange from WS$_2$ to MoS$_2$ and a step in the electrostatic potential of approximately 0.1~eV, which is consistent with the built-in potential observed experimentally~\cite{WangOneStepSynthesis2022,ChenLateralBuiltIn2015}. At contact with the electrode, the potential is smooth, which indicates that there are no obvious problem at the connection between the device and electrode regions. However, to unequivocally check whether the electrodes are appropriately screened from the interface it is necessary to perform the same calculation with a larger device region. Fig.~\ref{fig:05-TMD-converge} showcases a comparison of the electrostatic potential for two simulation cells with 56 and 40 atoms respectively. In both cases the electrostatic potential is near identical, thus demonstrating that the electrodes are appropriately screened from the interface. 

Near the Fermi level, the transmission function of the heterojunction follows the minimum of the transmission functions of the two electrodes. This indicates that the interface introduces very little scattering and that the limiting factor for transmission through this system is the availability of transport channels in the electrodes. At $E=-1$~eV backscattering is much more significant, and the transmission function of the junction almost drops to zero. Since this effect occurs well below the Fermi level, it should not affect the transport for reasonable hole doping levels. This behavior and the overall shape and amplitude are consistent between our SIESTA and \textsc{QunatumATK} calculations. 

\begin{figure}
    \centering
    \includegraphics[width=\textwidth]{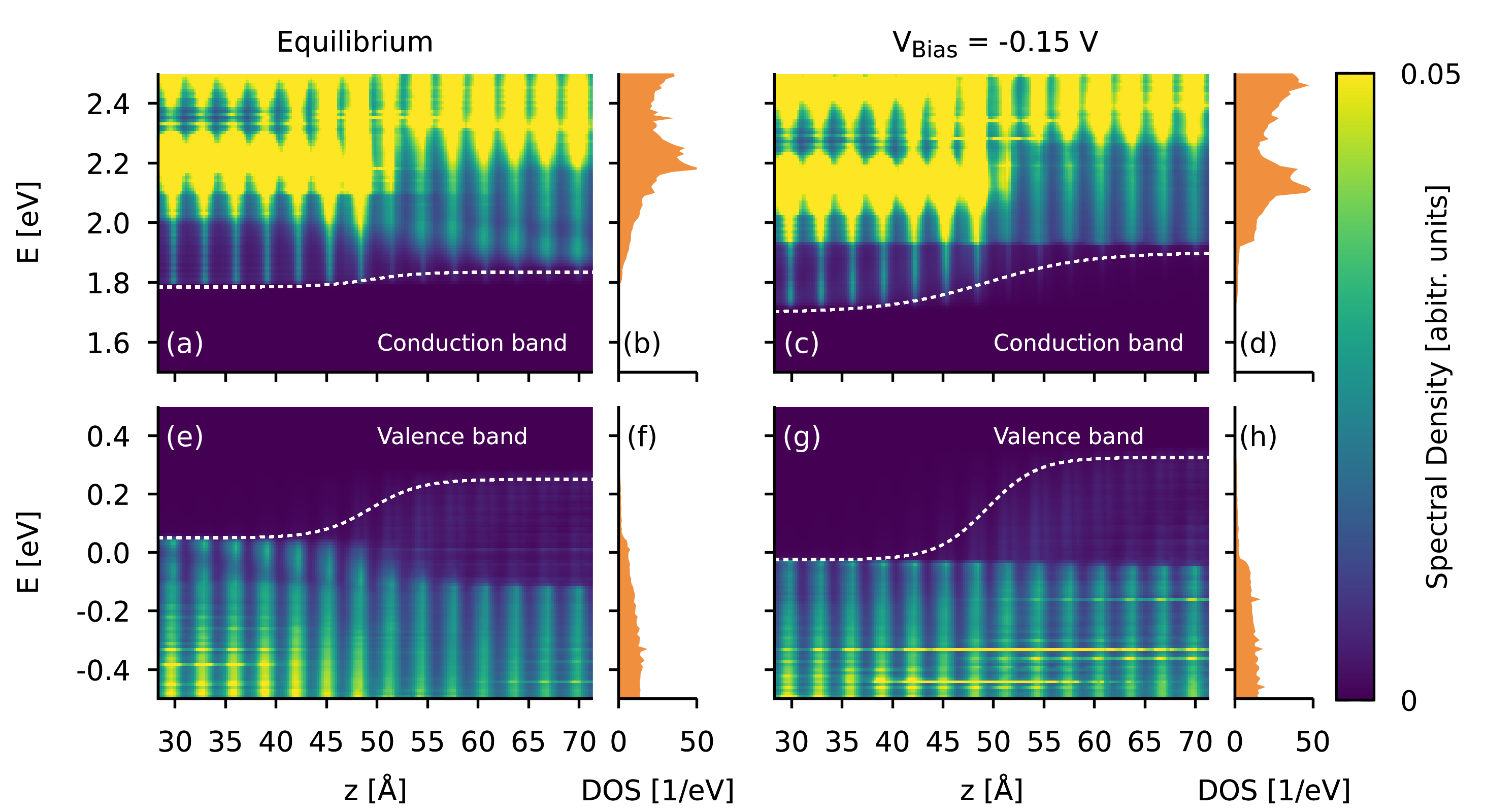}
    \caption{Local density of states averaged over the transverse directions of a lateral MoS$_2$-WS$_2$ heterjunction without applied bias (a-d) and with -0.15V bias (e-h). To highlight the band bending in the heterojunction the color scale is cut off. In the bright yellow regions in conduction bands the local density of states exceeds the maximum value of the color range and is, therefore, not accurately represented. }
    \label{fig:02-TMD-bending}
\end{figure}

After doping, the edge of the valence band in WS$_2$ (0.30~eV) is approximately 0.25~eV higher than that of MoS$_2$ (0.05~eV), which should lead to a band bending effect near the interface. To visualize this effect and test this aspect of our implementation, we calculate the local density of states $\rho(E,\vb{r})$ 
\begin{align}
    \rho^{\sigma\sigma'}(E,\vb{r}) &= \sum_{\mu\nu} \phi_\mu(\vb{r})\phi_\nu(\vb{r})\frac{1}{2}(\mathbf{G}-\mathbf{G}^\dagger)^{\sigma\sigma'}_{\mu\nu}\\
    \rho(E,\vb{r}) &= \Re{\rho^{\uparrow\uparrow}(E,\vb{r}) + \rho^{\downarrow\downarrow}(E,\vb{r})}
\end{align}
and integrate over the $x$ and $y$ directions. This density is visualized in Fig.~\ref{fig:02-TMD-bending} as a function of energy and position along the transport direction. The magnitude of the local density is indicated by the dark-blue-to-yellow color scale. The edge of the highest valence band of MoS$_2$ appears as a sharp feature $0.05$~eV above the Fermi level (Fig.~\ref{fig:02-TMD-bending} (e)). 
In contrast, the edge of the band of WS$_2$ is visible only as a faint halo. 
In WS$_2$ the valence band maximum lies 0.4~eV above the band maximum at $\Gamma$ and 0.45~eV above the next maximum at $K$. This results in a low density of states between  -0.1~eV and 0.25~eV.
The larger contrast on the MoS$_2$ side is a reflection of a comparatively sharp increase in the number of states at this energy. This effect is also visible in the transmission function of bulk MoS$_2$ which increases rapidly at the edge of the band (Fig.~\ref{fig:02-TMD-bulk-T} (inset)). In the conduction bands, the total density of states is significantly higher (Fig.~\ref{fig:02-TMD-bending} (b) and (f)), and the band edges are much clearer (Fig.~\ref{fig:02-TMD-bending} (a)). 
Upon application of a bias (Fig.~\ref{fig:02-TMD-bending} (c,h)) the bands are shifted by $\pm V_{\mathrm{Bias}}/2$ away from the Fermi level. 
These trends are consistent with the expected behavior of a junction with misaligned band edges. Furthermore, the intensity of the local density of states is consistent with the total density of states in the heterojunction. 

\begin{figure}
    \centering
    \includegraphics[width=\textwidth]{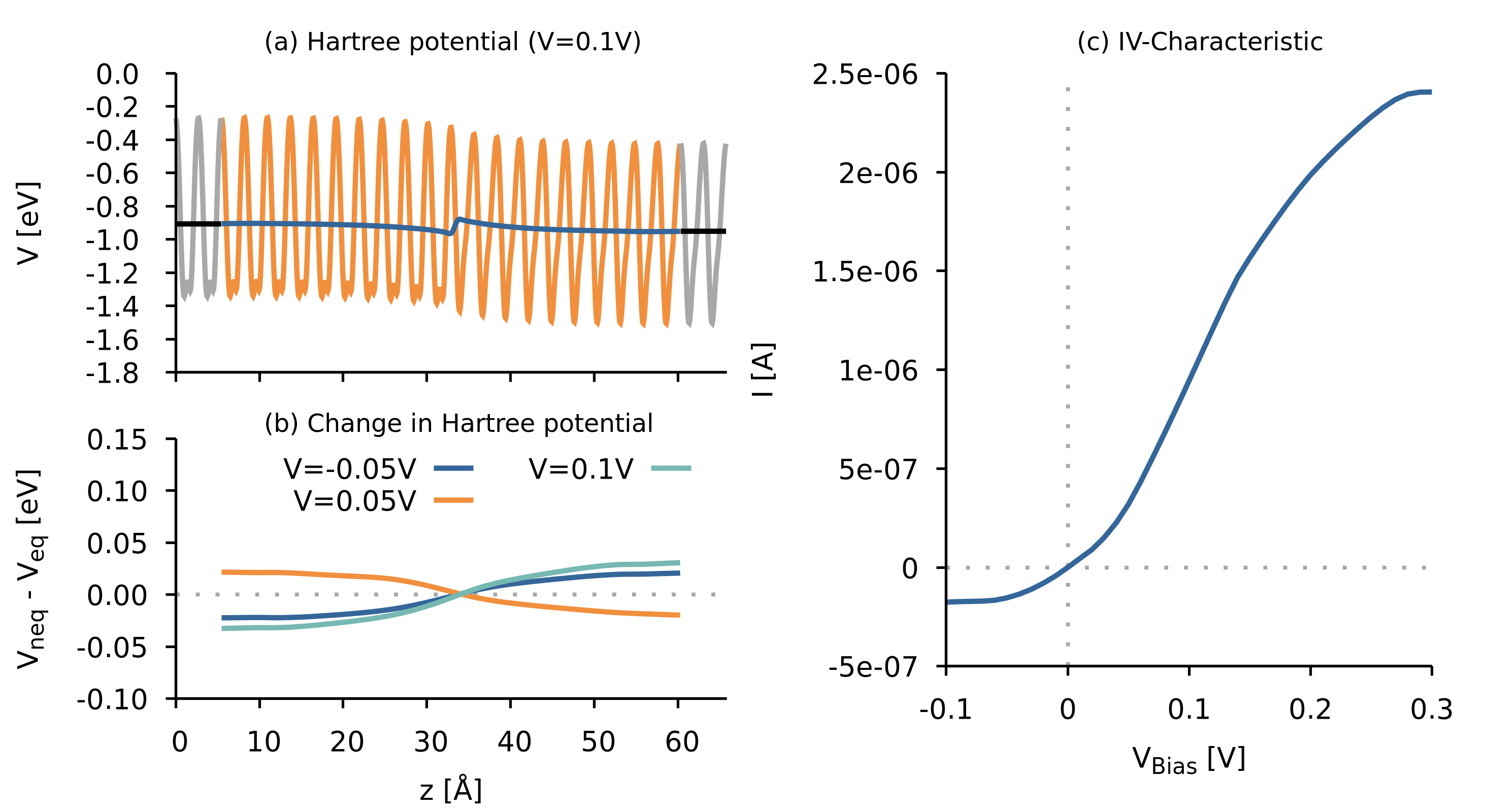}
    \caption{Zero-bias transmission of hole-doped ($5.76\cdot10^{-3}$~e/\AA{}$^2$) MoS$_2$ (a - blue line), WS$_2$ (a - orange line) and lateral MoS$_2$-WS$_2$ heterojunction with same doping concentration (b); planar- and macro averaged electrostatic potential (c) and charge density (d) of the same heterojunction. The band gap in WS$_2$ is approximately 0.02\,eV larger than in MoS$_2$ }
    \label{fig:02-TMD-bias}
\end{figure}

Fig.~\ref{fig:02-TMD-bias}(b) depicts the change in the planar averaged potential for different values of applied bias. The bias-induced potential is continuous and has an infliction point at the interface. Toward the electrodes the potential change becomes constant. For perfect three-dimensional metals, the screening length is vanishing small, and the potential should drop off sharply right at the interface. However, in our simulations, the electrodes are neither three-dimensional nor are they perfect metals. Therefore, the screening length becomes finite, and we observe this as a smoothed potential dropoff. 
In Fig.~\ref{fig:02-TMD-bias} (c) we show the IV characteristic of the heterojunction for bias voltages for which the chemical potential in the electrodes remains within the valence bands (-0.1~eV to 0.25~eV).
Depending on the direction of the applied bias, the valence bands of the two materials are either further misaligned (negative bias) or become more aligned (positive bias). As a result, the IV characteristic of the MoS$_2$-WS$_2$ heterojunction is asymmetric. For a small bias (between -10~meV and 10~meV) the current is proportional to the applied bias, as expected. For a larger positive bias, the conductivity increases until the band edges are aligned around $V_{\mathrm{Bias}}=0.15$\,eV at which point the conductivity decreases slightly. For negative bias, the conductivity decreases until the edge of the valence band of MoS$_2$ falls below the Fermi level at $V_{\mathrm{Bias}}=-0.1$\,eV and the electrode become semiconducting. This behavior corresponds to the diode-like behavior expected in this type of semiconductor junction. The relatively large leakage current for negative biases can be attributed to the high doping concentration in our calculations.

\subsubsection{Computational details}
We calculate the transmission function of the bulk monolayers along the primitive unit cell vector. 
In \textsc{TranSIESTA} we calculate the bulk transmission using the primitive unit cells of the monolayers.
Since \textsc{QuantumATK} only supports electrodes with rectangular unit cells, we, therefore, choose the conventional unit cell of the TMDs for the bulk calculations with \textsc{QuantumATK}. The width of the transversal dimension of the conventional unit cell is twice as large as in the primitive cell. The corresponding transmission function is also twice as large. To correct for this geometric factor, we divide the \textsc{QuantumATK} transmission function by a factor of 2. As a fail-safe, we also verified that using the same conventional unit cell in \textsc{TranSIESTA} yields a 2-times higher transmission function consistent with \textsc{QuantumATK}.

We perform the NEGF calculations in a unit cell containing layers of MoS$_2$ and WS$_2$, including an electrode layer on each side of the simulation cell. The in-plane lattice constants of the two monolayers are strained by 0.4\% to match the two materials ($a_{Lat}=3.165$~Å).
The distance between the Mo and S layers is slightly different on the two sides and amounts to 1.574~Å on the MoS$_2$ side and 1.607~Å on the WS$_2$ side. Monolayers are separated by 70~Å of vacuum in the perpendicular direction. We calculate the transport properties in the direction of the first lattice vector. We performed electrode calculations in the conventional unit cell on a 111\ttimes 15\ttimes 1 
$\mathbf{k}$ grid and use Bloch expansion along the second lattice vector to determine the self-energy in the larger supercell\cite{PapiorImprovementsNonequilibrium2017}. All calculations were performed with the PBE functional\cite{PerdewGeneralizedGradient1996} and norm-conversing pseudo-potentials from the PseudoDojo database\cite{vanSettenPseudoDojoTraining2018}(standard accuracy).

\subsection{CNT with magnetic molecule (1D) }
\begin{figure}
    \centering
    \begin{tikzpicture}
    \node (b) at (0,0) {\includegraphics[height=3.2cm]{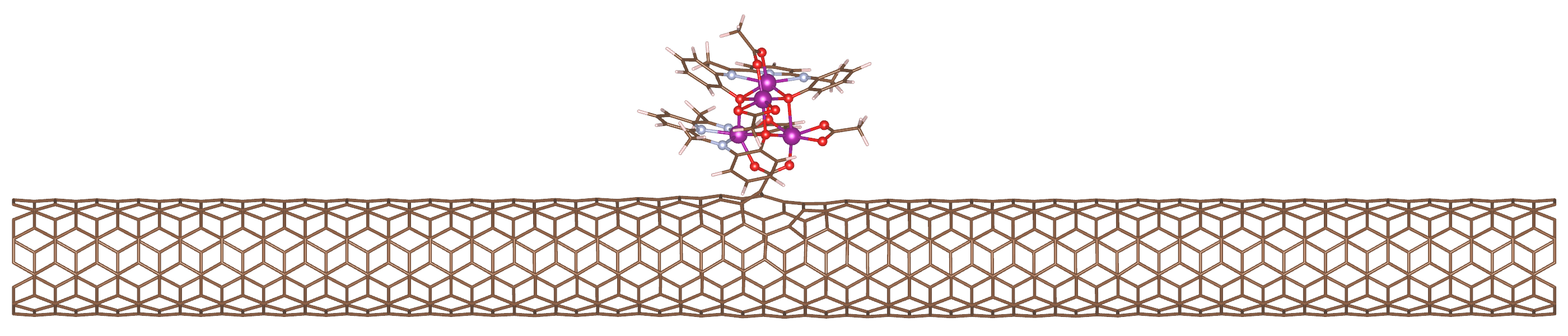}};
    \node[above right=0.1cm and 0.0cm of b.north west] (a) {\includegraphics[trim={42cm 0 0 0},clip,height=3.2cm]{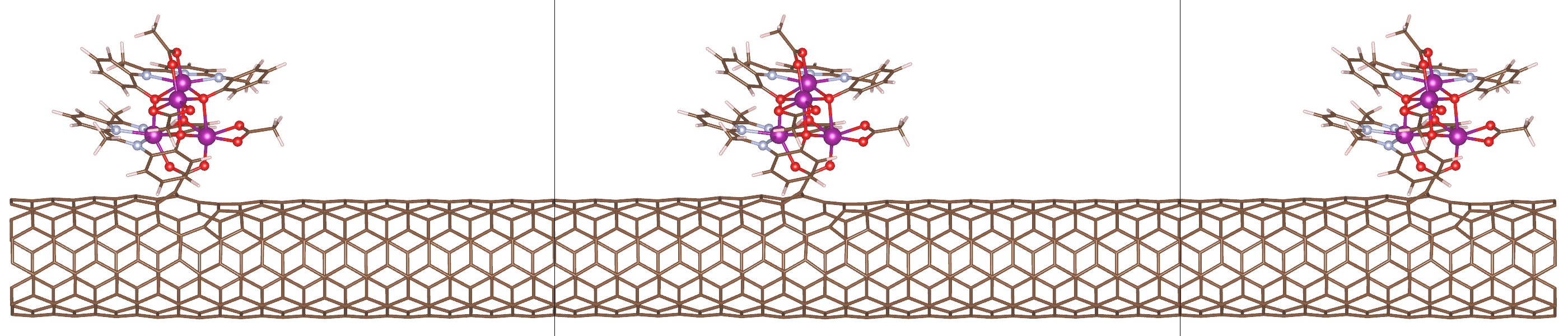}};
    \node[below right=0.0cm and 0.2cm of a.north east] (c) {\includegraphics[trim={0 4cm 0 0},clip,height=4.7cm]{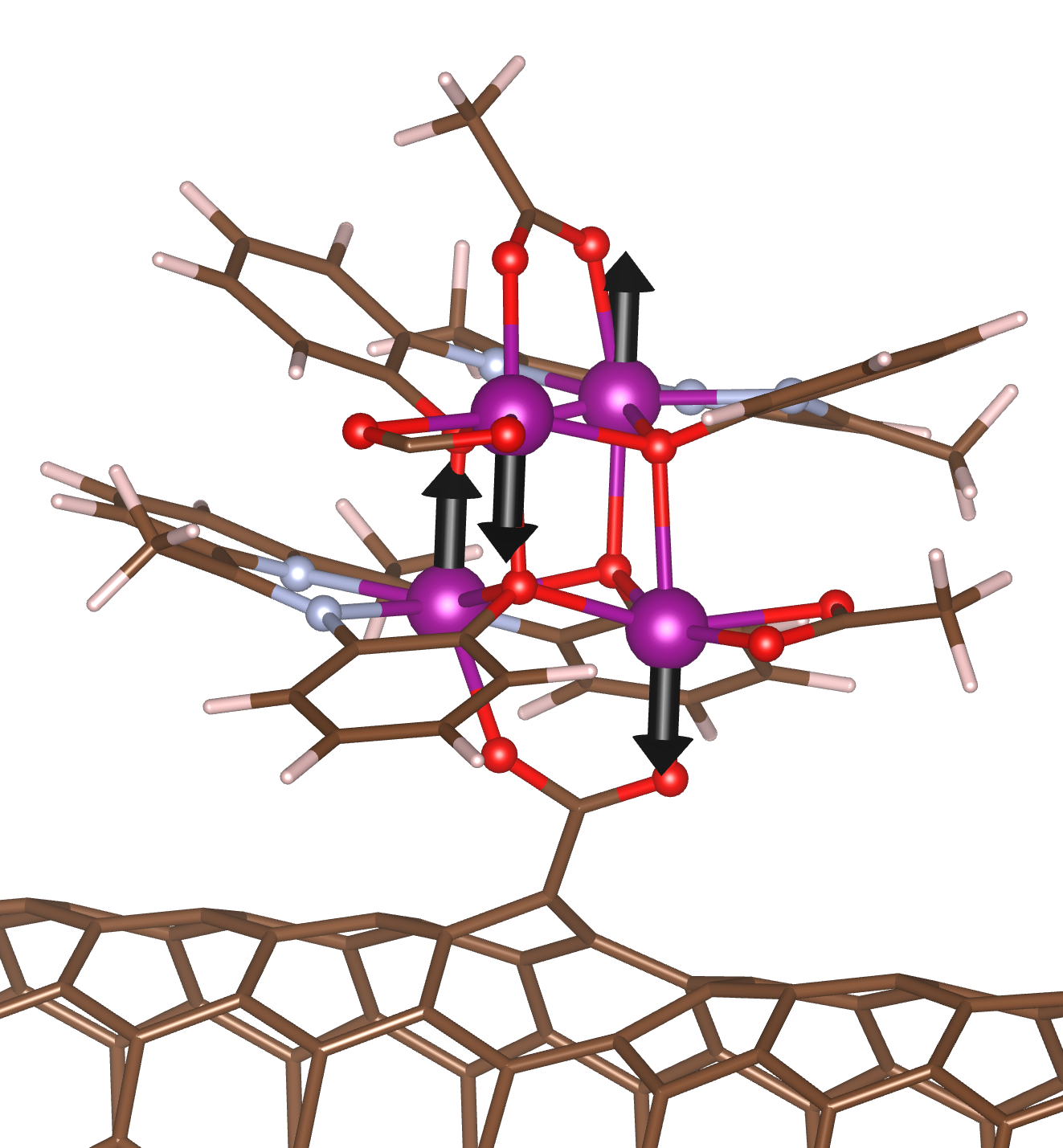}};
    
    \begin{scope}[on background layer]
        \fill[MainColor,opacity=0.3] (-7.5,-1.6) rectangle ++(0.8,1.5);
        \fill[MainColor,opacity=0.3] ( 7.5,-1.6) rectangle ++(-0.8,1.5);
        \node[single arrow, draw=MainColor, fill=MainColor, fill opacity=0.5,
                shape border rotate=180,
                single arrow head extend=3pt,
                minimum width = 8mm, 
                minimum height = 6mm] at (-8,-0.9) {}; 
        \node[single arrow, draw=MainColor, fill=MainColor, fill opacity=0.5,
                single arrow head extend=3pt,
                minimum width = 8mm, 
                minimum height = 6mm] at ( 8,-0.9) {}; 
        \node at (-6.9,-1.9) {\small Electrode};
        \node at ( 0.0,-1.9) {\small Scattering Region};
        \node at ( 6.9,-1.9) {\small Electrode};
    \end{scope}
    \end{tikzpicture}
    \caption{Periodic boundary conditions (top left) and open system (bottom) set-up for first-principles simulations of \{M$_4$\}-CNT hybrid systems. Zoom of the attached molecule and alignment of the magnetic moments of its magentic cores (top right).}
    \label{fig:06-CNT-setup}
\end{figure}
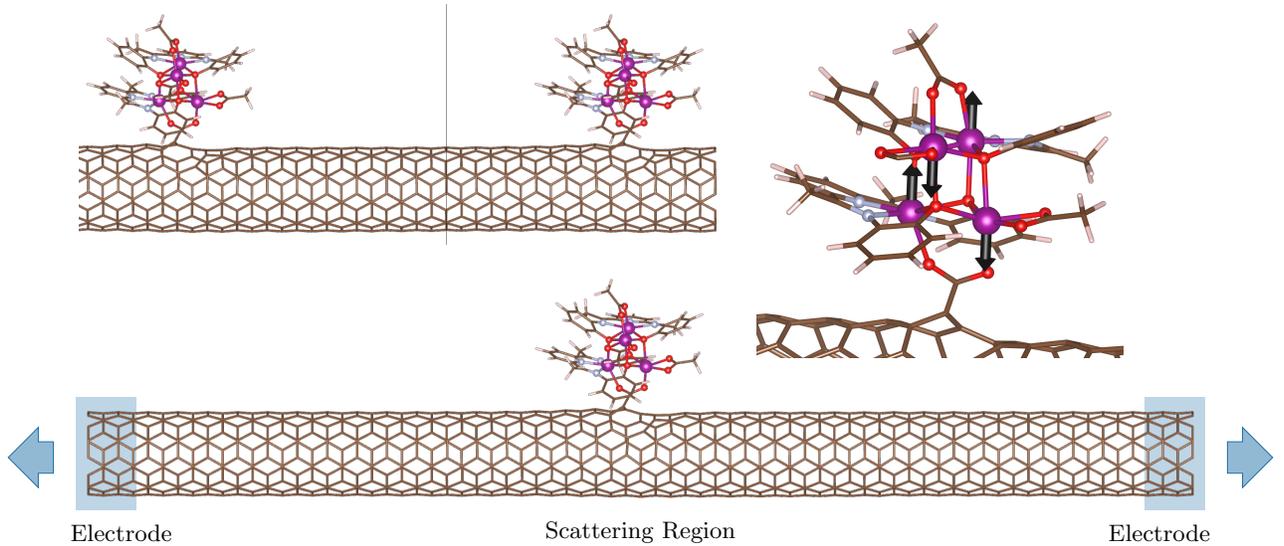

% \Nils{Should this section be cut?}
In this section, we study carbon nanotube functionalized with tetranuclear clusters \{M$_4$\}=[M$_4$(H$_2$L)$_2$(OAc)$_4$] (Fig.~\ref{fig:06-CNT-setup}). These tetranuclear clusters consist of 4 magnetic atoms linked by two different ligands H$_2$L (2,6-bis-(1-(2-hydroxyphenyl)iminoethyl)pyridine) and acetate C$_2$H$_3$O$_2^-$. The core of the cluster is formed by the four magnetic and four oxygen atoms which are arranged in nearly cubic structures. In the magnetic ground state of the cluster, the magnetic atoms with an acetate bridge are antiferromagnetically aligned and the pairs without the acetate bridge are aligned ferromagnetically~\cite{KampertLigandControlledMagnetic2009,AchilliMagneticProperties2022}. Thus, the whole cluster is a molecular antiferromagnet and could be used in antiferromagnetic spintronic devices such as magnetic random access memories or THz information technologies. To make such applications possible, multiple clusters need to be electrically contacted and linked together, for example, using metallic carbon nanotubes\cite{BessonIndependentCoherent2023}. Our collaborators have synthesized and characterized such \{M$_4$\} CNT devices with manganese (Mn) and cobalt (Co) as magnetic centers. Statistical analysis of the random telegraph signal in current passing through these devices indicates that \{Mn$_4$\} clusters exhibit long-lived coherent states, and \{Co$_4$\} do not. These excitation states correspond to excitation of non-degenerate $S_{\mathrm{total}}=0$ eigenstates of the cluster and do not involve spin flips. These excited states can not be modeled with our first-principles methods. However, we can study the interaction between the cluster and nanotube, to understand the difference in the coupling of the two clusters to the nanotube. For this purpose, we have performed DFT simulations of carbon nanotubes functionalized with a single \{Mn$_4$\} and \{Co$_4$\} clusters.

\begin{table}
    \centering
    \begin{tabular}{cccc}
         Cluster  & Cell size (\#CNT Rings) & $\abs{S}$ - PBC [$\mu_B$] & $\abs{S}$ Open System [$\mu_B$] \\\hline
         Mn$_4$ & 15 & 0.231 & --    \\
         Mn$_4$ & 23 & 0.173 & 0.028 \\
         Mn$_4$ & 27 & 0.083 & 0.028 \\
         Mn$_4$ & 31 & 0.057 & 0.030 \\ \hline
         Co$_4$ & 15 & 0.176 & --    \\
         Co$_4$ & 23 & 0.258 & 0.150 \\
         Co$_4$ & 27 & 0.447 & 0.147 \\
         Co$_4$ & 31 & 0.333 & 0.139 \\ \hline
    \end{tabular}
    \caption{Magnetic moments of carbon nanotubes functionalized with tetranuclear clusters (\{Mn$_4$\} and \{Co$_4$\}) computed in the LDA+U approximation with periodic boundary conditions (PBC) and with Green function techniques (open System) for different cell sizes. The units of the magnetic moment are Bohr magnetons (µB).}
    \label{tab:06-CNT-magn-moments}
\end{table}

We consider two simulation set-ups: a supercell with periodic boundary conditions (PBC), and an open system (OPEN) in which a single molecule is attached to an infinite tube (Fig.~\ref{fig:06-CNT-setup}). The latter allows us to ascertain the nature of the interaction between the molecule and the CNT by avoiding spurious interaction between the periodic replica of the molecule, mediated by the electronic states of the metallic CNT~\cite{CostaIndirectExchange2005,KirwanSuddenDecay2008,ZanolliSpinTransport2010}. In our simulations the \{M$_4$\} clusters are bound to the dangling carbon atom of a mono-vacancy in a metallic (5,5) carbon nanotube. The bond between the cluster and the nanotubes is formed by a -CO$_2^-$ group obtained by removing the C$H_3$ group from one of the acetate groups in the cluster. Previous studies have demonstrated that the mono-vacancy site is especially favorable for the functionalization of CNTs with molecules~\cite{ZanolliDefectiveCarbon2009} or magnetic nanoparticles~\cite{ZanolliSingleMoleculeSensing2012}.

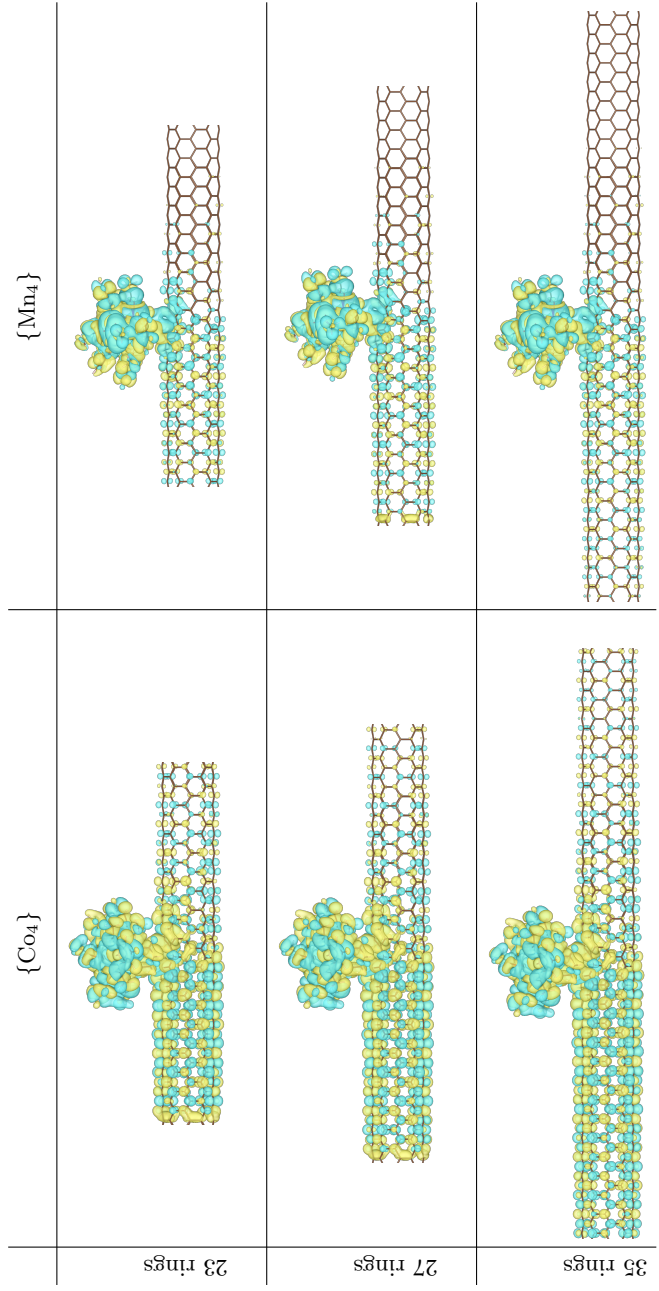
\begin{sidewaysfigure}
    \centering
    \begin{tikzpicture}
    \node[]        (a) at (0,0) {\includegraphics[height=0.85in]{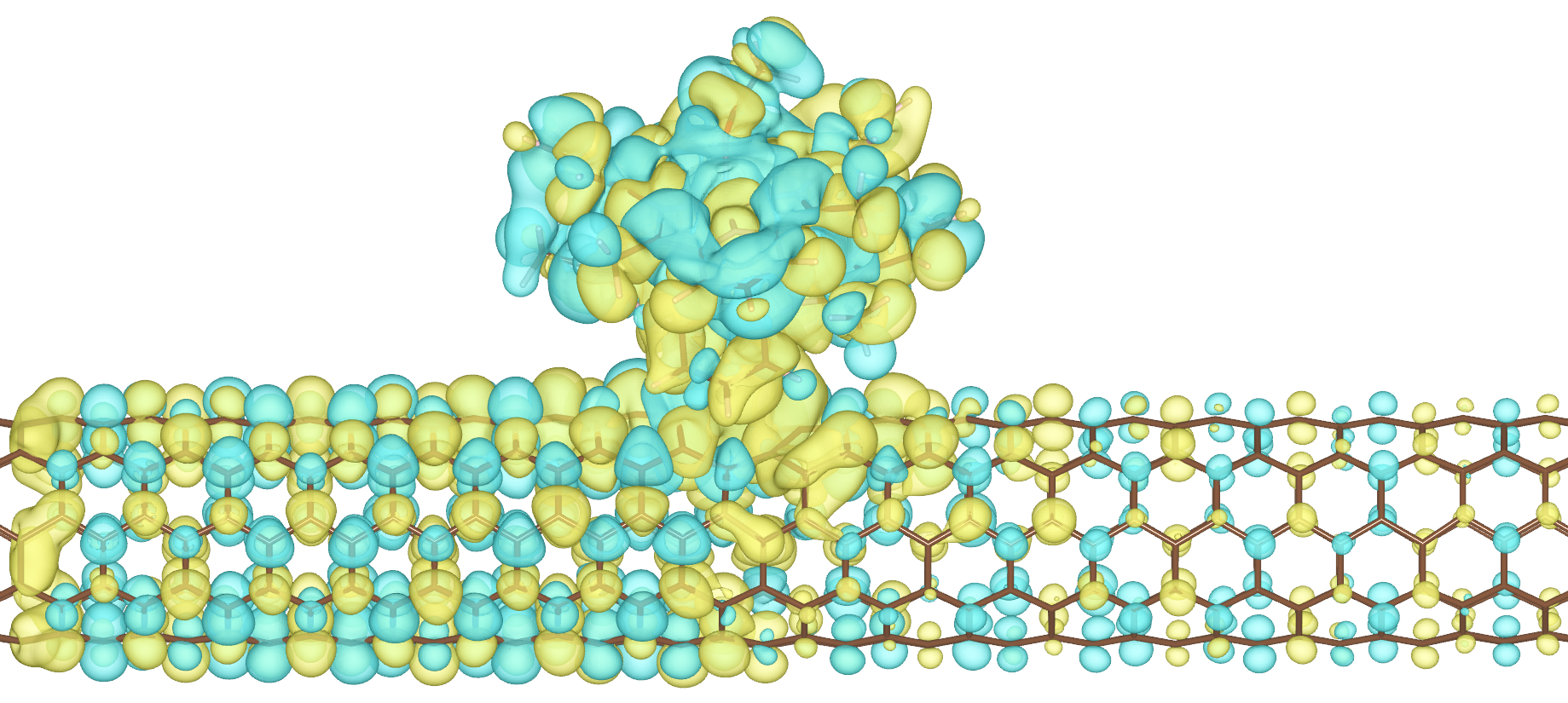}};
    \node[below=0.4cm of a] (c) {\includegraphics[height=0.85in]{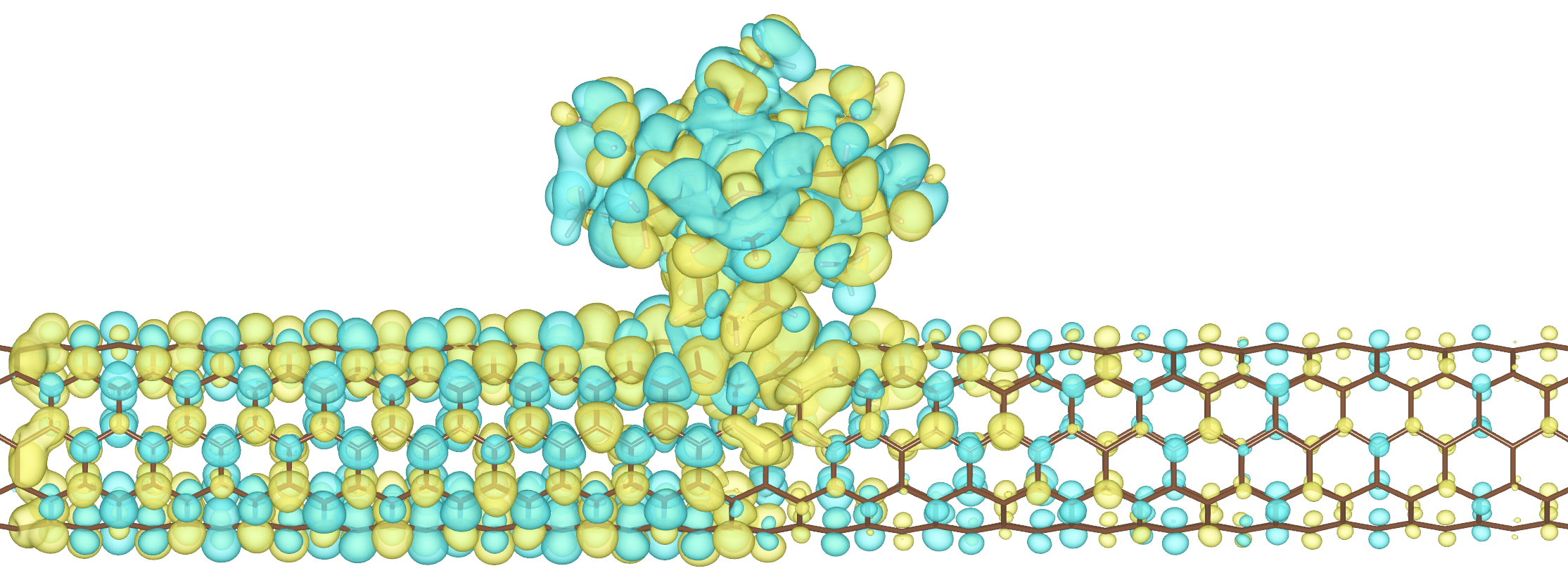}};
    \node[below=0.4cm of c] (e) {\includegraphics[height=0.85in]{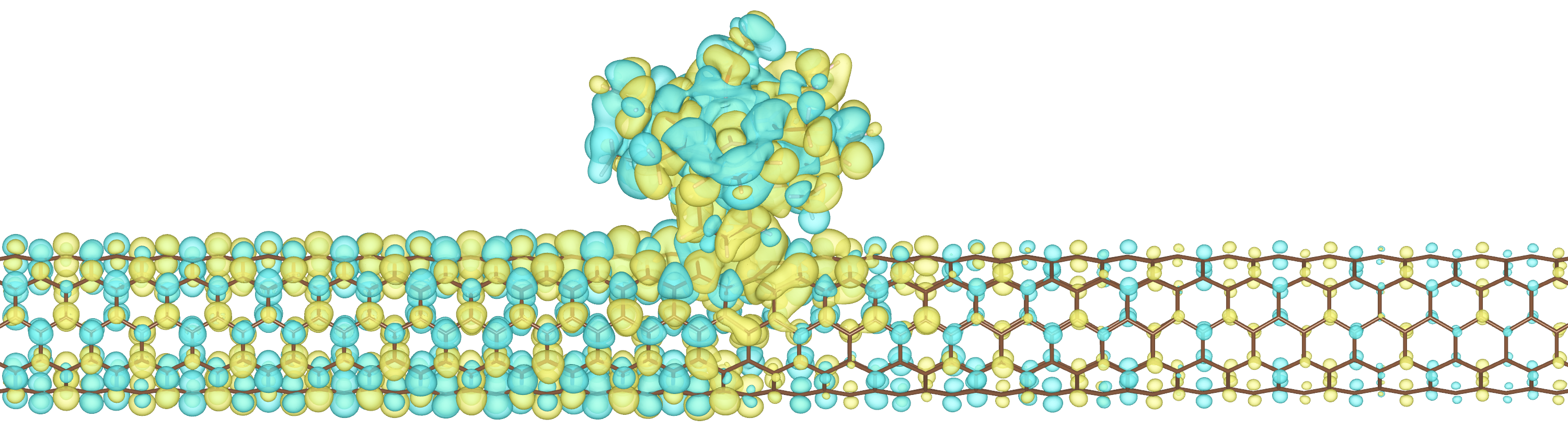}};
    \node[right=0.4cm of e] (f) {\includegraphics[height=0.85in]{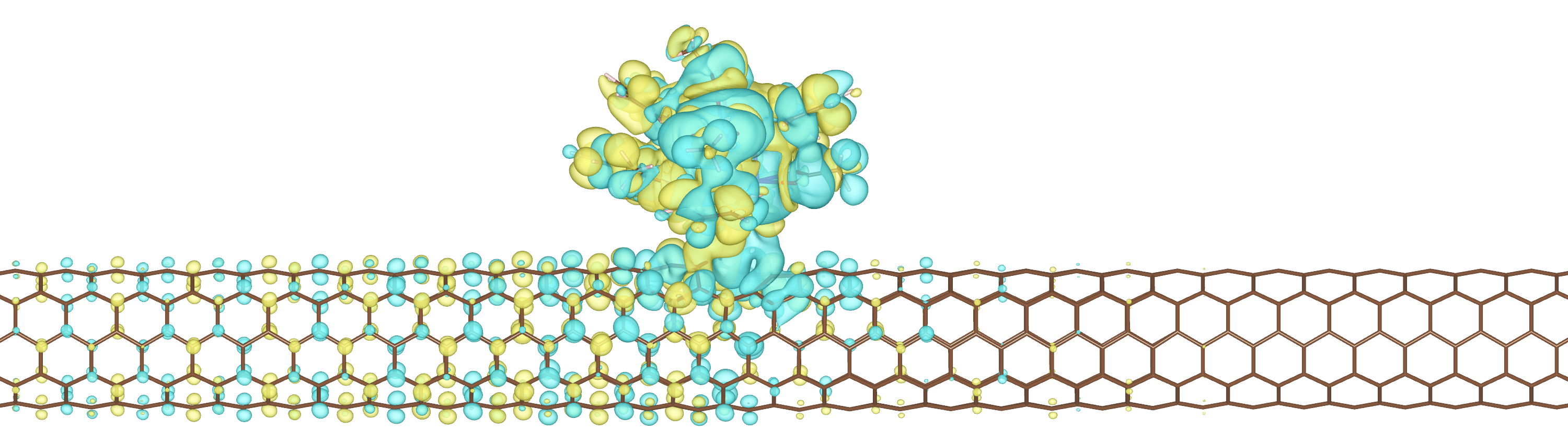}};
    \node[above=0.4cm of f] (d) {\includegraphics[height=0.85in]{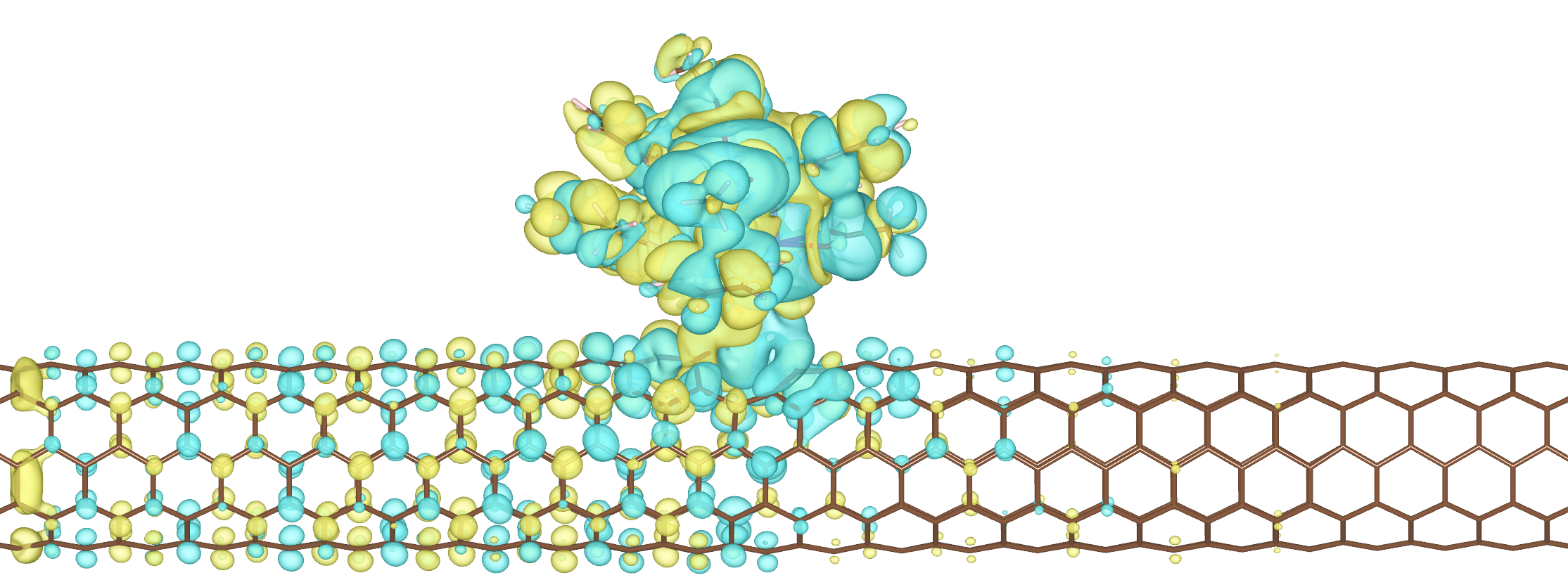}};
    \node[above=0.4cm of d] (b) {\includegraphics[height=0.85in]{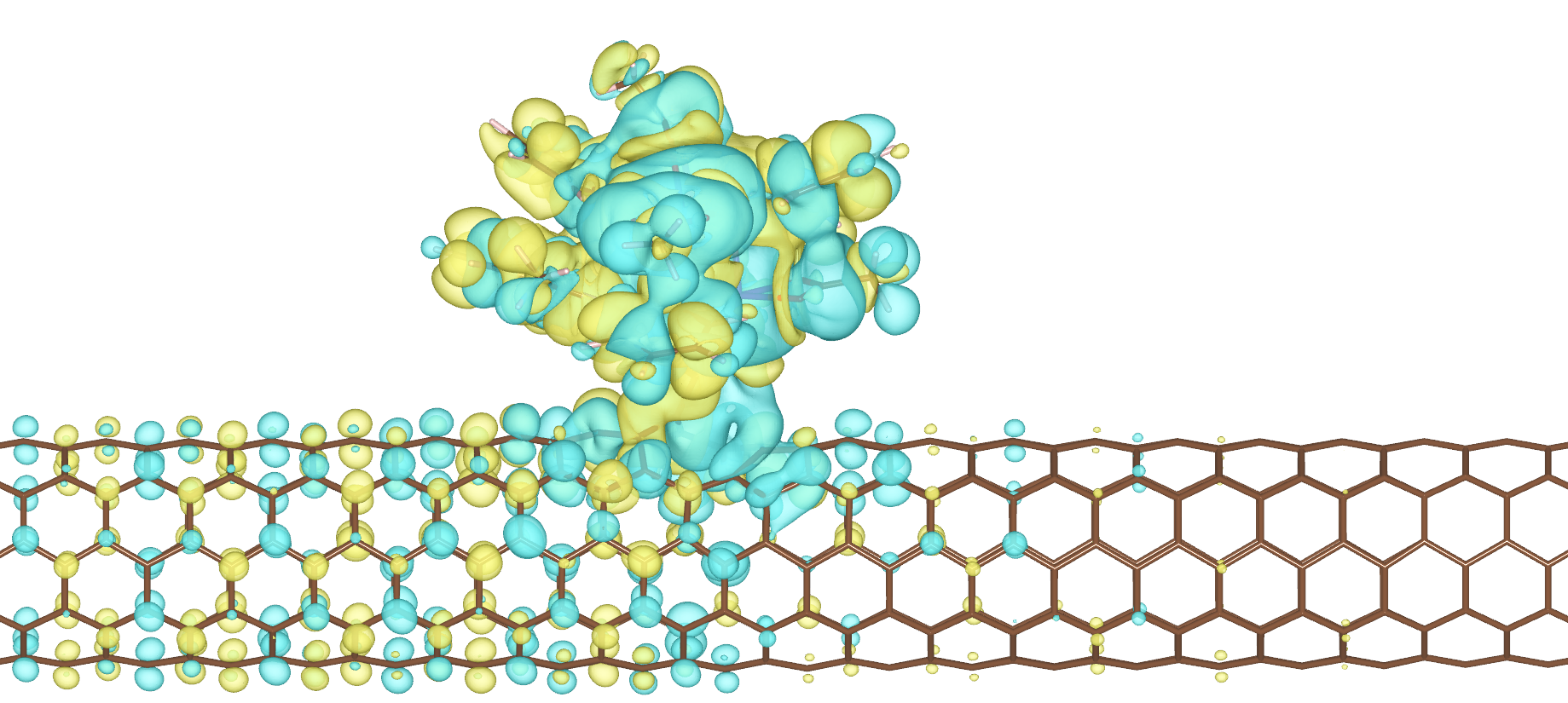}};
    \node[anchor=south west,rotate=90] (t1) at (e.south west) {\small 35 rings};
    \node[anchor=south west,rotate=90] (t2) at (t1.south |- c.south) {\small 27 rings};
    \node[anchor=south west,rotate=90] (t2) at (t1.south |- a.south) {\small 23 rings};
    \node[above=0.1cm of a] (Mn4) {\small \{Co$_4$\}};
    \node[above=0.1cm of b] (Co4) {\small \{Mn$_4$\}};
    \draw (t1.north |- a.north) -- (f.east |- a.north);
    \draw (t1.north |- c.north) -- (f.east |- c.north);
    \draw (t1.north |- e.north) -- (f.east |- e.north);
    \draw (e.west   |- Co4.north) -- (e.south west);
    \draw (f.west   |- Co4.north) -- (f.south west);
    \end{tikzpicture}
    \caption{Spin density of \{Co$_4$\}-CNT (left) and \{Mn$_4$\}-CNT (right) calculated with periodic boundary conditions; (top) LDA+U and collinear spin, (middle) LDA and collinear spin, and (bottom) LDA and spin-orbit coupling. Isosurface at 0.001 $\mu_B/$\AA$^3$.  }
    \label{fig:06-CNT-spin-densities}
\end{sidewaysfigure}

\begin{figure}
    \centering
    \begin{tikzpicture}
        \node[]        (a) at (0,0) {\includegraphics[width=0.45\textwidth]{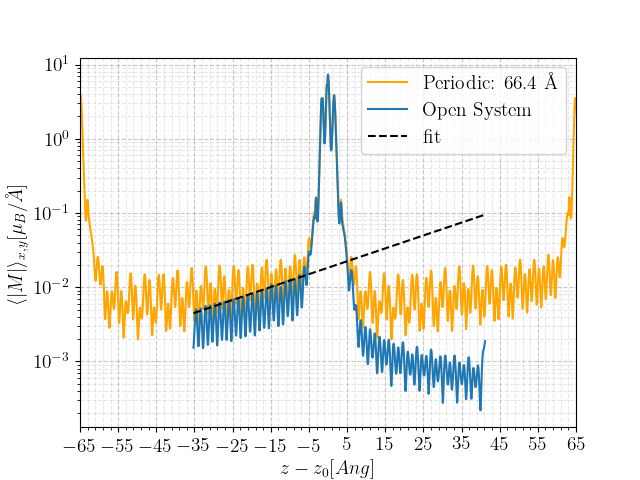}};
        \node[right=0.1cm of a] (c) {\includegraphics[width=0.45\textwidth]{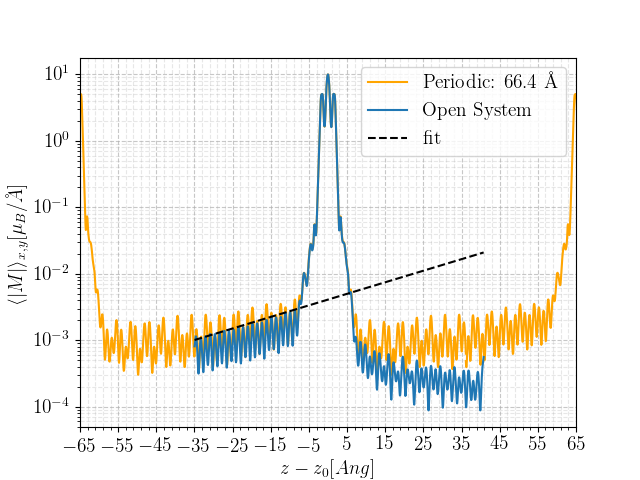}};
    \end{tikzpicture}
    \caption{(Open System) Absolute value of spin density integrated over x and y (z = 0 located at the molecule) for the \{Co$_4$\}-CNT (left) and \{Mn$_4$\}-CNT (right). The decay of the spin density in the open system is exponential: \{Co$_4$\}: $e^{(0.04 z - 4)}$ / \{Mn$_4$\}: $e^{(0.04 z - 5.5)}$. The asymmetry in the blue curve arises due to the asymmetry of the molecule-CNT link. This effect is not visible in PBC due to the periodic repetition of the grafted molecules.}
    \label{fig:06-CNT-decay}
\end{figure}

We compare the magnetic moments obtained in the open system setup with the values calculated for the periodic system by increasing the PBC cell size. We verified that the magnetic moment of \{Mn$_4$\}-CNT tends to the open system solution (0.03~$\mu_B$) while this is not the case for \{Co$_4$\}-CNT, where the PBC magnetic moment is not converged even in the for a simulation cell containing 35 unit cells of the nanotube (Table \ref{tab:06-CNT-magn-moments} and Fig.~\ref{fig:06-CNT-spin-densities}). This happens due to the long-range character of the indirect exchange coupling between magnetic clusters mediated by the conduction electrons of metallic carbon nanotubes~\cite{CostaIndirectExchange2005}. Therefore, we base our analysis on the spin and orbital moment in the OPEN system setup. From the latter, we find that the total magnetic moment of the \{Co$_4$\}-CNT system is 0.14~$\mu_B$, an order of magnitude larger than in the \{Mn$_4$\}-CNT case. The net magnetization of the cluster is almost one order of magnitude smaller than the total magnetization of the hybrid systems. The remaining magnetization is located on the linking carbon atom and nanotube. The spin moment induced in the tube decay exponentially from the vacancy site (Figure \ref{fig:06-CNT-decay}). The larger net magnetization of the \{Co$_4$\}-CNT indicates that the nature of the interaction between \{Co$_4$\} molecules is longer range than \{Mn$_4$\}, which is reflected in the interaction between the periodic replica. On the other hand, the decay of the magnetic perturbation in the two systems is similar (Fig.~S5), as it is determined by the conduction electrons of the nanotube. 

\begin{figure}
    \centering
    \includegraphics[width=\textwidth]{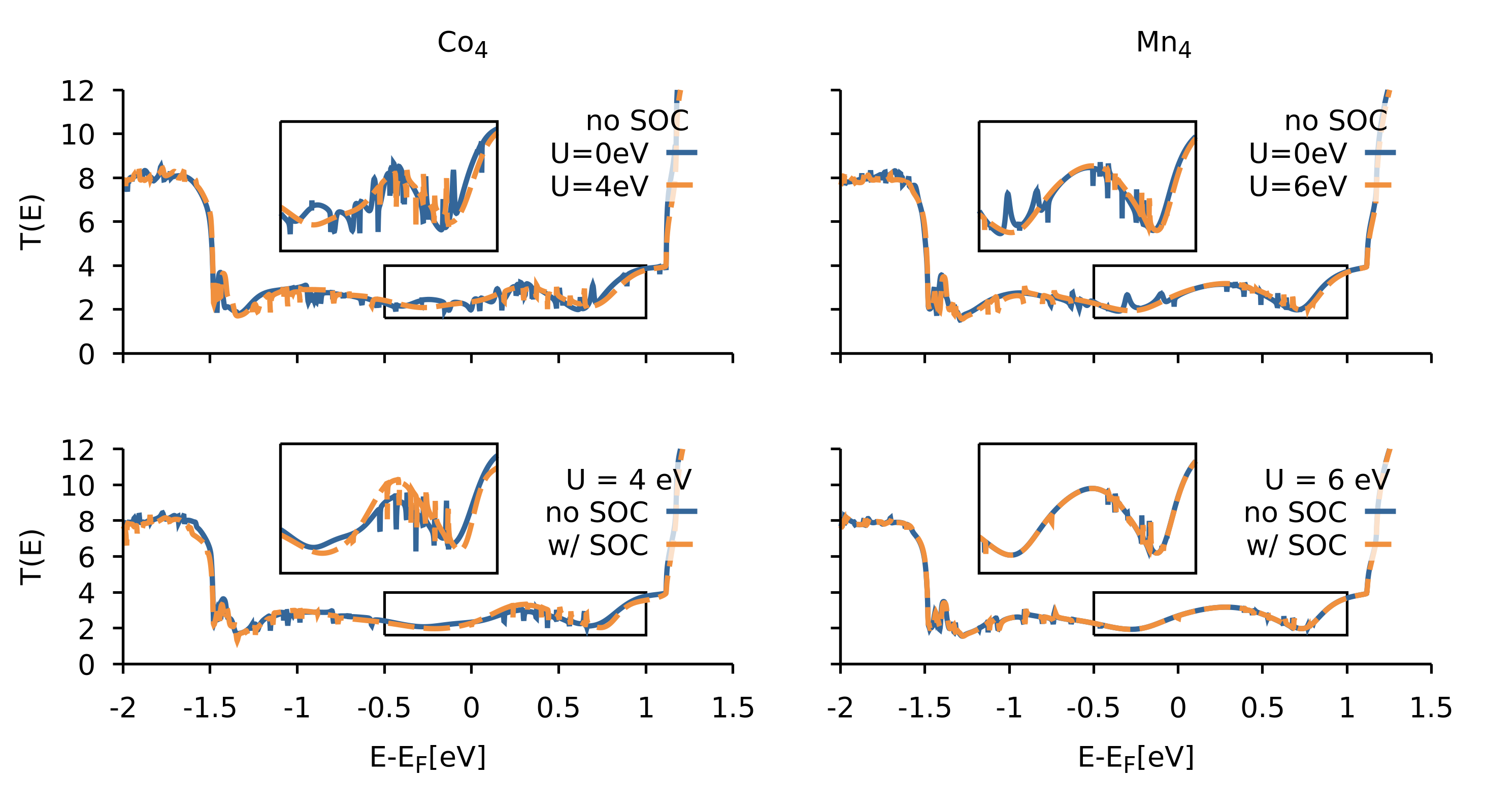}
    \caption{Transmission function of \{Co$_4$\}-CNT (a,c) and \{Mn$_4$\}-CNT (b,d); (a) and (b) comparison of the effect of Hubbard-like correction on the transmission function (without SOC); (c) and (d) comparison of the effect of SOC (without Hubbard U).}
    \label{fig:06-CNT-transmission}
\end{figure}

The transmission function of the functionalized nanotube (Figure \ref{fig:06-CNT-transmission}) closely resembles that of (5,5) carbon nanotubes with a tilted divacancy~\cite{ZanolliSpinTransport2010}. Around the Fermi level, the transmission is reduced by a factor of two compared to that of the pristine. Near the edge of the step, transmission increases and is slightly lower in the central part of the step. The transmission function is distinct from that of a (5,5) CNT with a monovacancy. The shape of the transmission function is largely insensitive to the type of cluster attached to it.
In Fig.~\ref{fig:06-CNT-transmission} we compare the transmission function of the two clusters calculated with and without SOC and Hubbard-like correction (Hubbard U). For both systems, the inclusion of the Hubbard U term affects the transmission function near the Fermi level. Since the Hubbard U is only included for the magnetic atoms in the cluster, this implies that states on the cluster are at least partially coupled to the electronic structure of the carbon nanotube. The effect of SOC on the transmission function is negligible for the \{Mn$_4$\}-CNT system, but significant in the \{Co$_4$\}-CNT, which is consistent with the expectation of larger orbital angular moment in Co. 

\begin{figure}
    \centering
    \includegraphics[width=\textwidth]{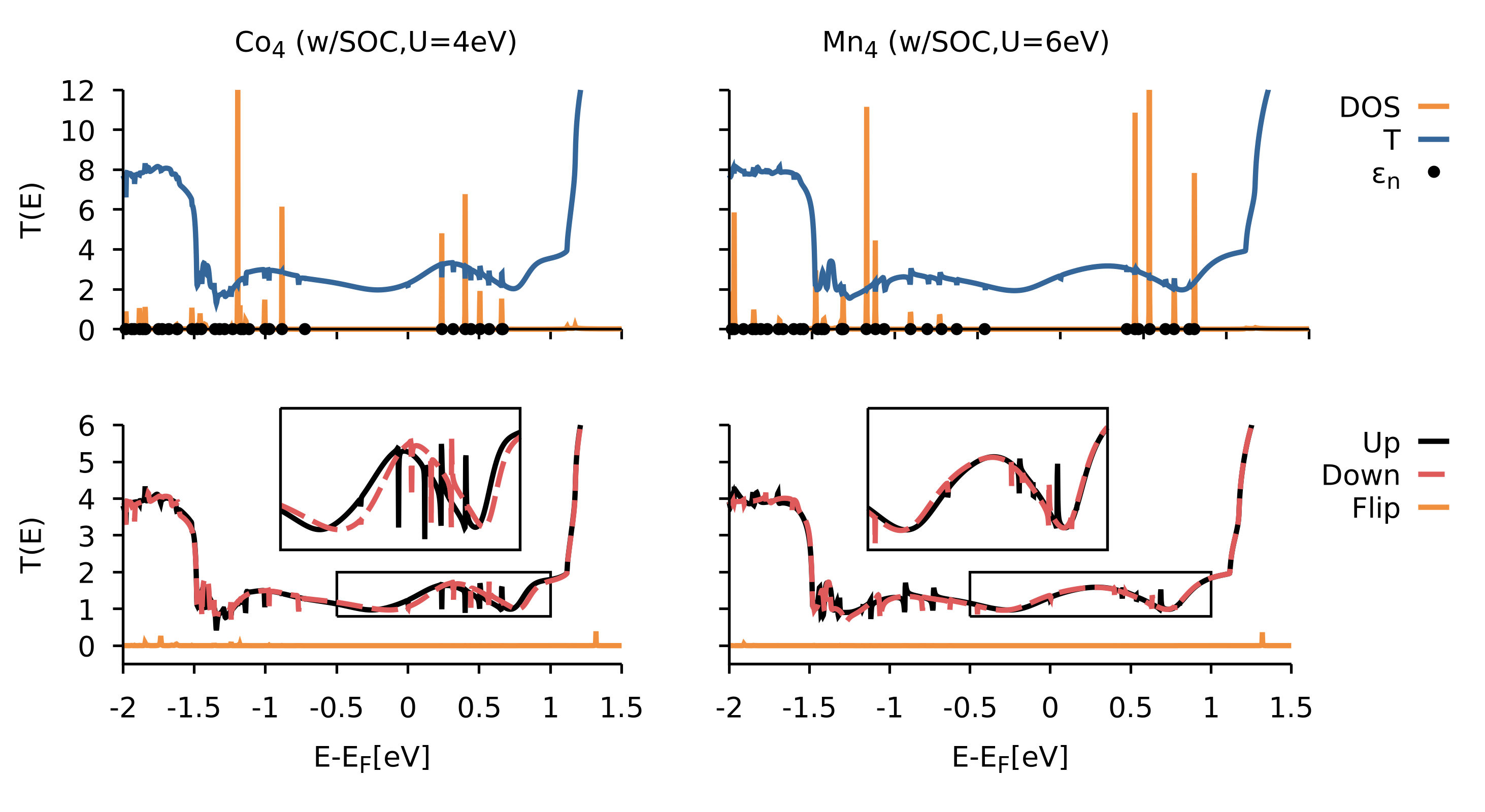}
    \caption{Bound density of states and spin-channel-projected transmission of \{Co$_4$\}-CNT (a,c) and \{Mn$_4$\}-CNT (b,d) calculated with SOC and with Hubbard U; (a) and (b) comparison of the eigenenergies $\epsilon_n$ of the isolated molecule (black dots), bound density of states (orange line) and the zero-bias transmission function $T$ (blue line); (c) and (d) spin-channel-projected transmission functions for the spin-up channel (black line), spin-down channel (red line) and spin-flip channels (orange line).}
    \label{fig:06-CNT-transmission-2}
\end{figure}

Both systems also exhibit a series of dips and a spikes transmission function. These dips are not predicted for the transmission function of (5,5) nanotubes with a divacancy\cite{ZanolliSpinTransport2010}, and provide further evidence of coupling between molecular states on the cluster and carbon nanotube. The number of positions of these features depends on the type of cluster attached to the nanotube and the extent of the inclusion of the Hubbard correction. 
Each of these features corresponds to an eigen state of the isolated molecule and a subset to bound states in the cluster (Fig.~\ref{fig:06-CNT-transmission-2}). We can identify bound states in the hybrid system by calculating the density of states from the Green function in the scattering region ($\rho(E)=\Tr{\mathbf{G}S}$) and the spectral functions for the two electrodes ($\rho^{\mathcal{A}}_{\mathfrak{e}}(E)=\Tr{\mathcal{A}_{\mathfrak{e}}S}$). The density of states calculated from the spectral functions only includes states that couple to the electrode. The density of states calculated from the Green function includes all states in the scattering region. As such, we can identify bound states from the difference of densities\cite{PapiorImprovementsNonequilibrium2017} ($\rho^{\mathrm{bound}}=\rho(E)-\sum_{\mathfrak{e}}\rho^{\mathcal{A}}_{\mathfrak{e}}(E)$). In Fig.~\ref{fig:06-CNT-transmission-2} we can see that the bound states are the coincides with some of the sharp features in the transmission function. 

Finally, we analyze the spin-dependent properties of the functionalized nanotubes~(Fig.~\ref{fig:06-CNT-transmission}). The transmission function for the spin-up and down channels is similar for both systems. The \{Co$_4$\}-CNT presents small, slightly larger, difference particularly near the Fermi level.
Above the Fermi level, the transmission function for the spin-down channel is shifted to higher energies. This difference in the transmission of the up and down channels arises due to the net magnetization of the system. The \{Mn$_4$\}-CNT does not exhibit this effect because its net magnetization is significantly lower. In both systems, the position of the sharp feature in the transmission is distinct for the two spin channels.  This indicates that electric manipulations of the magnetic states of the molecule could be possible. Scattering spin flip transmission is negligible in both systems. 

\subsubsection{Computational Details}
To account for the strong correlation of the 3d electrons and avoid the excessive delocalization of the d-states predicted in the Local Density Approximation, Hubbard-like corrections with U=6~eV and U=4~eV were used for Mn and Co, respectively. The same value was used by \textcite{KampertLigandControlledMagnetic2009} in their calculations on \{Mn$_4$\}. A standard double-zeta polarized (DZP) basis set was used for carbon, nitrogen, and hydrogen, and an optimized double-zeta (DZ) for Mn, Co, and O. Calculations were spin-polarized and performed assuming collinear spins. Convergence of electronic structure and magnetic properties was achieved for a real-space grid cut-off of 400 Ry, and a Fermi-Dirac smearing of 100 K in the LDA+U calculation, while with SOC a cut-off of 650 Ry and electronic temperature of 1 K was adopted. The atomic positions were relaxed using standard periodic boundary conditions with a 1 × 1 × 12 $\mathbf{k}$ points sampling of the Brillouin zone for 15 cells of {M4}-CNT (shifted grid), and the conjugate gradient algorithm. The maximum force on atoms was smaller than 0.04 eV/Å for the CNT+{M4} system. 

Simulations with PBC were performed for 15, 23, 27, and 35 armchair (5, 5) supercells to analyze the long-range decay of the induced spin polarization in the system. The supercells correspond to a distance between the graft points of 36.9, 56.6, 66.5, 86.2~Å, and a separation between the molecules of ~23.5, 43.0, 53.0, 73.0~Å, as the lateral size of the molecule is about 13.5~Å. We used a sampling of the Brillouin zone equivalent to 1\ttimes 1 \ttimes 12 k points for 15 cells of \{M$_4$\}-CNT (shifted grid). The smallest cell was adopted to find the magnetic configuration in the ground state of the grafted molecules (Table S1 and S2).

\section{Conclusion}
We have presented the non-equilibrium Green function approach to quantum transport and its implementation at the DFT+NEGF level for systems with noncollinear spins and spin-orbit coupling. This makes TranSIESTA an open-source DFT+NEGF code capable of performing non-equilibrium, multi-terminal transport in the presence of noncollinear spin configuration and general spin-orbit phenomena. This new implementation can utilize all highly scaleable and effective algorithms that were recently added to the code~\cite{PapiorImprovementsNonequilibrium2017}, and is suitable for large-scale transport simulations. As such, this implementation represents a step forward for simulations of quantum materials and spintronics devices at large. 

We applied our new implementation to a series of systems to demonstrate possible applications and to show the validity of the code. In particular, we calculate the anisotropic magnetoresistance in one-dimensional iron chains and a Fe/MgO/Fe tunneling junction. Furthermore, we study quantum transport in TMD nanodevices, which are characterized by strong spin-orbit coupling.

\section{Acknowledgements}

Work supported by the EU-H2020 MaX ``Materials Design at the Exascale'' CoE (Grant No. 824143), the Spanish MINECO, MICIU, AEI and EU FEDER (Grant No. PGC2018-096955-C43). ICN2 is funded by the CERCA Program (Generalitat de Catalunya) and the Severo Ochoa Program from Spanish MINECO (Grant No. SEV-2017-0706). NW has received funding from the EU-H2020 research and innovation programme under the Marie Sklodowska-Curie programme (Grant No. 754558). 
ZZ acknowledges financial support by the Ramon y
Cajal program MINECO AEI and FSE-UE (RYC-2016-19344), the Netherlands Sector Plan program 2019–2023, 
the research program “Materials for the Quantum Age (QuMat)”   financed by the Dutch Ministry of Education, Culture and Science (OCW) with registration nr. 024.005.006, and the European Union’s Horizon Europe research and innovation programme under grant agreement No 101130384 (QUONDENSATE).
We acknowledge computing resources on MareNostrum4 at Barcelona Supercomputing Center (BSC) and Discoverer in Sofia, Bulgaria, provided through the Tier-0 PRACE Research Infrastructure Project Access (OptoSpin project id. 2020225411). We also acknowledge computing time from RES (activity FI-2020-1-0014) and technical support provided by the Barcelona Supercomputing Center.
This work was sponsored by NWO-Domain
Science for the use of supercomputer facilities through projects "Quantum Materials by Design" with nr 2022.012, and EINF-1858.

% \nocite{*}
\bibliographystyle{Frontiers-Vancouver}
\bibliography{main}

\end{document}